\documentclass[12pt, a4paper]{article}
\usepackage{graphicx}
\usepackage{subeqnarray}
\begin{document}
%
%
%
%
\def\astrobj#1{#1}
\newenvironment{lefteqnarray}{\arraycolsep=0pt\begin{eqnarray}}
{\end{eqnarray}\protect\aftergroup\ignorespaces}
\newenvironment{lefteqnarray*}{\arraycolsep=0pt\begin{eqnarray*}}
{\end{eqnarray*}\protect\aftergroup\ignorespaces}
\newenvironment{leftsubeqnarray}{\arraycolsep=0pt\begin{subeqnarray}}
{\end{subeqnarray}\protect\aftergroup\ignorespaces}
\newcommand{\diff}{{\rm\,d}}
\newcommand{\img}{{\rm i}}
\newcommand{\sV}{\mskip 3mu /\mskip-10mu V}
\newcommand{\sP}{\mskip 3mu /\mskip-10mu p}
\newcommand{\sT}{\mskip 3mu /\mskip-08mu T}
\newcommand{\sX}{\mskip 3mu /\mskip-12mu X}
\newcommand{\appleq}{\stackrel{<}{\sim}}
\newcommand{\appgeq}{\stackrel{>}{\sim}}
\newcommand{\Int}{\mathop{\rm Int}\nolimits}
\newcommand{\Nint}{\mathop{\rm Nint}\nolimits}
\newcommand{\arcsinh}{\mathop{\rm arcsinh}\nolimits}
\newcommand{\range}{{\rm -}}
\newcommand{\sgn}{\mathop{\rm sgn}\nolimits}
\newcommand{\displayfrac}[2]{\frac{\displaystyle #1}{\displaystyle #2}}
\def\astrobj#1{#1}
%
\title{A principle of corresponding states for two-component,
self-gravitating fluids}
\author{{R.~Caimmi}\footnote{
{\it Astronomy Department, Padua Univ., Vicolo Osservatorio 2,
I-35122 Padova, Italy}
email: roberto.caimmi@unipd.it~~~
fax: 39-049-8278212}
%
%
\phantom{agga}}
%
%
\maketitle
\begin{quotation}
\section*{}
\begin{Large}
\begin{center}

Abstract

\end{center}
\end{Large}
\begin{small}

\noindent\noindent

Macrogases are defined as two-component,
large-scale celestial objects where the
subsystems interact only via gravitation.
The macrogas equation of state is formulated
and compared to the van der Waals (VDW)
equation of state for ordinary gases.
By analogy, it is assumed that real
macroisothermal curves in macrogases
occur as real isothermal curves in
ordinary gases, where a phase transition
(vapour-liquid observed in ordinary gases and
gas-stars assumed in macrogases) takes
place along a horisontal line
in the macrovolume-macropressure
$({\sf O}\sX_{\rm V}\sX_{\rm p})$ plane.
The intersections between real and theoretical
(deduced from the equation of state) macroisothermal
curves, makes two regions
of equal surface as for ordinary gases
obeying the VDW equation of state.
A numerical algorithm is developed
for determining the following points
of a selected theoretical macroisothermal curve
on the $({\sf O}\sX_{\rm V}\sX_{\rm p})$ plane:
the three intersections with the related real
macroisothermal curve,
and the two extremum points (one maximum and
one minimum).   Different kinds of macrogases are
studied in detail: UU, where U density
profiles are flat, to be conceived as a simple
guidance case; HH, where H density profiles
obey the Hernquist (1990) law, which satisfactorily
fits to observed spheroidal
components of galaxies; HN/NH, where N density
profiles obey the Navarro-Frenk-White (1995,
1996, 1997) law, which satisfactorily fits to
simulated nonbaryonic dark matter haloes.
A different trend is shown by theoretical
macroisothermal curves on the
$({\sf O}\sX_{\rm V}\sX_{\rm p})$ plane,
according if density profiles are sufficiently
mild (UU) or sufficiently steep (HH, HN/NH).
In the former alternative, no critical
macroisothermal curve exists, below or above which
the trend is monotonic.   In the latter
alternative, a critical macroisothermal
curve exists as shown by VDW gases, where the
critical point may be defined as the horisontal
inflexion point.   In any case, by analogy with
VDW gases, the first quadrant of the
$({\sf O}\sX_{\rm V}\sX_{\rm p})$ plane may be
divided into three parts, namely (i) the G
region, where only gas exists; (ii) the S region,
where only stars exist; (iii) the GS region,
where both gas and stars exist.   With regard to
HH and HN/NH macrogases, an application is made
to a subsample ($N=16$) of elliptical galaxies
extracted from larger samples $(N=25,~N=48)$
of early type galaxies investigated within the
SAURON project (Cappellari et al. 2006, 2007).
Under the simplifying assumption of universal
mass ratio of the two subsystems, $m$, different
models characterized by different scaled
truncation radii i.e. concentrations, $\Xi_i$,
$\Xi_j$, are considered and the related position
of sample objects on the
$({\sf O}\sX_{\rm V}\sX_{\rm p})$ plane is
determined.   Macrogases fitting to elliptical
galaxies are expected to lie within the S region
or slightly outside the boundary between the
S and the GS region at most.  Accordingly,
models where sample objects lie outside the S
region and far from its boundary, or cannot be
positioned on the
$({\sf O}\sX_{\rm V}\sX_{\rm p})$ plane, are
rejected.   For each macrogas, twenty models
are considered for different values of $(\Xi_i,
\Xi_j, m)$, namely $\Xi_i, \Xi_j=5,$ 10, 20,
$+\infty$ $(\Xi_i, \Xi_j,$ both either finite
or infinite), and $m=10,$ 20.   Acceptable
models are (10, 10, 20), (10, 20, 20),
(20, 10, 20), (20, 20, 20), for HH macrogases,
and (10, 5, 10), (10, 10, 20), (20, 10, 20),
for HN/NH macrogases.   Tipically, fast rotators
are found to lie within the S region, while
slow rotators are close (from both sides) to
the boundary between the S and the GS region.   The net
effect of the uncertainty affecting observed
quantities, on the position of sample objects on
the $({\sf O}\sX_{\rm V}\sX_{\rm p})$ plane, is
also investigated.   Finally, a principle of
corresponding states is formulated for macrogases
with assigned density profiles and scaled
truncation radii.

\noindent
{\it keywords -
galaxies: evolution - dark matter: haloes.}
\end{small}
\end{quotation}

\section{Introduction} \label{intro}

Tidal interactions between neighbouring objects span across
the whole admissible range of lenghts in nature: from, say,
atoms to cluster of galaxies i.e. from micro to macrocosmos.
The role of tidal interactions is of basic importance in
driving a wide variety of physical phenomena.   In dealing
with microscosmos, tidal forces between molecules are
responsible for the occurrence of the liquid and solid phase,
and the presence of a triple point where the gas, liquid,
and solid phase coexist, for an assigned homogeneous substance
(e.g., Landau \& Lifchitz, 1967, Chaps.\,VII-VIII, hereafter
quoted as LL67).   In dealing
with ordinary cosmos, the tidal action of a white dwarf star
on a sufficiently close (filling the whole volume enclosed by
the Roche equipotential surface) red giant companion, makes
mass transfer into the white dwarf until a critical mass is
attained and the star ends its life into a catastrophic SnIa
supernova explosion (e.g., Burrows, 2000).   In dealing with
macrocosmos, the tidal action induced by massive haloes on
hosted galaxies affects their formation and evolution process,
due to a larger depth of the potential well, resulting in a
different correlation of observables with respect to galaxies
in absence of massive halos (e.g., D'Onofrio et al., 2006).

Ordinary fluids are collisional, which makes the stress
tensor be isotropic and the velocity distribution obey
the Maxwell law.   Tidal interactions therein act between
colliding particles (e.g., LL67, Chap.\,VII, \S74).
Astrophysical fluids (leaving aside extremely dense
environments such as galactic nuclei) are collisionless,
which makes the stress tensor be anisotropic and the
velocity distribution do not obey the Maxwell law.
Tidal interactions therein act between a single particle
and the system as a whole.

Given that tidal interactions are at work in both
collisional and collisionless fluids, the existence
of an analogy between the two may be the subject of a
legitimate question.   To this respect, an investigation
must necessarily be restricted to theoretical considerations,
as astrophysical fluids (conceived as macrogases) cannot
be tested in laboratory.   More specifically, a macrogas
equation of state has to be formulated in terms of three
variables (macrovolume, macropressure, macrotemperature),
and the related macroisothermal curves (i.e. the macropressure
as a function of the macrovolume for selected constant
macrotemperatures) has to be compared with their counterparts
deduced from the van der Waals (hereafter quoted as VDW)
equation of state for ordinary gases.
If some analogy exists, it can be extended to (undetectable)
real macroisothermal curves as a working hypothesis.
Finally, an application can be made to galaxies or
clusters of galaxies.

The VDW equation of state can be expressed in (dimensionless)
reduced volume, reduced pressure, and reduced temperature.
Similarly to the Lane-Emden equation for polytropes (e.g.,
Chandrasekhar 1939, Chap.\,IV, \S4; Caimmi, 1986), the
reduced VDW equation holds for a class of fluids instead
of a single fluid.   In general, the states of two systems
with equal values of the reduced variables, are defined as
corresponding states.   According to the principle of
corresponding states, two fluids which obey the reduced VDW
equation of state and exhibit equal values of two among three
reduced variables, necessarily exhibit equal values of the
remaining reduced variable.   For further details refer to
classical textbooks (e.g., LL67, Chap.\,VIII, \S85).

In the light of an analogy between ordinary gases and macrogases,
the formulation of a principle of corresponding states in the
latter case could be highly rewarding.
A macrogas equation of state was formulated in earlier
attempts (Caimmi and Secco, 1990, hereafter quoted as CS90;
Caimmi and Valentinuzzi, 2008, hereafter quoted as CV08),
where the analogy between macrogases and VDW gases was
only mentioned, and isofractional mass $(m={\rm const})$
curves were plotted for a few selected density profiles.
The current paper aims to establish a closer analogy, where
the macrovolume is related to the fractional radius, $y$,
the macropressure to the fractional mass, $m$, and the
macrotemperature to the fractional energy, $\phi$.   The
basic assumptions and the formalism remain unchanged with
respect to the last parent paper (CV08).

The present investigation is mainly devoted to
the following points: (i) expression of an equation
of state for two-component astrophysical fluids,
conceived as macrogases; (ii) comparison between
macroisothermal curves and isothermal curves
related to VDW gases, with regard to a simple
guidance case and two cases which satisfactorily
fit to observations or simulations; (iii)
application to a subsample $(N=16)$ of elliptical
galaxies (CV08), extracted from larger samples
$(N=25,~N=48)$ of early-type galaxies investigated
within the SAURON project (Cappellari et al., 2006,
2007, hereafter quoted as S\,IV, S\,X, respectively).

The work is organized as follows.   The equation
of state of ideal and VDW gases are reviewed,
and related isothermal curves are shown, in Section
\ref{vande}.   A macrogas equation of state is
formulated in terms of macrovolume, macropressure,
macrotemperature, and related macroisothermal
curves are shown for flat and steep density profiles,
in Section \ref{macro}.   An application to elliptical
galaxies for which masses, radii, and rms velocities
can be determined, is performed in Section \ref{aell},
where the selection of acceptable models is made and
an interpretation of the results is outlined.   The
conclusion is drawn in Section \ref{conc}.   Further
details on two specific points are reported in the
Appendix.

\section{Ordinary fluids}\label{vande}

Let ordinary fluids be conceived as fluids where
the effects of gravitation on the equation of state
may safely be neglected
e.g., on the surface of the Earth.   The simplest
description is provided by the theory of ideal gas.

Ideal gases are collisional fluids defined by the
following properties: (i) particles are identical
spheres; (ii) the number of particles is extremely
large; (iii) the motion of particles is random;
(iv) collisions between particles or with the wall
of the box are perfectly elastic; (v) interactions
between particles or with the wall of the box are
null.

The equation of state of ideal gases may be written
under the form (e.g., LL67, Chap.\,IV, \S42):
\begin{equation}
\label{eq:gid}
pV=kNT~~;
\end{equation}
where $p$ is the pressure, $V$ the volume, $T$ the
temperature, $N$ the particle number, and $k$ the
Boltzmann constant.   The product, $pV$, has the
dimensions of an energy and, in fact, the mean kinetic
energy per degree of freedom equals the product,
$(1/2)kNT$, and the mean kinetic energy of motions
along the $X_{\rm p}$ axis reads:
\begin{leftsubeqnarray}
\slabel{eq:kTa}
&& \overline{(E_{\rm kin})_{pp}}=\frac12N\overline{m}\sigma_{pp}^2=
\frac12kNT~~; \\
\slabel{eq:kTb}
&& \overline{m}\sigma_{pp}^2=kT~~;
\label{seq:kT}
\end{leftsubeqnarray}
where $\overline{m}$ is the mean particle mass and
$\sigma_{pp}$ the rms velocity component along the
$X_{\rm p}$ axis.   In the light of the theory of ideal
gases, Eqs.\,(\ref{eq:gid}) and (\ref{seq:kT})
disclose the meaning of the Boltzmann constant:
for fixed pressure, volume, and particle number,
the mean kinetic energy remains unchanged, regardless
of the nature of the gas.

In getting a better description of real gases,
the above assumption (v) is relaxed and interactions
between particles are taken into consideration.
The VDW generalization
of the equation of state of ideal gases, Eq.\,(\ref{eq:gid}),
reads (van der Waals, 1873):
\begin{equation}
\label{eq:VdW}
\left(p+A\frac{N^2}{V^2}\right)(V-NB)=kNT~~;
\end{equation}
where $A$ and $B$ are constants which depend
on the nature of the particles.   More specifically,
the presence of an attractive interaction between
particles reduces both the force and the frequency
of particle-wall collisions: the net effect is a reduction
of the pressure, proportional to the square numerical
density, expressed as $A(N/V)^2$.   
On the other hand, the whole volume of the box,
$V$, is not accessible to particles, in that they are
conceived as identical spheres: the free volume within the
box is $V-NB$, where $B$ is the volume of a single sphere.
For further details refer to specific textbooks (e.g., LL67,
Chap.\,VII, \S74).

\begin{figure*}[t]
\begin{center}
\includegraphics[scale=0.8]{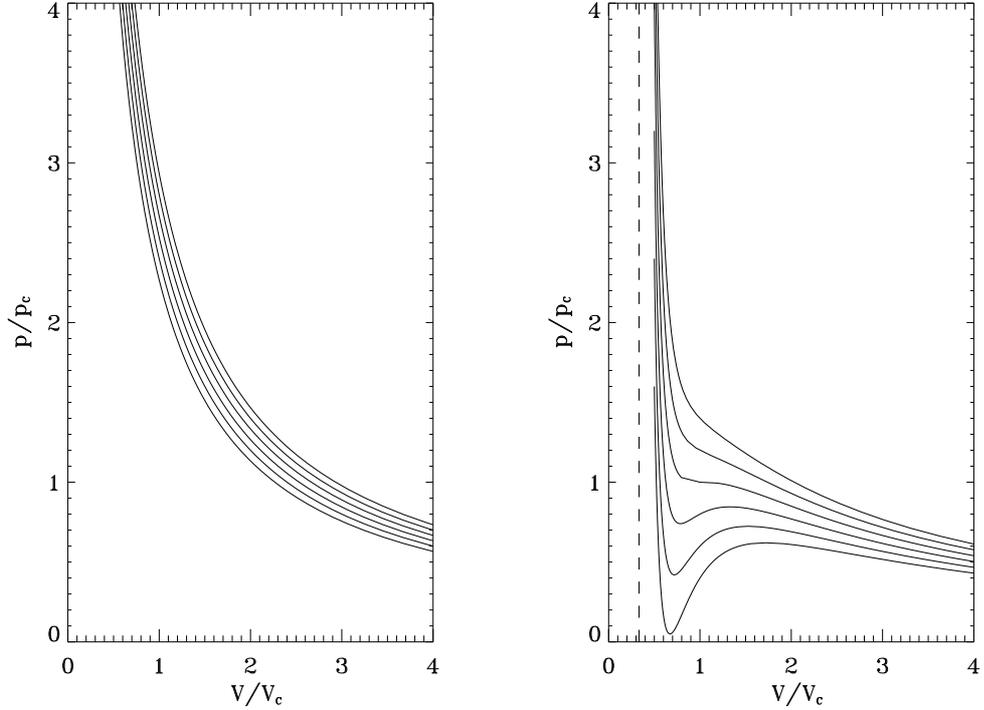}
\caption{Isothermal curves related to ideal (left panel)
and VDW (right panel) gases, respectively.   Isothermal
curves (from bottom to top) correspond to $T/T_{\rm c}=$0.85,
0.90, 0.95, 1.00, 1.05, 1.10.   No extremum point exists
above the critical isothermal curve, $T/T_{\rm c}=1$.}
\label{f:viso}
\end{center}
\end{figure*}

\begin{figure*}[t]
\begin{center}
\includegraphics[scale=0.8]{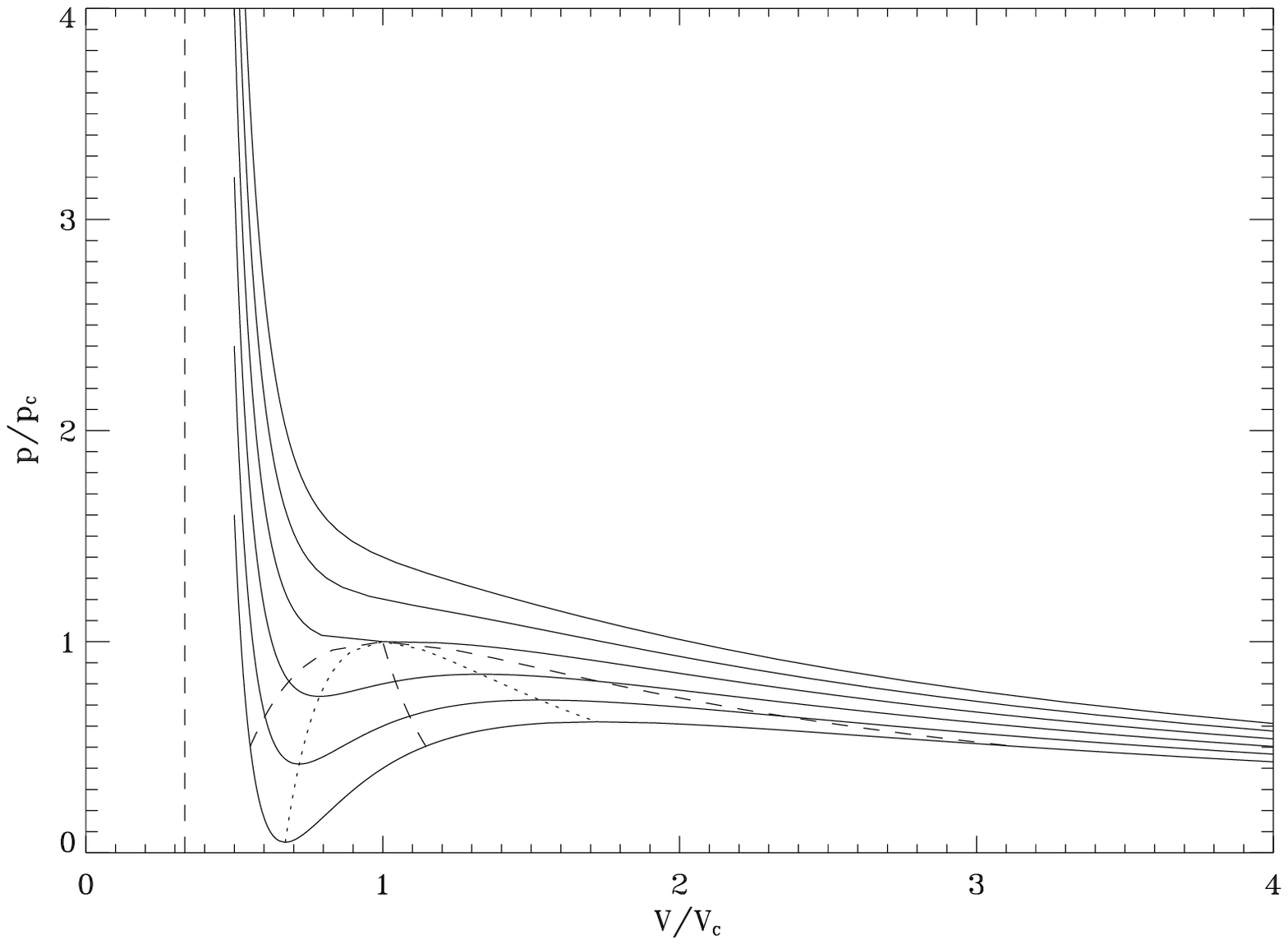}
\caption{Same as in Fig.\,\ref{f:viso} (right panel), where
the occurrence (within the bell-shaped area bounded by the
dashed curve) of saturated vapour is considered.
Above the critical isothermal
curve $(T=T_{\rm c})$ the trend is similar with respect to
ideal gases.   Below the critical isothermal curve and
on the right of the dashed curve, the gas
still behaves as an ideal gas.   Below the critical
isothermal curve and on the left of the dashed curve, the
liquid shows little change in volume as the pressure rises.
Within the bell-shaped area bounded by the dashed curve,
the liquid phase is in equilibrium with the saturated vapour
phase.   A reduced volume implies
smaller saturated vapour fraction and larger liquid fraction
at constant pressure, and vice versa.   The VDW equation of
state is
no longer valid in this region.   The dashed curve (including
the central branch) is the locus of intersection between VDW
and real isothermal curves, the latter being related to constant
pressure where liquid and vapour phases coexist.   The dotted
curve is the locus of VDW isothermal extremum points.}
\label{f:vris}
\end{center}
\end{figure*}

The isothermal ($T=$ const) curves for ideal gases are
hyperbolas with axes, $p=\mp V$, conformly to Eq.\,(\ref
{eq:gid}).   In VDW theory of real gases, the isothermal
curves exhibit two extremum points, which reduce to a
single horisontal inflexion point when a critical temperature
is attained,
as shown in
Fig.\,\ref{f:viso}.
Well above the critical isothermal curve, $T\gg T_{\rm c}$,
the trends exhibited by ideal and VDW gases look very
similar.  Below the critical isothermal curve, $T<T_{\rm c}$,
the behaviour of VDW gases is different with respect to
ideal gases and, in addition, the related isothermal
curves provide a wrong description within a specific
region where saturated vapour and liquid phases coexist.
Further details are shown in Fig.\,\ref{f:vris}.
Above the critical isothermal
curve $(T=T_{\rm c})$ the trend is similar with respect to
ideal gases.   Below the critical isothermal curve and
on the right of the dashed curve, the supersaturated
vapour still behaves as an ideal gas.   Below the critical
isothermal curve and on the left of the dashed curve, the
liquid shows little change in volume as the pressure rises.
Within the bell-shaped area bounded by the dashed curve,
the liquid phase is in equilibrium with the saturated vapour
phase.   A reduced volume implies
smaller saturated vapour fraction and larger liquid fraction
at constant pressure, and vice versa.   The VDW equation of
state is
no longer valid in this region.   The dashed curve (including
the central branch) is the locus of intersections between VDW
and real isothermal curves, the latter being related to constant
pressure where liquid and vapour phases coexist.   The dotted
curve is the locus of VDW isothermal extremum points.

\begin{figure*}[t]
\begin{center}
\includegraphics[scale=0.8]{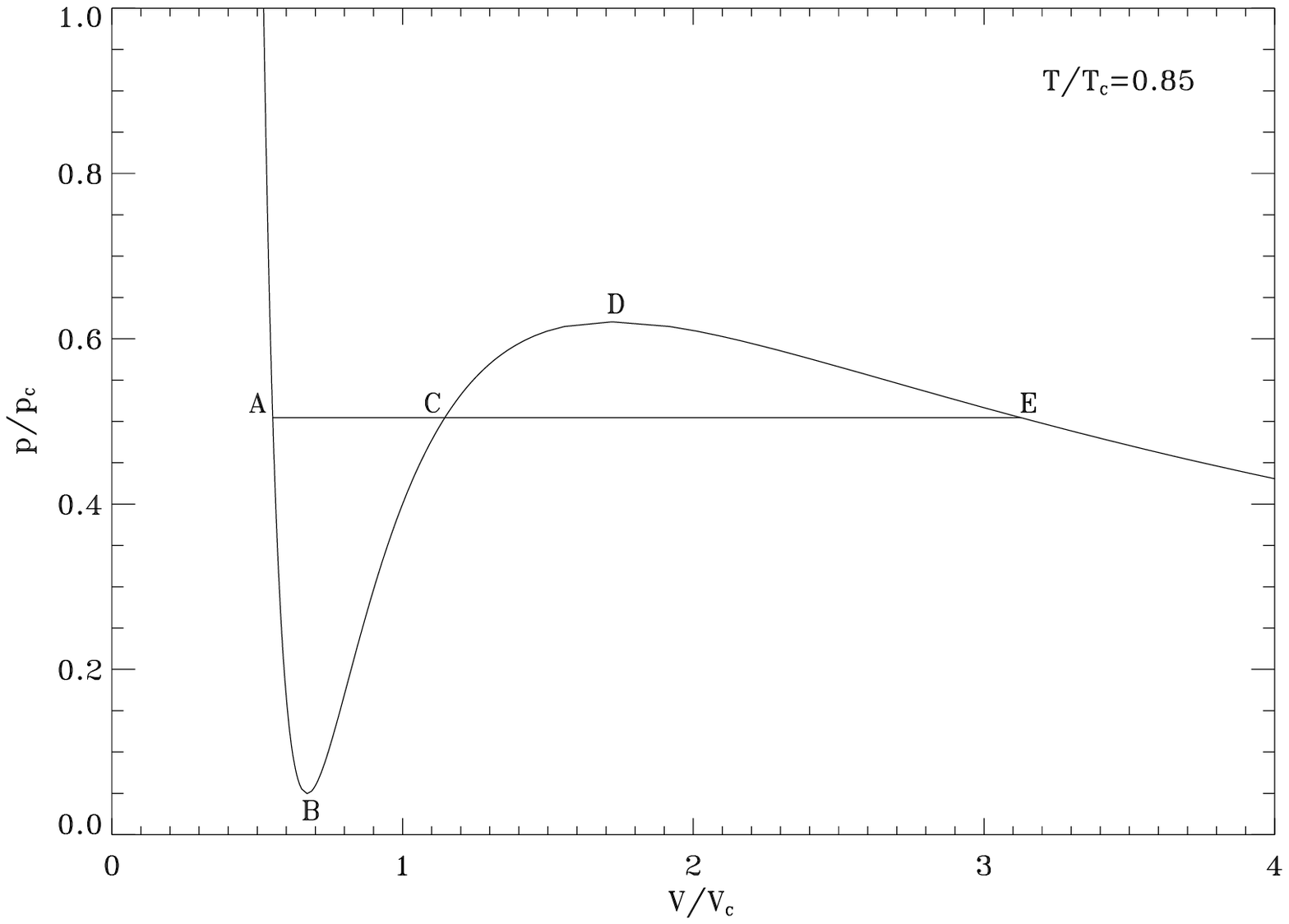}
\caption{A specific $(T/T_{\rm c}=0.85)$ VDW and corresponding
real isothermal curve.   The above mentioned curves
coincide within the range, $V\le V_{\rm A}$
and $V\ge V_{\rm E}$.    The VDW isothermal curve  exhibits
two extremum points: a minimum, ${\sf B}$, and a maximum,
${\sf D}$, while the real isothermal curve is flat within
the range, $V_{\rm A}\le V\le V_{\rm E}$.  Configurations related to
the VDW isothermal curve within the range, $V_{\rm A}\le V\le V_{\rm B}$
(due to tension forces acting on the particles yielding
superheated liquid), and
$V_{\rm D}\le V\le V_{\rm E}$ (due to the occurrence of undercooled
vapour), may be obtained under special conditions, while
configurations within the range, $V_{\rm B}\le V\le V_{\rm D}$, are
always unstable.   The volumes, $V_{\rm A}$ and $V_{\rm E}$, correspond
to the maximum value in presence of the sole liquid phase
and the minimum value in presence of the sole
vapour phase, respectively.   The regions, {\sf ABC} and
{\sf CDE}, have equal area.   For further details refer to
the text.}
\label{f:vrar}
\end{center}
\end{figure*}

A specific $(T/T_{\rm c}=0.85)$ VDW and corresponding real
isothermal curve, are represented in Fig.\,\ref{f:vrar}.
The VDW isothermal curve and the
real isothermal curve coincide within the range, $V\le V_{\rm A}$
and $V\ge V_{\rm E}$.    The VDW isothermal curve  exhibits
two extremum points: a minimum, ${\sf B}$, and a maximum,
${\sf D}$, while the real isothermal curve is flat, within
the range, $V_{\rm A}\le V\le V_{\rm E}$.  Configurations related to
the VDW isothermal curve within the range, $V_{\rm A}\le V\le V_{\rm B}$
(due to tension forces acting on the particles yielding
superheated liquid), and
$V_{\rm D}\le V\le V_{\rm E}$ (due to the occurrence of undercooled
vapour), may be obtained under special conditions, while
configurations within the range, $V_{\rm B}\le V\le V_{\rm D}$, are
always unstable.   The volumes, $V_{\rm A}$ and $V_{\rm E}$, correspond
to the maximum value in presence of the sole liquid phase
and the minimum value in presence of the sole
vapour phase, respectively.

The surfaces, {\sf ABC} and {\sf CDE}, are equal, as first
inferred by Maxwell (e.g., Rostagni, 1957, Chap.\,XII,
\S19).   The VDW and real isothermal curves represented
in Fig.\,\ref{f:vrar} being related to the same temperature,
$T$, the cycle, {\sf ABCDECA}, is completely both isothermal
and reversible, and the work, $W$, performed therein cannot be
positive to avoid violation of the second law of the
thermodynamics.   The cycles, {\sf ABCA} and {\sf CDEC},
occurring in counterclockwise and clockwise sense, respectively,
are also completely both isothermal and reversible.
Accordingly, $W_{\sf ABCDECA}=W_{\sf ABCA}-W_{\sf CDEC}\le0$.
A similar procedure, related to the reversed cycle,
{\sf ACEDCBA}, yields $W_{\sf ACEDCBA}=W_{\sf CEDC}-W_{\sf CBAC}
\le0$.   Then $W_{\sf ABCDECA}=W_{\sf ACEDCBA}=0$, which implies
$W_{\sf ABCA}=W_{\sf CDEC}=
W_{\sf CEDC}=W_{\sf CBAC}$ and, in turn, the equality between
the related surfaces.   For further details refer to specific
textbooks (e.g., LL67, Chap.\,VIII, \S85).

In order to simplify both notation and calculations, it is
convenient to deal with (dimensionless) reduced variables (e.g.,
Rostagni, 1957, Chap.\,XII, \S16; LL67, Chap.\,VIII, \S85).
To this aim, the first step is the knowledge of the parameters
related to the critical point, $V_{\rm c}, p_{\rm c}, T_{\rm c}$.   Using the
VDW equation of state, Eq.\,(\ref{eq:VdW}), the pressure and
its first and second partial derivatives, with respect to
the volume, read:
\begin{lefteqnarray}
\label{eq:pW}
&& p=\frac{kNT}{V-NB}-A\frac{N^2}{V^2}~~;\qquad N={\rm const}~~; \\
\label{eq:p1W}
&& \left(\frac{\partial p}{\partial V}\right)_{V,T}=-\frac{kNT}{(V-NB)^2}+
2A\frac{N^2}{V^3}~~; \\
\label{eq:p2W}
&& \left(\frac{\partial^2 p}{\partial V^2}\right)_{V,T}=\frac{2kNT}
{(V-NB)^3}-6A\frac{N^2}{V^4}~~;
\end{lefteqnarray}
where the domain is $V>NB$, $V=NB$ is a vertical
asymptote, and $p=0$ is a horisontal asymptote.
The critical isothermal corresponds to the highest
temperature allowing a liquid phase, which occurs
therein only at the critical point.   The critical
isothermal curve exhibits neither a minimum nor a
maximum, which are replaced by a horisontal inflexion
point coinciding with the critical point.  Accordingly,
$(\partial p/\partial V)_{V_{\rm c},T_{\rm c}}=0$,
$(\partial^2p/\partial V^2)_{V_{\rm c},T_{\rm c}}=0$, and
$p_{\rm c}=kNT_{\rm c}/(V_{\rm c}-NB)-AN^2/V_{\rm c}^2$.   The solution
of the related system is:
\begin{lefteqnarray}
\label{eq:Vc}
&& V_{\rm c}=3NB~~; \\
\label{eq:Tc}
&& T_{\rm c}=\frac8{27}\frac AB\frac1k~~; \\
\label{eq:pc}
&& p_{\rm c}=\frac1{27}\frac A{B^2}~~; \\
\label{eq:Zc}
&& Z_c=\frac{p_{\rm c}V_{\rm c}}{kNT_{\rm c}}=\frac38~~;
\end{lefteqnarray}
where, in general, the compressibility factor,
$Z=pV/(kNT)$, defines the degree of departure
from the behaviour of ideal gases, for which
$Z=1$, according to Eq.\,(\ref{eq:gid}).
For further details refer to specific textbooks
(e.g., Rostagni, 1957, Chap.\,XII, \S20; LL67,
Chap.\,VIII, \S85).

With regard to the reduced variables:
\begin{equation}
\label{eq:rv}
\sV=\frac V{V_{\rm c}}~~;\qquad\sP=\frac p{p_{\rm c}}~~;
\qquad\sT=\frac T{T_{\rm c}}~~;
\end{equation}
the ideal gas equation of state, Eq.\,(\ref{eq:gid}),
and the VDW equation of state, Eq.\,(\ref{eq:VdW}),
reduce to:
\begin{lefteqnarray}
\label{eq:ri}
&& \sP\sV=\frac83\sT~~; \\
\label{eq:rW1}
&& \left(\sP+\frac3{\sV^2}\right)\left(\sV-\frac13\right)=\frac83\sT~~;\qquad
\sV>\frac13~~;
\end{lefteqnarray}
and Eqs.\,(\ref{eq:pW}), (\ref{eq:p1W}), and
(\ref{eq:p2W}), reduce to:
\begin{lefteqnarray}
\label{eq:rW2}
&& \sP=\frac{8\sT}{3\sV-1}-\frac3{\sV^2}~~; \\
\label{eq:rW3}
&& \left(\frac{\partial\sP}{\partial\sV}\right)_{\sV,\sT}=-\frac{24\sT}
{(3\sV-1)^2}+\frac6{\sV^3}~~; \\
\label{eq:rW4}
&& \left(\frac{\partial^2\sP}{\partial\sV^2}\right)_{\sV,\sT}=\frac
{144\sT}{(3\sV-1)^3}-\frac{18}{\sV^4}~~;
\end{lefteqnarray}
where, for assigned $\sT$, the domain of the function,
$\sP(\sV)$, is $\sV>1/3$, $\sV=1/3$ is a vertical
asymptote, and $\sP=0$ is a horisontal asymptote.
In the special case of the critical point, $\sV=1$,
$\sT=1$, $\sP=1$, the partial derivatives are null,
as expected.

The extremum points, via Eq.\,(\ref{eq:rW3}), are
defined by the relation:
\begin{equation}
\label{eq:ext}
f(\sV)=\frac{(3\sV-1)^2}{4\sV^3}=\sT~~;
\end{equation}
which is satisfied at the critical point, as
expected.   The function on the left-hand side
of Eq.\,(\ref{eq:ext}) has two extremum points:
a minimum at $\sV=1/3$ (outside the physical
domain) and a maximum at $\sV=1$, where $\sT=1$.
Accordingly, Eq.\,(\ref{eq:ext}) is never
satisfied for $\sT>1$, which implies no extremum
point for related isothermal curves, as expected.
The contrary holds for $\sT<1$, where it can be
seen that the third-degree equation associated
to Eq.\,(\ref{eq:rW3}) has three real solutions,
related to extremum points.   One lies outside
the physical domain, which implies $\sV\le1/3$.
The remaining two are obtained as the intersections
between the curve, $f(\sV)$, expressed by Eq.\,(\ref
{eq:ext}), and the straight line, $y=\sT$, keeping
in mind that $f(1/3)=0$, $f(1)=1$, and $\lim_{\sV\to
+\infty}f(\sV)=0$.

The third-degree equation associated to Eq.\,(\ref
{eq:rW3}), may be ordered as:
\begin{leftsubeqnarray}
\slabel{eq:3dea}
&& \sV^3-9a\sV^2+6a\sV-a=0~~; \\
\slabel{eq:3deb}
&& a=\frac1{4\sT}~~;
\label{seq:3de}
\end{leftsubeqnarray}
with regard to the standard formulation (e.g.,
Spiegel, 1968, Chap.\,9):
\begin{equation}
\label{eq:3dx}
x^3+a_1x^2+a_2x+a_3=0~~;
\end{equation}
the discriminants of Eq.\,(\ref{eq:3dea}) are:
\begin{lefteqnarray}
\label{eq:Q}
&& Q=\frac{3a_2-a_1^2}9=a(2-9a)~~; \\
\label{eq:R}
&& R=\frac{9a_1a_2-27a_3-2a_1^3}{54}=\frac{a(1-18a+54a^2)}2~~; \\
\label{eq:D}
&& D=Q^3+R^2=\frac{a^2(1-4a)}4~~;
\end{lefteqnarray}
where $D=0$ in the special case of the critical
isothermal curve $(\sT=1, a=1/4)$, $D<0$ for
$\sT<1$, and $D>0$ for $\sT>1$.   Accordingly,
three coincident real solutions exist if $D=0$,
three (at least two) different real solutions
if $D<0$, one real (outside the physical domain)
and two complex coniugate if $D>0$.

The three real solutions $(D\le0)$ may be expressed as
(e.g., Spiegel, 1968, Chap.\,9):
\begin{leftsubeqnarray}
\slabel{eq:rsola}
&& \sV_1=2\sqrt{-Q}\cos\left(\pi+\frac\theta3\right)-\frac13a_1~~; \\
\slabel{eq:rsolb}
&& \sV_2=2\sqrt{-Q}\cos\left(\pi+\frac\theta3+\frac{2\pi}3\right)-
\frac13a_1~~; \\
\slabel{eq:rsolc}
&& \sV_3=2\sqrt{-Q}\cos\left(\pi+\frac\theta3+\frac{4\pi}3\right)-
\frac13a_1~~; \\
\slabel{eq:rsold}
&& \theta=\arctan\frac{\sqrt{-D}}R~~;
\label{seq:rsol}
\end{leftsubeqnarray}
where $a_1=-9a$ and, in the special case of the
critical isothermal curve, $a=1/4$, $Q=-1/16$,
$D=0$, which implies $\sV_0=\min(\sV_1,\sV_2,
\sV_3)$, $\sV_{\rm A}=\sV_{\rm B}=\sV_{\rm C}=\sV_{\rm D}=\sV_{\rm E}=\max
(\sV_1,\sV_2,\sV_3)$.   In the special case,
$\sT\to0$, Eq.\,(\ref{eq:3dea}) reduces to a
second-degree equation whose solutions are
$\sV_{01}=\sV_{02}=1/3$, while the related
function is otherwise divergent as $a\to+\infty$.
In general, the extremum points of VDW isothermal
curves $(\sT\le1)$ occur at $\sV=\sV_{\rm B}$ (minimum) and
$\sV=\sV_{\rm D}$ (maximum), $\sV_{\rm B}\le\sV_{\rm D}$.   As
$\sT\to0$, $\sV_{\rm B}\to1/3$, $\sV_{\rm D}\to+\infty$,
where, in all cases, $1/3<\sV_{\rm B}\le1\le\sV_{\rm D}$.

The two areas defined by the intersection of a
generic VDW isothermal curve $(\sT\le1)$ and
related real isothermal curves (see Fig.\,\ref
{f:vrar}), are expressed as:
\begin{leftsubeqnarray}
\slabel{eq:S1a}
&& W_1=\int_{V_{\rm A}}^{V_{\rm C}}p_{\rm C}\diff V-\int_{V_{\rm A}}^{V_
{\rm C}}p\diff V=p_{\rm C}V_{\rm C}
\left[\sP_C(\sV_{\rm C}-\sV_{\rm A})-\int_{\sV_{\rm A}}^{\sV_{\rm C}}\sP
\diff\sV\right];\qquad \\
\slabel{eq:S1b}
&& W_2=\int_{V_{\rm C}}^{V_{\rm E}}p\diff V-\int_{V_{\rm C}}^{V_{\rm E}}p_
{\rm C}\diff V=p_{\rm C}V_{\rm C}
\left[\int_{\sV_{\rm C}}^{\sV_{\rm E}}\sP\diff\sV-\sP_C(\sV_{\rm E}-\sV_
{\rm C})\right];\qquad
\label{seq:S1}
\end{leftsubeqnarray}
and the substitution of Eq.\,(\ref{eq:rW2}) into
(\ref{seq:S1}) allows explicit expressions for
the integrals.   The result is:
\begin{leftsubeqnarray}
\slabel{eq:S2a}
&& \frac{W_1}{p_{\rm C}V_{\rm C}}=\sP_{\rm C}(\sV_{\rm C}-\sV_{\rm A})-\frac
83\sT\ln\frac{3\sV_{\rm C}-1}{3\sV_{\rm A}-1}+
\frac{3(\sV_{\rm C}-\sV_{\rm A})}{\sV_{\rm A}\sV_{\rm C}}~~; \\
\slabel{eq:S2b}
&& \frac{W_2}{p_{\rm C}V_{\rm C}}=\frac83\sT\ln\frac{3\sV_{\rm E}-1}{3\sV_
{\rm C}-1}-
\frac{3(\sV_{\rm E}-\sV_{\rm C})}{\sV_{\rm C}\sV_{\rm E}}-\sP_C(\sV_{\rm E}-
\sV_{\rm C})~~;
\label{seq:S2}
\end{leftsubeqnarray}
and the condition, $W_1=W_2$, after some algebra reads:
\begin{equation}
\label{eq:S12}
\sP_C=\frac83\frac{\sT}{\sV_{\rm E}-\sV_{\rm A}}\ln\frac{3\sV_{\rm E}-1}{3\sV_{\rm A}-1}-\frac3
{\sV_{\rm A}\sV_{\rm E}}~~;
\end{equation}
where, for a selected isothermal curve, the unknowns
are $\sP_C=\sP_A=\sP_E$, $\sV_{\rm A}$, and $\sV_{\rm E}$.

The reduced volumes, $\sV_{\rm A}$, $\sV_{\rm C}$, $\sV_{\rm E}$,
see Fig.\,\ref{f:vrar}, may be considered as
intersections between a VDW isothermal curve
$(\sT<1)$ and a horisontal straight line, $\sP=\sP_C$,
in the $({\sf O}\sV\sP)$ plane.   In other words,
$\sV_{\rm A}$, $\sV_{\rm C}$, $\sV_{\rm E}$, are the real solutions
of the third-degree equation:
\begin{equation}
\label{eq:3Wrr}
\sV^3-\left(\frac13+\frac83\frac{\sT}{\sP_C}\right)\sV^2+\frac3{\sP_C}\sV-
\frac1{\sP_C}=0~~;
\end{equation}
which has been deduced from Eq.\,(\ref{eq:rW2}),
particularized to $\sP=\sP_C$.   The related
solutions may be calculated using Eqs.\,(\ref{seq:rsol}).
The last unknown, $\sP_C$, is determined from Eq.\,(\ref
{eq:S12}).

An inspection of Fig.\,\ref{f:vrar} shows that
the points, {\sf A} and {\sf E}, are located
on the left of the minimum, {\sf B}, and on
the right of the maximum, {\sf D}, respectively.
Keeping in mind the above results, the following
inequality holds: $\sV_{\rm A}\le\sV_{\rm B}\le1\le\sV_{\rm D}\le
\sV_{\rm E}$, which implies further investigation on
the special case, $\sV_{\rm C}=1$.   The particularization
of the VDW equation of state, Eq.\,(\ref{eq:rW2}), to the
point, ${\sf C}={\sf C_1}$, assuming $\sV_{C_1}=1$,
yields:
\begin{equation}
\label{eq:TVC1}
\sT=\frac{\sP_{C_1}+3}4~~;
\end{equation}
and Eq.\,(\ref{eq:3Wrr}) reduces to:
\begin{leftsubeqnarray}
\slabel{eq:3dba}
&& \sV^3-(1+2b)\sV^2+3b\sV-b=0~~; \\
\slabel{eq:3dbb}
&& b=\frac1{\sP_{C_1}}~~;
\label{seq:3db}
\end{leftsubeqnarray}
with regard to the generic third-degree equation,
Eq.\,(\ref{eq:3dx}), the three solutions, $x_1$,
$x_2$, $x_3$, satisfy the relations (e.g., Spiegel,
1968, Chap.\,9):
\begin{leftsubeqnarray}
\slabel{eq:x123a}
&& x_1+x_2+x_3=-a_1~~; \\
\slabel{eq:x123b}
&& x_1x_2+x_2x_3+x_3x_1=a_2~~; \\
\slabel{eq:x123c}
&& x_1x_2x_3=-a_3~~;
\label{seq:x123}
\end{leftsubeqnarray}
where, in the case under discussion:
\begin{leftsubeqnarray}
\slabel{eq:b123a}
&& a_1=-1-2b~~;\qquad a_2=3b~~;\qquad a_3=-b~~; \\
\slabel{eq:b123b}
&& x_1=\sV_{\rm A}~~;\qquad x_2=\sV_{C_1}=1~~;\qquad x_3=\sV_{\rm E}~~;
\label{seq:b123}
\end{leftsubeqnarray}
and the substitution of Eqs.\,(\ref{seq:b123})
into two among (\ref{seq:x123}) yields:
\begin{leftsubeqnarray}
\slabel{eq:VAEa}
&& \sV_{\rm A}=b-\sqrt{b^2-b}~~; \\
\slabel{eq:VAEb}
&& \sV_{\rm E}=b+\sqrt{b^2-b}~~;
\label{seq:VAE}
\end{leftsubeqnarray}
and the combination of Eqs.\,(\ref{eq:TVC1}),
(\ref{eq:3dbb}), and (\ref{seq:VAE}) produces:
\begin{leftsubeqnarray}
\slabel{eq:VAETa}
&& \sV_{\rm A}=\frac{1-2\sqrt{1-\sT}}{4\sT-3}~~;\qquad\sT\le1~~; \\
\slabel{eq:VAETb}
&& \sV_{\rm E}=\frac{1+2\sqrt{1-\sT}}{4\sT-3}~~;\qquad\sT\le1~~;~~;
\label{seq:VAET}
\end{leftsubeqnarray}
which, together with $\sV_{C_1}=1$, are the abscissae
of the intersection points between a selected VDW
isothermal curve in the $({\sf O}\sV\sP)$ plane and
the straight line, $\sP=\sP_{C_1}$, in the special case
under discussion.

The substitution of Eqs.\,(\ref{seq:VAET}) into
(\ref{eq:S12}), the last being related to the real
isothermal curve, yields:
\begin{equation}
\label{eq:S12C}
\frac{\sT}{\sqrt{1-\sT}}\ln\frac{3-2\sT+3\sqrt{1-\sT}}{3-2\sT-3\sqrt{1-\sT}}
=6~~;
\end{equation}
which holds only for the critical isothermal curve,
$\sT=1$.   Accordingly, the abscissa of the intersection
point, {\sf C}, between a selected VDW isothermal curve
and related real isothermal curve, see Fig.\,\ref{f:vrar},
cannot occur at $\sV_{\rm C}=1$ unless the critical isothermal
curve is considered.  Then the third-degree equation,
Eq.\,(\ref{eq:3Wrr}), must be solved in the general case
by use of  Eqs.\,(\ref{seq:rsol}).    The results are shown
in Tab.\,\ref{t:vispo}, where the following parameters
(in reduced variables) are listed for each VDW
isothermal curve, see Fig.\,\ref{f:vrar}: the temperature, $\sT$, the lower
volume limit, $\sV_{\rm A}$, for which the liquid and vapour phase
coexist; the extremum point (minimum) volume, $\sV_{\rm B}$;
the intermediate volume, $\sV_{\rm C}$, for which the pressure equals
its counterpart related to the corresponding lower and
upper volume limit, for which the liquid and vapour phase coexist;
the extremum point (maximum) volume, $\sV_{\rm D}$;
the upper volume limit, $\sV_{\rm E}$, for which the liquid and
vapour phase coexist; the extremum point (minimum) pressure,
$\sP_B$; the pressure, $\sP_A=\sP_C=\sP_E$, related to the
horisontal real isothermal curve; the extremum point (maximum)
pressure, $\sP_D$.
\begin{table}
\caption{Values of parameters, $\sT$, $\sV_{\rm A}$, $\sV_{\rm B}$,
$\sV_{\rm C}$, $\sV_{\rm D}$, $\sV_{\rm E}$, $\sP_B$, $\sP_C$,
$\sP_D$, within the range, $0.85\le\sT\le0.99$, using a step,
$\Delta\sT=0.01$.   All values equal unity at the critical point.
Index captions: A, C, E - intersections
between VDW and real isothermal curves; B - extremum point
of minimum; D - extremum point of maximum.   Extremum points
are related to VDW isothermal curves, while their real
counterparts are flat in presence of both liquid and vapour
phase.   For aesthetical reasons, 01 on head columns stands for unity.}
\label{t:vispo}
\begin{center}
\begin{tabular}{|l|l|l|l|l|l|l|l|l|} \hline
$\sT$ & $10\sV_{\rm A}$ & $10\sV_{\rm B}$ & $01\sV_{\rm C}$ & $01\sV_{\rm D}$ & $01\sV_{\rm E}$ & $10\sP_B$ &
$10\sP_C$ & $10\sP_D$ \\
\hline
 0.85 & 5.5336 & 6.7168 & 1.1453 & 1.7209 & 3.1276 & 0.4963 & 5.0449 & 6.2055 \\
 0.86 & 5.6195 & 6.8003 & 1.1337 & 1.6821 & 2.9545 & 1.2750 & 5.3125 & 6.4005 \\
 0.87 & 5.7116 & 6.8883 & 1.1225 & 1.6436 & 2.7909 & 2.0346 & 5.5887 & 6.6011 \\
 0.88 & 5.8106 & 6.9814 & 1.1116 & 1.6052 & 2.6360 & 2.7752 & 5.8736 & 6.8076 \\
 0.89 & 5.9176 & 7.0804 & 1.1009 & 1.5669 & 2.4889 & 3.4965 & 6.1674 & 7.0205 \\
 0.90 & 6.0340 & 7.1860 & 1.0905 & 1.5285 & 2.3488 & 4.1984 & 6.4700 & 7.2401 \\
 0.91 & 6.1615 & 7.2994 & 1.0804 & 1.4900 & 2.2151 & 4.8807 & 6.7816 & 7.4669 \\
 0.92 & 6.3022 & 7.4221 & 1.0706 & 1.4511 & 2.0869 & 5.5430 & 7.1021 & 7.7014 \\
 0.93 & 6.4593 & 7.5561 & 1.0610 & 1.4117 & 1.9634 & 6.1849 & 7.4318 & 7.9443 \\
 0.94 & 6.6369 & 7.7040 & 1.0516 & 1.3715 & 1.8438 & 6.8058 & 7.7707 & 8.1963 \\
 0.95 & 6.8412 & 7.8697 & 1.0425 & 1.3300 & 1.7271 & 7.4049 & 8.1188 & 8.4584 \\
 0.96 & 7.0819 & 8.0593 & 1.0336 & 1.2867 & 1.6118 & 7.9811 & 8.4762 & 8.7319 \\
 0.97 & 7.3756 & 8.2830 & 1.0249 & 1.2404 & 1.4960 & 8.5328 & 8.8429 & 9.0185 \\
 0.98 & 7.7554 & 8.5611 & 1.0164 & 1.1892 & 1.3761 & 9.0576 & 9.2191 & 9.3209 \\
 0.99 & 8.3091 & 8.9461 & 1.0081 & 1.1278 & 1.2430 & 9.5510 & 9.6048 & 9.6437 \\
\hline
\end{tabular}
\end{center}
\end{table}
The locus of the intersections between VDW and real
isothermal curves is represented in Fig.\,\ref{f:vris}
as a trifid curve, where the left, the right, and the
middle branch correspond to $\sV_{\rm A}$, $\sV_{\rm E}$, and
$\sV_{\rm C}$, respectively.   The common starting point
coincides with the critical point.   The locus of the
VDW isothermal curve extremum points is represented
in Fig.\,\ref{f:vris} as a dotted curve starting from
the critical point, where the left and the right
branch corresponds to minimum and maximum points,
respectively.

A fluid state can be represented in reduced variables
as ($\sV$, $\sP$, $\sT$), where one variable may be
expressed as a function of the remaining two, by use
of the reduced ideal gas equation of state, Eq.\,(\ref
{eq:ri}), or the reduced VDW equation of state,
Eq.\,(\ref{eq:rW1}).   The formulation in terms of
reduced variables, Eqs.\,(\ref{eq:rv}), makes the
related equation of state universal i.e. it holds
for any fluid.   Similarly, the Lane-Emden equation
expressed in polytropic (dimensionless) variables,
describes the whole class of polytropic gas spheres
with assigned polytropic index, in hydrostatic
equilibrium (e.g., Chandrasekhar 1939, Chap.\,IV,
\S4).

The states of two fluids with equal ($\sV$, $\sP$,
$\sT$), are defined as corresponding states.
The mere existence of an equation of state
yields the following result.
\begin{trivlist}
\item[\hspace\labelsep{\bf Law of corresponding
states.}] \sl
Given two fluids, the equality between two among
three reduced variables, $\sV$, $\sP$, $\sT$,
implies the equality between the remaining related
reduced variables i.e. the two fluids are in
corresponding states.
\end{trivlist}
The law was first formulated by van der Waals in
1880.   For further details refer to specific
textbooks (e.g., LL67, Chap.\,VIII, \S85).

\section{Astrophysical fluids}\label{macro}

Let macrogases be defined as two-component fluids
which interact only gravitationally.   The virial
theorem for subsystems reads (Caimmi et al., 1984;
Caimmi and Secco, 1992; CV08):
\begin{leftsubeqnarray}
\slabel{eq:viruva}
&& 2(E_u)_{\rm kin}+(E_{uv})_{\rm vir}=0~~; \qquad u=i,j~~;\qquad v=j,i~~;\\
\slabel{eq:viruvb}
&& (E_{uv})_{\rm vir}=(E_u)_{\rm sel}+(E_{uv})_{\rm tid}~~;
\label{seq:viruv}
\end{leftsubeqnarray}
where $i$ and $j$ denote the inner and outer subsystem,
respectively, $E_{\rm kin}$ is the kinetic energy,
$E_{\rm sel}$, $E_{\rm tid}$, and $E_{\rm vir}$, are
the self, tidal, and virial potential energy,
respectively.   The related definitions are:
\begin{lefteqnarray}
\label{eq:Eking}
&& (E_u)_{\rm kin}=\frac12\int_{S_u}\rho_u(x_1,x_2,x_3)\sum_{s=1}^3(v_u)_s^2
\diff^3S_u~~; \\
\label{eq:Eselg}
&& (E_u)_{\rm sel}=\int_{S_u}\rho_u(x_1,x_2,x_3)\sum_{s=1}^3x_s\frac
{\partial{\cal V}_u}{\partial x_s}\diff^3S_u \nonumber \\
&& \phantom{(E_u)_{\rm sel}}=-\frac12\int_{S_u}\rho_u(x_1,x_2,x_3){\cal V}_u
(x_1,x_2,x_3)\diff^3S_u~~; \\
\label{eq:Etidg}
&& (E_{uv})_{\rm tid}=\int_{S_u}\rho_u(x_1,x_2,x_3)\sum_{s=1}^3x_s\frac
{\partial{\cal V}_v}{\partial x_s}\diff^3S_u~~;
\end{lefteqnarray}
where $\rho$ is the density, $v_s$ the velocity component,
$S$ the volume, and ${\cal V}$ the gravitational potential.

The virial theorem makes a necessary (but not sufficient)
condition for dynamical or hydrostatic equilibrium, which
implies the parameters of the virialized configuration
must be considered as averaged on a sufficiently long time.
For further details refer to specific textbooks (e.g.,
Landau and Lifchitz, 1966, Chap.\,II, \S10) and to an earlier
paper (Caimmi, 2007).   On the other
hand, general trends exhibited by virialized configurations
hold, in particular, for self-consistent density profiles
implying nonnegative distribution functions.   Accordingly,
density profiles shall be chosen regardless from their
self-consistency, aiming to investigate general trends
instead of local properties.
To avoid the determination of the gravitational potential,
which is the most difficult step towards an explicit
formulation of potential energies, future considerations
shall be restricted to homeoidally striated ellipsoids
(Roberts, 1962).   The following results are
taken from earlier attempts (CV08, and further references
therein), to which an interested reader is addressed.

The isopycnic (i.e. constant density) surfaces are defined
by the following law:
\begin{leftsubeqnarray}
\slabel{eq:rhoa}
&& \rho_u=\rho_u^\dagger f_u(\xi_u)~~;\qquad f_u(1)=1~~;\qquad u=i,j~~; \\
\slabel{eq:rhob}
&& \xi_u=\frac{r_u}{r_u^\dagger}~~;\qquad0\le\xi_u\le\Xi_u~~;\qquad
\Xi_u=\frac{R_u}{r_u^\dagger}~~;
\label{seq:rho}
\end{leftsubeqnarray}
where the dagger denotes a selected reference
isopycnic surface, $r$ is the radial coordinate
along a selected direction, $R$ the related truncation
radius, $\xi$ a related scaled radial coordinate,
and $\Xi$ the related scaled truncation radius.

The mass and the self potential energy read:
\begin{leftsubeqnarray}
\slabel{eq:Ma}
&& M_u=(\nu_u)_{\rm mas}M_u^\dagger~~;\qquad M_u^\dagger=\frac{4\pi}3
\rho_u^\dagger(a_u)_1^\dagger(a_u)_2^\dagger(a_u)_3^\dagger~~; \\
\slabel{eq:Mb}
&& (\nu_u)_{\rm mas}=\frac32\int_0^{\Xi_u} F_u(\xi_u)\diff\xi_u~~; \qquad
F_u(\xi_u)=2\int_{\xi_u}^{\Xi_u}f_u(\xi_u)\xi_u\diff\xi_u~~;
\label{seq:M}
\end{leftsubeqnarray}
\begin{lefteqnarray}
\label{eq:Esel}
&& (E_u)_{\rm sel}=-(\nu_u)_{\rm sel}
\frac{G(M_u^\dagger)^2}{(a_u^\dagger)_1}B_u~~;\qquad
(\nu_u)_{\rm sel}=\frac9{16}\int_0^{\Xi_u}F_u^2(\xi_u)\diff\xi_u~~; 
\end{lefteqnarray}
where $a_\ell^\dagger$ are semiaxes of the
reference isopycnic surface, $\nu_{\rm mas}$
and $\nu_{\rm sel}$ are profile factors
which depend only on the mass distribution,
and $B$ is a shape factor.   For homogeneous
configurations, $\nu_{\rm mas}=\Xi^3$ and
$\nu_{\rm sel}=(3/10)\,\Xi^5$.  For spherical
shapes, $B=2$.

Under the further restriction of similar and similarly
placed boundaries, the following relations hold:
\begin{leftsubeqnarray}
\slabel{eq:yma}
&& \xi_i=y^\dagger\xi_j~~;\qquad\frac{\Xi_j}{\Xi_i}=\frac y{y^\dagger}~~;
\qquad\frac{(\nu_j)_{\rm mas}}{(\nu_i)_{\rm mas}}=\frac m{m^\dagger}~~; \\
\slabel{eq:ymb}
&& y=\frac{R_j}{R_i}~~;\qquad y^\dagger=\frac{r_j^\dagger}{r_i^\dagger}~~;
\qquad m=\frac{M_j}{M_i}~~;\qquad m^\dagger=\frac{M_j^\dagger}{M_i^\dagger}~~;
\label{seq:ym}
\end{leftsubeqnarray}
which makes tidal and virial potential energy reduce to:
\begin{leftsubeqnarray}
\slabel{eq:Exxxa}
&& (E_{uv})_{\rm xxx}=-\frac{G(M_u^\dagger)^2}{(a_u^\dagger)_1}
(\nu_{uv})_{\rm xxx}B~~; \\
\slabel{eq:Exxxb}
&& u=i,j~~;\qquad v=j,i~~;\qquad{\rm xxx}={\rm tid},\,{\rm vir}~~;
\label{seq:Exxx}
\end{leftsubeqnarray}
and the explicit expression of the profile factors reads:
\begin{leftsubeqnarray}
\slabel{eq:nuuva}
&& (\nu_{ij})_{\rm tid}=-\frac98m^\dagger w^{({\rm ext})}(\eta)~~;\qquad
(\nu_{ji})_{\rm tid}=-\frac98\frac{y^\dagger}{m^\dagger}w^{({\rm int})}
(\eta)~~; \\
\slabel{eq:nuuvb}
&& (\nu_{uv})_{\rm vir}=(\nu_u)_{\rm sel}+(\nu_{uv})_{\rm tid}~~;\qquad
u=i,j~~;\qquad v=j,i~~; \\
\slabel{eq:nuuvc}
&& \eta=\frac{\Xi_i}{y^\dagger}=\frac{\Xi_j}y~~;\qquad y\ge1~~;
\label{seq:nuuv}
\end{leftsubeqnarray}
where the functions, $w^{({\rm int})}$ and
$w^{({\rm ext})}$, are defined as:
\begin{leftsubeqnarray}
\slabel{eq:wiea}
&& w^{({\rm int})}(\eta)=\int_0^\eta F_j(\xi_j)\frac
{\diff F_i}{\diff\xi_j}\xi_j\diff\xi_j~~; \\
\slabel{eq:wieb}
&& w^{({\rm ext})}(\eta)=\int_0^\eta F_i(\xi_i)\frac
{\diff F_j}{\diff\xi_j}\xi_j\diff\xi_j~~;
\label{seq:wie}
\end{leftsubeqnarray}
for further details refer to Appendix \ref{a:prof}.
In conclusion, Eqs.\,(\ref{seq:rho})-(\ref
{seq:wie}) allow the calculation of the
virial potential energy for homeoidally
striated ellipsoids related to 
similar and similarly placed boundaries.

The fractional virial potential energy, reads:
\begin{equation}
\label{eq:phic}
\phi=\frac{(E_{ji})_{\rm vir}}{(E_{ij})_{\rm vir}}=\frac{(m^\dagger)^2}
{y^\dagger}\frac{(\nu_{ji})_{\rm vir}}{(\nu_{ij})_{\rm vir}}
=\frac{m^2}y\frac{\Xi_j}{\Xi_i}\left[\frac{(\nu_i)_{\rm mas}}
{(\nu_j)_{\rm mas}}\right]^2\frac{(\nu_{ji})_{\rm vir}}{(\nu_{ij})_
{\rm vir}}~~;
\end{equation}
which, for assigned density profiles,
depends on either the reference fractional
mass, $m^\dagger$, and the fractional scaling
radius, $y^\dagger$, or the fractional mass, $m$,
and the fractional truncation radius, $y$.

Strictly speaking, Eq.\,(\ref{eq:phic}) is
valid provided the indices,
$i$ and $j$, denote the embedded and the
embedding subsystem, respectively, which
implies $y\ge1$.
If the role of the two subsystems is
reversed, $0\le y\le1$, it has to be
kept in mind that the
inner and the outer component are
denoted by the indices, $j$ and $i$,
respectively.    Then the quantities
of interest must be calculated
according to the changes, $m\to
m^{-1}$, $m^\dagger\to(m^\dagger)^
{-1}$, $y\to y^{-1}$, $y^\dagger\to
(y^\dagger)^{-1}$, $i\leftrightarrow
j$, which yields $\phi=(E_{ij})_{\rm
vir}/(E_{ji})_{\rm vir}$ in the domain,
$y\ge1$.   Following the above mentioned
procedure where, in addition, $\phi\to
\phi^{-1}$, allows the explicit
expression of the fractional virial
energy, $\phi=(E_{ji})_{\rm  vir}/
(E_{ij})_{\rm vir}$, in the domain,
$0\le y\le1$, which extends the whole
domain to $0\le y<+\infty$.

In absence of truncation radius,
$\Xi\to+\infty$, $\eta\to+\infty$,
the reversion occurs when the density
drops to zero and nothing changes
except in infinitesimal terms of
higher order and infinite terms of
lower order.   Accordingly, there
is no need to perform the reversion
in this case.

The combination of  Eqs.\,(\ref{seq:nuuv})
and (\ref{eq:phic}) yields:
\begin{leftsubeqnarray}
\slabel{eq:phiea}
&& \phi=\frac{(m^\dagger)^2}{y^\dagger}\displayfrac{(\nu_j)_{\rm sel}-\frac98
\frac{y^\dagger}{m^\dagger}w^{({\rm int})}(\eta)}{(\nu_i)_{\rm sel}-\frac98
m^\dagger w^{({\rm ext})}(\eta)}~~; \\
\slabel{eq:phieb}
&& \phi=\frac{\Xi_j}{\Xi_i}\left[\frac{(\nu_i)_{\rm mas}}{(\nu_j)_{\rm mas}}
\right]^2\frac{m^2}y\displayfrac{(\nu_j)_{\rm sel}-\frac98\frac{\Xi_i}{\Xi_j}
\frac{(\nu_j)_{\rm mas}}{(\nu_i)_{\rm mas}}\frac ymw^{({\rm int})}(\eta)}
{(\nu_i)_{\rm sel}-\frac98\frac{(\nu_i)_{\rm mas}}{(\nu_j)_{\rm mas}}m
w^{({\rm ext})}(\eta)}~~;
\label{seq:phie}
\end{leftsubeqnarray}
where, for assigned scaled truncation radii,
$\Xi_i$ and $\Xi_j$, the independent variables
are $m^\dagger$, $y^\dagger$, or $m$, $y$.
In terms of a second-degree equation in $m^\dagger$
or $m$, Eqs.\,(\ref{seq:phie}) read:
\begin{leftsubeqnarray}
\slabel{eq:eq2a}
&& \hat{A}x^2+\hat{B}x+\hat{C}=0~~; \\
\slabel{eq:eq2b}
&& \hat{A}=k_{\rm A}(\nu_j)_{\rm sel}~~;\qquad k_{\rm A}=1,~\frac{\Xi_j}
{\Xi_i}\left[\frac{(\nu_i)_{\rm mas}}{(\nu_j)_{\rm mas}}\right]^2~~; \\
\slabel{eq:eq2c}
&& \hat{B}=-\frac98k_{\rm B}\left[w^{({\rm int})}(\eta)-\phi w^{({\rm ext})}
(\eta)\right]~~;\qquad k_{\rm B}=y^\dagger,~\frac{(\nu_i)_{\rm mas}}{(\nu_j)_
{\rm mas}}y~~; \\
\slabel{eq:eq2d}
&& \hat{C}=-k_{\rm C}(\nu_i)_{\rm sel}\phi~~;\qquad k_{\rm C}=y^\dagger,~y~~;
\label{seq:eq2}
\end{leftsubeqnarray}
where the positive solution is:
\begin{equation}
\label{eq:seq2}
x=\frac{\sqrt{\hat{B}^2-4\hat{A}\hat{C}}-\hat{B}}{2\hat{A}}~~;\qquad x=m^
\dagger,~m~~;
\end{equation}
and the negative solution has been disregarded
due to the lack of physical meaning.

In the special case of coinciding density profiles,
$f_i=f_j$, $F_i=F_j$, and scaled truncation radii,
$\Xi_i=\Xi_j$, fractional masses and truncation
radii also coincide, $m^\dagger=m$, $y^\dagger=y$, 
via Eqs.\,(\ref{seq:ym}), and the same holds for
the profile factors, $(\nu_i)_{\rm mas}=(\nu_j)_
{\rm mas}$, $(\nu_i)_{\rm sel}=(\nu_j)_{\rm sel}$,
which depend on the scaled truncation radii.
Accordingly, Eqs.\,(\ref{eq:phiea}), (\ref{eq:phieb}),
also coincide.

To get a closer analogy with real gases, let
Eqs.\,(\ref{seq:phie}) be rewritten after a
change of variables, as:
\begin{leftsubeqnarray}
\slabel{eq:Xra}
&& X_{\rm p}^\dagger X_{\rm V}^\dagger\displayfrac{1-\frac98\frac1{(\nu_j)_{\rm sel}}
\frac{w^{({\rm int})}(\eta)}{(X_{\rm p}^\dagger)^{1/2}X_{\rm V}^\dagger}}{1-\frac98\frac1
{(\nu_i)_{\rm sel}}(X_{\rm p}^\dagger)^{1/2}w^{({\rm ext})}(\eta)}=K^\dagger(\Xi_i,
\Xi_j)X_{\rm T}^\dagger~~; \\
\slabel{eq:Xrb}
&& X_{\rm p} X_{\rm V}\displayfrac{1-\frac98\frac{\Xi_i}{\Xi_j}\frac{(\nu_j)_{\rm mas}}
{(\nu_i)_{\rm mas}}\frac1{(\nu_j)_{\rm sel}}
\frac{w^{({\rm int})}(\eta)}{X_{\rm p}^{1/2}X_{\rm V}}}{1-\frac98\frac1
{(\nu_i)_{\rm sel}}X_{\rm p}^{1/2}w^{({\rm ext})}(\eta)}=K(\Xi_i,\Xi_j)X_{\rm T}~~; \\
\slabel{eq:Xrc}
&& \eta=\Xi_iX_{\rm V}^\dagger=\Xi_jX_{\rm V}~~;\qquad y\ge1~~; \\
\slabel{eq:Xrd}
&& X_{\rm p}^\dagger=(m^\dagger)^2~~;\quad X_{\rm V}^\dagger=\frac1{y^\dagger}~~;\quad
X_{\rm T}^\dagger=\phi~~;\quad K^\dagger(\Xi_i,\Xi_j)=\frac{(\nu_i)_{\rm sel}}
{(\nu_j)_{\rm sel}}~~; \\
\slabel{eq:Xre}
&& X_{\rm p}=m^2~;\quad X_{\rm V}=\frac1y~;\quad X_{\rm T}=
\phi~;\quad K(\Xi_i,\Xi_j)=\frac{\Xi_i}{\Xi_j}\left[\frac{(\nu_j)_
{\rm mas}}{(\nu_i)_{\rm mas}}\right]^2\frac{(\nu_i)_{\rm sel}}{(\nu_j)_
{\rm sel}}~;\qquad
\label{seq:Xr}
\end{leftsubeqnarray}
which can be conceived as an equation of state for macrogases.
The variables, $X_{\rm V}$, $X_{\rm p}$, $X_{\rm T}$, play a similar role as
the volume, the pressure, and the temperature, for ordinary
gases.   Accordingly, $X_{\rm V}$, $X_{\rm p}$, $X_{\rm T}$, shall be defined
as macrovolume, macropressure, and macrotemperature, respectively.

The combination of Eqs.\,(\ref{seq:ym}), (\ref{eq:Xrd}), and
(\ref{eq:Xre}) yields:
\begin{equation}
\label{eq:XXC}
X_{\rm V}^\dagger=\frac{\Xi_i}{\Xi_j}X_{\rm V}~~;\qquad X_{\rm p}^\dagger=\left[
\frac{(\nu_i)_{\rm mas}}{(\nu_j)_{\rm mas}}\right]^2X_{\rm p}~~;\qquad
X_{\rm T}^\dagger=X_{\rm T}~~;
\end{equation}
which links the variables, $(X_{\rm V}^\dagger,X_{\rm p}^\dagger,X_{\rm T}^\dagger)$,
to $(X_{\rm V},X_{\rm p},X_{\rm T})$, and vice versa.

Strictly speaking, the macrogas equation of state 
should be deduced from dimensional (instead of dimensionless)
virial equations for an assigned subsystem, as outlined in
Appendix \ref{a:dime}.   On the other hand, a description
in terms of dimensionless variables turns out to be more
useful.

If the interaction terms are omitted, $w^{({\rm int})}=
w^{({\rm ext})}=0$, Eqs.\,(\ref{eq:Xra}) and (\ref{eq:Xrb})
reduce to:
\begin{leftsubeqnarray}
\slabel{eq:Xia}
&& X_{\rm p}^\dagger X_{\rm V}^\dagger=K^\dagger(\Xi_i,\Xi_j)X_{\rm T}^\dagger~~; \\
\slabel{eq:Xib}
&& X_{\rm p} X_{\rm V}=K(\Xi_i,\Xi_j)X_{\rm T}~~;
\label{seq:Xi}
\end{leftsubeqnarray}
which may be considered as equation of state of
ideal macrogases, where ``ideal'' means ``the
interaction terms are omitted''.

The parameters, $K^\dagger$ and $K$, appearing
in either macrogas equation of state, depend
on the scaled truncation radii, $\Xi_i$ and
$\Xi_j$, and on the selected density profiles.
In other words, the macrogas equation of state
is not universal, but takes a different form
for different density profiles.   A restricted
number of special cases shall be studied below,
grounding on earlier results (CV08), to which
an interested reader is addressed for further
details.   In any case, the following method
shall be used: (i) select two density profiles;
(ii) fix related scaled truncation radii, $\Xi_i$
and $\Xi_j$; (iii) choose a macrotemperature,
$\phi$; (iv) plot related macroisothermal curves,
by solving Eq.\,(\ref{eq:Xra}) or (\ref{eq:Xrb}).

\subsection{UU macrogases}\label{UU}

The related density profiles maintain uniform,
which is equivalent to polytropes with index,
$n=0$ (e.g., Chandrasekhar, 1939, Chap.\,IV,
\S4; Caimmi, 1986), but implies negative
distribution functions for stellar fluids
(Vandervoort, 1980).   The particularization
of the general expressions to the case under
discussion, yields for the quantities of
interest (CV08):
\begin{lefteqnarray}
\label{eq:fU}
&& f_u(\xi_u)=1~~;\qquad0\le\xi_u\le\Xi_u~~;\qquad u=i,j~~; \\
\label{eq:FU}
&& F_u(\xi_u)=\Xi_u^2-\xi_u^2~~;\qquad u=i,j~~; \\
\label{eq:numU}
&& (\nu_u)_{\rm mas}=\Xi_u^3~~;\qquad u=i,j~~; \\
\label{eq:nusU}
&& (\nu_u)_{\rm sel}=\frac3{10}\Xi_u^5~~;\qquad u=i,j~~; \\
\label{eq:wiU}
&& w^{({\rm int})}(\eta)=-\frac4{15}\Xi_i^2\eta^3
\left(\frac52y^2-\frac32\right)~~; \\
\label{eq:weU}
&& w^{({\rm ext})}(\eta)=-\frac4{15}\Xi_i^2\eta^3~~;
\end{lefteqnarray}
where Eqs.\,(\ref{eq:wiU}) and (\ref{eq:weU})
hold under the additional restriction of similar
and similarly placed boundaries (CV08).

In the case under consideration of uniform
density profiles, without loss of generality,
it can be assumed a scaled truncation radius,
$\Xi_u=R_u/r_u^\dagger=1$, which implies $y=y^
\dagger$, $m=m^\dagger$, due to Eqs.\,(\ref{seq:ym})
and (\ref{eq:numU}).   Accordingly, Eq.\,(\ref{eq:phic})
reduces to (CV08):
\begin{leftsubeqnarray}
\slabel{eq:phlU}
&& \phi=\frac{(m^\dagger)^2}{y^\dagger}\left(\frac y{y^\dagger}\right)^5
\displayfrac{1+\frac{({y^\dagger})^3}{m^\dagger}\frac1{y^5}\left(
\frac52y^2-\frac32\right)}{1+\frac{m^\dagger}{(y^\dagger)^3}}~~;\qquad
y\ge1~~; \\
\slabel{eq:phuU}
&& \phi=\left(\frac y{y^\dagger}\right)^5
\displayfrac{m^\dagger(y^\dagger)^2\left[1+\frac{m^\dagger}{(y^\dagger)^3}
\right]}{1+m^\dagger\left(\frac y{y^\dagger}\right)^3\left(\frac52-\frac32
y^2\right)}~~;\qquad0\le y\le1~~;
\label{seq:phiU}
\end{leftsubeqnarray}
where $y$ is the outer to inner ellipsoid axis
ratio, $R_j/R_i$, according to Eq.\,(\ref{eq:ymb}).
In addition, Eqs.\,(\ref{seq:eq2}) and (\ref{eq:seq2})
reduce to:
\begin{leftsubeqnarray}
\slabel{eq:ABCa}
&& m=\sqrt{\beta^2+y\phi}-\beta~~; \\
\slabel{eq:ABCb}
&& \beta=\frac12\frac1{y^2}\left(\frac52y^2-\frac32-\phi\right)~~;\qquad
y\ge1~~; \\
\slabel{eq:ABCc}
&& \beta=\frac12y^3\left[1-\left(\frac52\frac1{y^2}-\frac32\right)\phi
\right]~~;\qquad0\le y\le1~~; \\
\slabel{eq:ABCd}
&& A=\frac3{10}~~;\qquad C=-\frac3{10}y\phi~~;\qquad B=\frac35\beta~~;
\label{seq:ABC}
\end{leftsubeqnarray}
where the negative solution is not considered due
to the lack of physical meaning.

The explicit expression of the square fractional
mass, $m^2$, extracted from Eq.\,(\ref{eq:ABCa})
in dimensionless variables, $X_{\rm p}=X_{\rm p}^\dagger=m^2=(m^\dagger)^2$,
$X_{\rm V}=X_{\rm V}^\dagger=1/y=1/y^\dagger$, $X_{\rm T}=X_{\rm T}^\dagger=\phi$,
Eqs.\,(\ref{eq:Xrd}), (\ref{eq:Xre}), reads:
\begin{leftsubeqnarray}
\slabel{eq:sopia}
&& X_{\rm p}=2\beta^2+\frac{X_{\rm T}}{X_{\rm V}}-2\beta\sqrt{\beta^2+\frac{X_{\rm T}}{X_{\rm V}}}~~; \\
\slabel{eq:sopib}
&& 2\beta=X_{\rm V}^2\left(\frac52\frac1{X_{\rm V}^2}-\frac32-X_{\rm T}\right)~~;\qquad
0<X_{\rm V}\le1~~; \\
\slabel{eq:sopic}
&& 2\beta=\frac1{X_{\rm V}^3}\left[1-\left(\frac52X_{\rm V}^2-\frac32\right)X_{\rm T}
\right]~~;\qquad X_{\rm V}\ge1~~;
\label{seq:sopi}
\end{leftsubeqnarray}
or, more explicitly:
\begin{leftsubeqnarray}
\slabel{eq:sopea}
&& X_{\rm p}=2\frac14X_{\rm V}^4\left(\frac52\frac1{X_{\rm V}^2}-\frac32-X_{\rm T}\right)^2+\frac
{X_{\rm T}}{X_{\rm V}}-X_{\rm V}^2\left(\frac52\frac1{X_{\rm V}^2}-\frac32-X_{\rm T}\right) \nonumber \\
&& \phantom{X_{\rm p}=}\cdot\left[\frac14X_{\rm V}^4\left(\frac52\frac1{X_{\rm V}^2}-\frac32-
X_{\rm T}\right)^2+\frac{X_{\rm T}}{X_{\rm V}}\right]^{1/2}~~;\qquad0<X_{\rm V}\le1~~; \\
\slabel{eq:sopeb}
&& X_{\rm p}=2\frac14\frac1{X_{\rm V}^6}\left[1-\left(\frac52X_{\rm V}^2-\frac32\right)X_{\rm T}
\right]^2+\frac{X_{\rm T}}{X_{\rm V}}-\frac1{X_{\rm V}^3}\left[1-\left(\frac52X_{\rm V}^2-\frac32
\right)X_{\rm T}\right]\qquad \nonumber \\
&& \phantom{X_{\rm p}=}\cdot\left\{\frac14\frac1{X_{\rm V}^6}\left[1-\left(\frac52X_{\rm V}^2-
\frac32\right)X_{\rm T}\right]^2+\frac{X_{\rm T}}{X_{\rm V}}\right\}^{1/2}~~;\qquad X_{\rm V}\ge1~~;
\label{seq:sope}
\end{leftsubeqnarray}
which is the actual UU (AUU) macrogas equation of state.

The ideal UU (IUU) macrogas equation of state is obtained
by the combination of Eqs.\,(\ref{eq:Xre}), (\ref{eq:Xib}),
(\ref{eq:numU}), and (\ref{eq:nusU}).
The result is:
\begin{equation}
\label{eq:ph1U}
X_{\rm p}=\frac{X_{\rm T}}{X_{\rm V}}~~;
\end{equation}
which represents a hyperbola with equal axes, for fixed
$X_{\rm T}$.

\begin{figure*}[t]
\begin{center}
\includegraphics[scale=0.8]{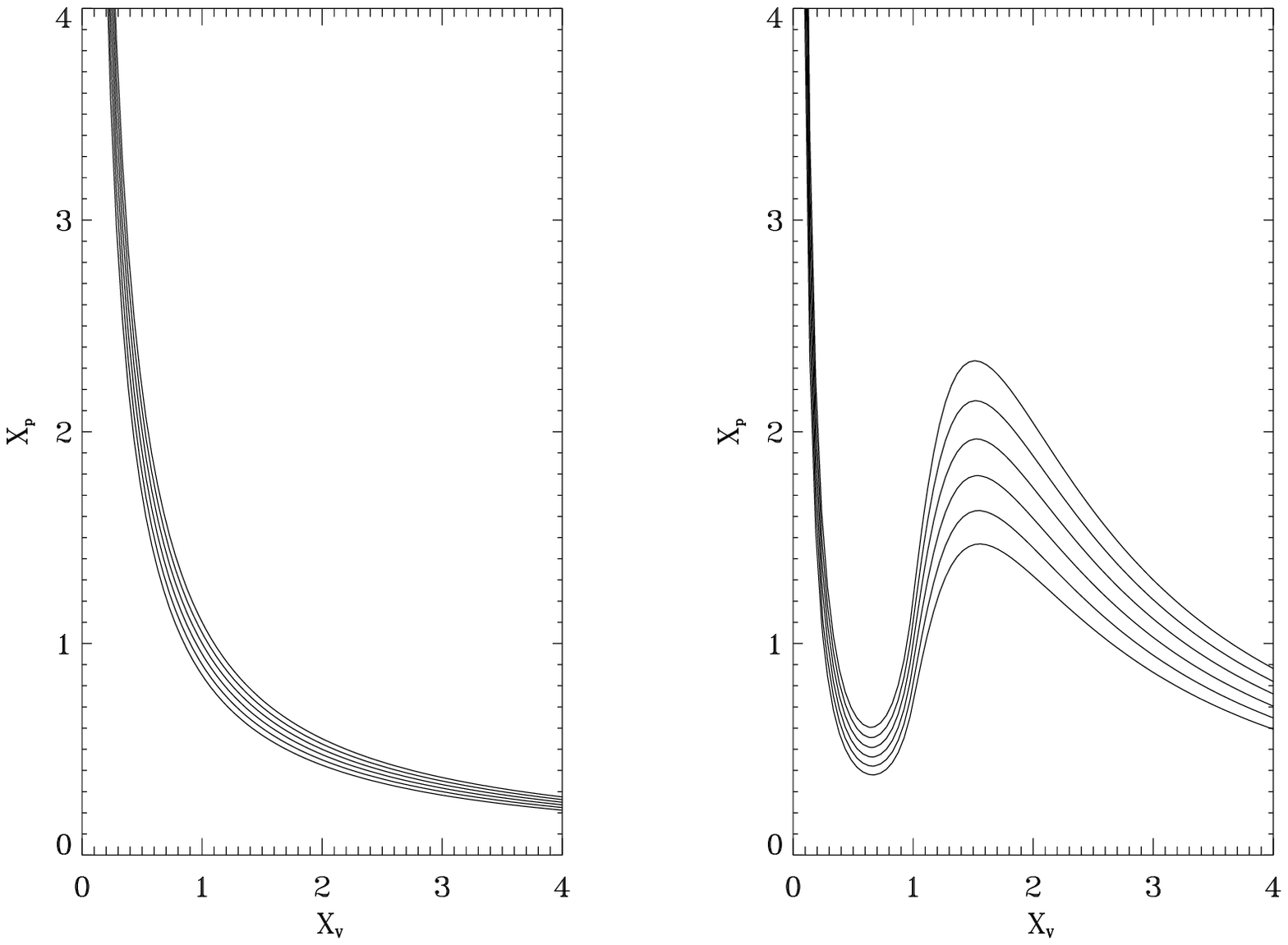}
\caption{Macroisothermal curves related to IUU (left panel)
and AUU (right panel) macrogases, respectively.
Macroisothermal
curves (from bottom to top) correspond to $X_{\rm T}=$ 0.85,
0.90, 0.95, 1.00, 1.05, 1.10.   No critical
macroisothermal curve exists, above or below or above which the extremum
points disappear.   The coordinates, $X_{\rm V}=X_{\rm V}^
\dagger$, $X_{\rm p}=X_{\rm p}^\dagger$, $X_{\rm T}=X_{\rm T}^\dagger$, may be
conceived as normalized to their fictitious critical
counterparts, $X_{V_{\rm c}}=X_{V_{\rm c}}^\dagger=1$, $X_{p_{\rm c}}=
X_{p_{\rm c}}^\dagger=1$, $X_{T_{\rm c}}=X_{T_{\rm c}}^\dagger=1$.}
\label{f:uuso}
\end{center}
\end{figure*}

Macroisothermal curves related to IUU (tidal potential energy
excluded) and AUU (tidal potential energy included)
macrogases, are plotted in Fig.\,\ref{f:uuso}, left and
right panel, respectively, for values of the
macrotemperature, $X_{\rm T}=0.85$, 0.90, 0.95, 1.00, 1.05,
1.10, from bottom to top.   The coordinates, $X_{\rm V}=X_{\rm V}^
\dagger$, $X_{\rm p}=X_{\rm p}^\dagger$, $X_{\rm T}=X_{\rm T}^\dagger$, may be
conceived as normalized to their fictitious critical
counterparts, $X_{V_{\rm c}}=X_{V_{\rm c}}^\dagger=1$, $X_{p_{\rm c}}=
X_{p_{\rm c}}^\dagger=1$, $X_{T_{\rm c}}=X_{T_{\rm c}}^\dagger=1$, as
$\phi=m=m^\dagger$ for $y=y^\dagger=1$, according to
Eqs.\,(\ref{seq:phiU}) or (\ref{eq:phim}), which implies
$\phi=1$ for $m=m^\dagger=1$.   The comparison with ideal and
VDW gases, plotted in Fig.\,\ref{f:viso}, shows a
similar trend, except the absence of a critical
macroisothermal curve, above which the extremum
points disappear.

\begin{figure*}[t]
\begin{center}
\includegraphics[scale=0.8]{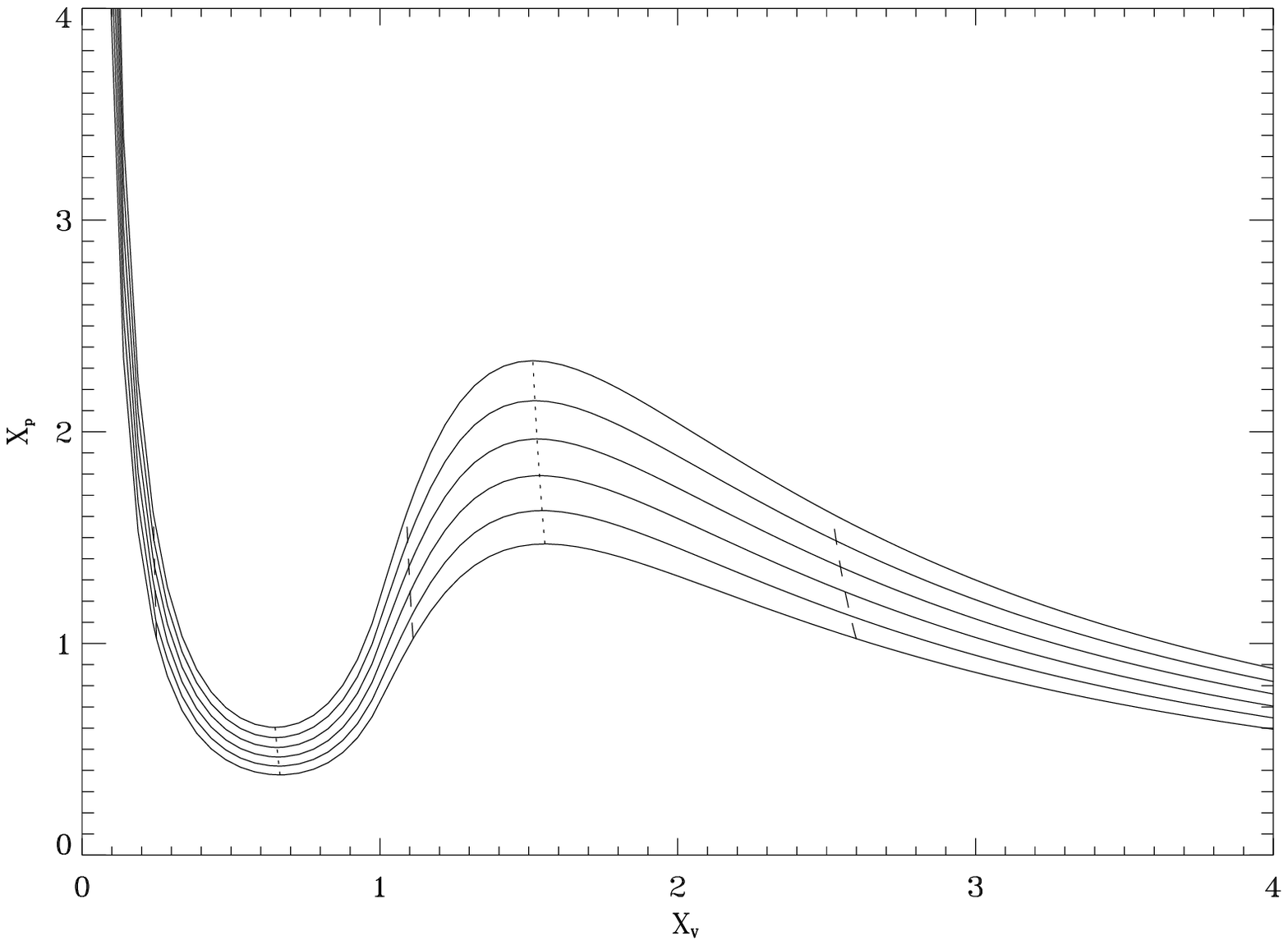}
\caption{Same as in Fig.\,\ref{f:uuso} (right panel).
The loci of the extremum points are represented as
dotted lines.   The loci of the intersections
between actual and  real macroisothermal
curves are represented as dashed lines.   The
absence of the critical macroisothermal curve
makes a band-like instead of a bell-shaped region exist on the plane.
The coordinates, $X_{\rm V}=X_{\rm V}^
\dagger$, $X_{\rm p}=X_{\rm p}^\dagger$, $X_{\rm T}=X_{\rm T}^\dagger$, may be
conceived as normalized to their fictitious critical
counterparts, $X_{V_{\rm c}}=X_{V_{\rm c}}^\dagger=1$, $X_{p_{\rm c}}=
X_{p_{\rm c}}^\dagger=1$, $X_{T_{\rm c}}=X_{T_{\rm c}}^\dagger=1$.}
\label{f:uuis}
\end{center}
\end{figure*}

Contrary to ordinary gases, no experiment can be
performed on macrogases to ascertain the existence
of a phase transition moving along a selected
macroisothermal curve, where the path is a horisontal
line instead of a curve including the extremum points.
Then the existence of the above mentioned phase
transition and flat real macroisothermal curves,
must necessarily be assumed as a working
hypothesis, by analogy with VDW isothermal curves
(below the critical one).   The loci of extremum
points of AUU macroisothermal curves plotted in
Fig.\,\ref{f:uuso} (right panel), are represented as
dotted lines in Fig.\,\ref{f:uuis}.

Unlike the VDW equation of state, Eq.\,(\ref{eq:rW2}),
the AUU macrogas equation of state,
Eq.\,(\ref{seq:sope}), is not analytically integrable.
Then the procedure used for determining a selected
macroisothermal curve, must be numerically performed.
The main steps are (i) calculate the intersections,
$X_{V_{\rm A}}$, $X_{V_{\rm C}}$, $X_{V_{\rm E}}$,
$X_{V_{\rm A}}< X_{V_{\rm C}}
< X_{V_{\rm E}}$, between the generic horisontal line in
the $({\sf O}X_{\rm V} X_{\rm p})$ plane, $X_{\rm p}=$ const, and the
AUU macrogas equation of state, within the range,
$X_{p_{\rm B}}< X_{\rm p}< X_{p_{\rm D}}$, where ${\sf B}$ and
${\sf D}$ denote the extremum points of minimum and
maximum, respectively; (ii) calculate the area of
the regions, ${\sf ABC}$ and ${\sf CDE}$; (iii) find
the special value, $X_{\rm p}=X_{p_0}$, which makes the
two areas equal; (iv) trace the real UU (RUU)
macroisothermal curve as a horisontal line connecting
the points, $(X_{V_{\rm A}},X_{p_{\rm A}})$, $(X_{V_{\rm C}},X_{p_{\rm C}})$,
$(X_{V_{\rm E}},X_{p_{\rm E}})$, $X_{p_{\rm A}}=X_{p_{\rm C}}=X_{p_{\rm E}}=
X_{p_0}$.

\begin{figure*}[t]
\begin{center}
\includegraphics[scale=0.8]{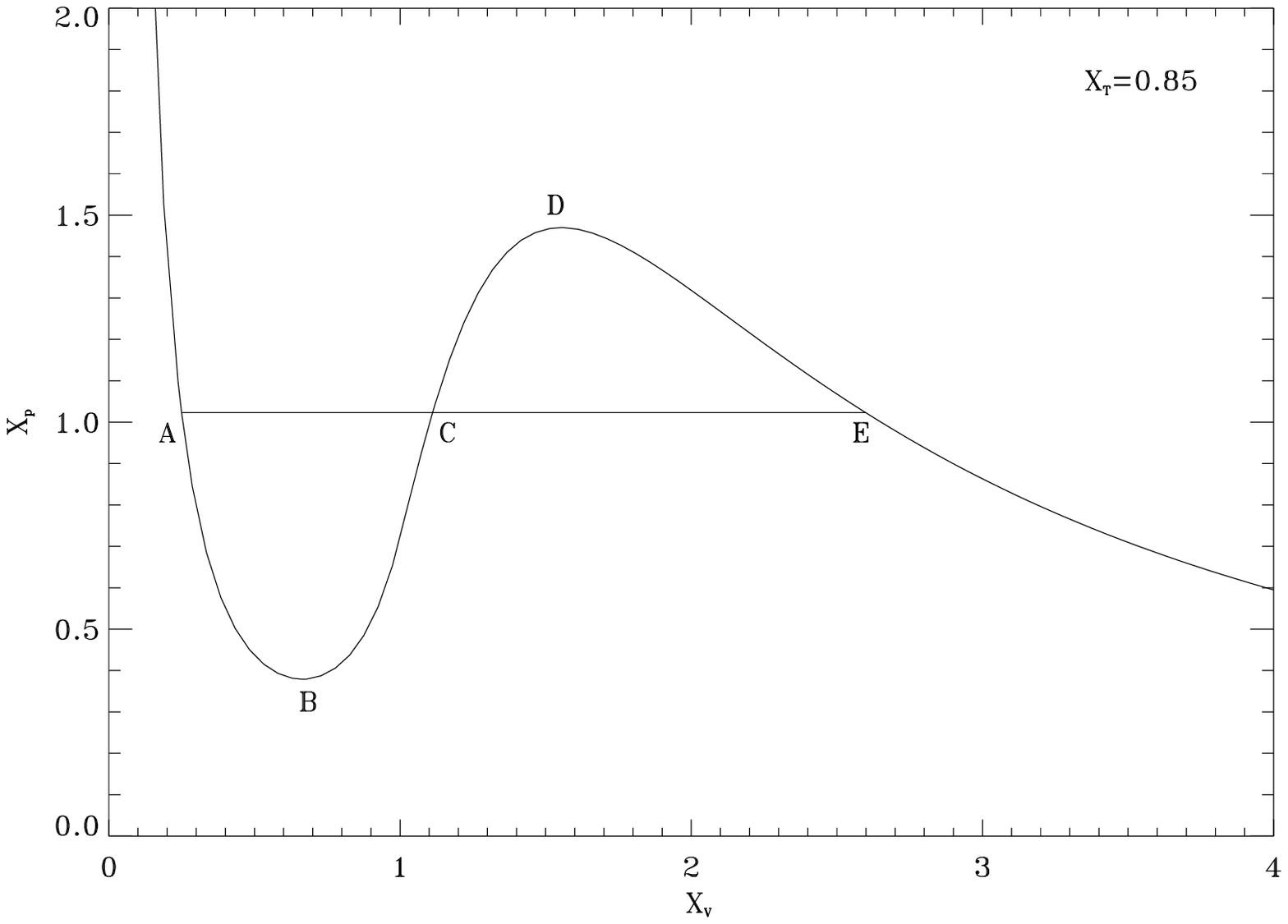}
\caption{A specific $(X_{\rm T}=0.85)$ AUU and corresponding
RUU macroisothermal curve.   The above mentioned curves 
coincide within the range, $X_{\rm V}\le X_{V_{\rm A}}$, $X_{\rm V}\ge X_
{V_{\rm E}}$.   The AUU macroisothermal curve  exhibits
two extremum points: a minimum, ${\sf B}$, and a maximum,
${\sf D}$, while the RUU macroisothermal curve is flat within
the range, $X_{V_{\rm A}}\le X_{\rm V}\le X_{V_{\rm E}}$.
The regions, {\sf ABC} and {\sf CDE}, have equal area.
The coordinates, $X_{\rm V}=X_{\rm V}^
\dagger$, $X_{\rm p}=X_{\rm p}^\dagger$, $X_{\rm T}=X_{\rm T}^\dagger$, may be
conceived as normalized to their fictitious critical
counterparts, $X_{V_{\rm c}}=X_{V_{\rm c}}^\dagger=1$, $X_{p_{\rm c}}=
X_{p_{\rm c}}^\dagger=1$, $X_{T_{\rm c}}=X_{T_{\rm c}}^\dagger=1$.}
\label{f:uuar}
\end{center}
\end{figure*}

The loci of the intersections between
AUU and RUU macroisothermal curves, are represented
as dashed lines in Fig.\,\ref{f:uuis}.  The absence
of a critical macroisothermal curve makes a band-like instead of a
bell-shaped region exist on the $({\sf O}X_{\rm V} X_{\rm p})$
plane, as shown by comparison with Fig.\,\ref{f:vris}.
The AUU and RUU macroisothermal curves
are plotted in Fig.\,\ref{f:uuar}, with regard to the
special case, $X_{\rm T}=0.85$.
Values of parameters, $X_{\rm T}$, $X_{V_{\rm A}}$, $X_{V_{\rm B}}$,
$X_{V_{\rm C}}$, $X_{V_{\rm D}}$, $X_{V_{\rm E}}$, $X_{p_{\rm B}}$,
$X_{p_{\rm C}}$,
$X_{p_{\rm D}}$, are listed in Tab.\,\ref{t:uuspo} within
the range, $0.85\le X_{\rm T}\le1.10$, using a step, $\Delta
X_{\rm T}=0.01$.
\begin{table}
\caption{Values of parameters, $X_{\rm T}$, $X_{V_{\rm A}}$, $X_{V_{\rm B}}$,
$X_{V_{\rm C}}$, $X_{V_{\rm D}}$, $X_{V_{\rm E}}$, $X_{p_{\rm B}}$,
$X_{p_{\rm C}}$,
$X_{p_{\rm D}}$ (to be conceived as normalized to their fictitious critical
counterparts, $X_{V_{\rm c}}=1$, $X_{p_{\rm c}}=1$,  $X_{T_{\rm c}}=1$),
within the range, $0.85\le X_{\rm T}\le1.10$, using a step,
$\Delta X_{\rm T}=0.01$.   Index captions: A, C, E - intersections
between AUU and RUU macroisothermal curves; B - extremum point
of minimum; D - extremum point of maximum.   Extremum points
are related to AUU macroisothermal curves, while their RUU counterparts
are flat within the range, $X_{V_{\rm A}}\le X_{\rm V}\le X_{V_{\rm E}}$.
For aesthetical reasons, 01 on head columns stands for unity.}
\label{t:uuspo}
\begin{center}
\begin{tabular}{|l|l|l|l|l|l|l|l|l|} \hline
$X_{\rm T}$ & $10X_{V_{\rm A}}$ & $10X_{V_{\rm B}}$ & $01X_{V_{\rm C}}$ & $01X_{V_{\rm D}}$
& $01X_{V_{\rm E}}$ & $10X_{p_{\rm B}}$ & $01X_{p_{\rm C}}$ & $01X_{p_{\rm D}}$ \\
\hline
0.85 & 2.4980 & 6.6478 & 1.1115 & 1.5541 & 2.5993 & 3.7895 & 1.0228 & 1.4699 \\
0.86 & 2.4939 & 6.6407 & 1.1105 & 1.5521 & 2.5952 & 3.8711 & 1.0440 & 1.5008 \\
0.87 & 2.4899 & 6.6337 & 1.1094 & 1.5501 & 2.5912 & 3.9534 & 1.0654 & 1.5320 \\
0.88 & 2.4858 & 6.6267 & 1.1084 & 1.5482 & 2.5873 & 4.0364 & 1.0870 & 1.5635 \\
0.89 & 2.4818 & 6.6197 & 1.1073 & 1.5463 & 2.5836 & 4.1201 & 1.1087 & 1.5953 \\
0.90 & 2.4776 & 6.6127 & 1.1064 & 1.5445 & 2.5798 & 4.2046 & 1.1307 & 1.6275 \\
0.91 & 2.4735 & 6.6059 & 1.1054 & 1.5427 & 2.5762 & 4.2897 & 1.1529 & 1.6599 \\
0.92 & 2.4693 & 6.5991 & 1.1044 & 1.5409 & 2.5725 & 4.3755 & 1.1753 & 1.6927 \\
0.93 & 2.4653 & 6.5922 & 1.1034 & 1.5392 & 2.5692 & 4.4620 & 1.1979 & 1.7258 \\
0.94 & 2.4610 & 6.5854 & 1.1025 & 1.5374 & 2.5657 & 4.5492 & 1.2207 & 1.7591 \\
0.95 & 2.4569 & 6.5786 & 1.1016 & 1.5358 & 2.5623 & 4.6371 & 1.2437 & 1.7928 \\
0.96 & 2.4526 & 6.5719 & 1.1007 & 1.5341 & 2.5590 & 4.7256 & 1.2669 & 1.8268 \\
0.97 & 2.4483 & 6.5653 & 1.0998 & 1.5325 & 2.5557 & 4.8149 & 1.2903 & 1.8611 \\
0.98 & 2.4442 & 6.5587 & 1.0989 & 1.5309 & 2.5527 & 4.9048 & 1.3138 & 1.8958 \\
0.99 & 2.4398 & 6.5520 & 1.0981 & 1.5293 & 2.5495 & 4.9954 & 1.3377 & 1.9307 \\
1.00 & 2.4356 & 6.5455 & 1.0972 & 1.5278 & 2.5465 & 5.0866 & 1.3616 & 1.9659 \\
1.01 & 2.4313 & 6.5389 & 1.0964 & 1.5263 & 2.5435 & 5.1785 & 1.3858 & 2.0015 \\
1.02 & 2.4271 & 6.5328 & 1.0955 & 1.5248 & 2.5407 & 5.2711 & 1.4101 & 2.0374 \\
1.03 & 2.4228 & 6.5260 & 1.0947 & 1.5233 & 2.5378 & 5.3643 & 1.4347 & 2.0735 \\
1.04 & 2.4184 & 6.5195 & 1.0939 & 1.5219 & 2.5349 & 5.4582 & 1.4595 & 2.1100 \\
1.05 & 2.4142 & 6.5131 & 1.0931 & 1.5204 & 2.5323 & 5.5528 & 1.4844 & 2.1468 \\
1.06 & 2.4098 & 6.5068 & 1.0924 & 1.5191 & 2.5295 & 5.6480 & 1.5097 & 2.1839 \\
1.07 & 2.4055 & 6.5004 & 1.0916 & 1.5177 & 2.5269 & 5.7438 & 1.5350 & 2.2213 \\
1.08 & 2.4011 & 6.4942 & 1.0908 & 1.5163 & 2.5242 & 5.8403 & 1.5606 & 2.2590 \\
1.09 & 2.3969 & 6.4879 & 1.0901 & 1.5150 & 2.5217 & 5.9374 & 1.5863 & 2.2971 \\
1.10 & 2.3925 & 6.4816 & 1.0894 & 1.5137 & 2.5192 & 6.0351 & 1.6123 & 2.3354 \\
\hline
\end{tabular}
\end{center}
\end{table}

\subsection{HH macrogases}\label{HH}

The related density profiles exhibit a central
cusp and null value at infinite distances
(Hernquist, 1990), and have been proved to be
consistent with nonnegative distribution functions,
in an acceptable parameter range (Ciotti, 1996).
The particularization of the general expressions
to the case under discussion, yields for the
quantities of interest (CV08):
\begin{lefteqnarray}
\label{eq:fH}
&& f_u(\xi_u)=\frac8{\xi_u(1+\xi_u)^3}~~;\qquad0\le\xi_u\le\Xi_u~~;\qquad
u=i,j~~; \\
\label{eq:FH}
&& F_u(\xi_u)=\frac8{(1+\xi_u)^2}-\frac8{(1+\Xi_u)^2}~~;\qquad u=i,j~~; \\
\label{eq:numH}
&& (\nu_u)_{\rm mas}=\frac{12\Xi_u^2}{(1+\Xi_u)^2}~~;\qquad u=i,j~~; \\
\label{eq:nusH}
&& (\nu_u)_{\rm sel}=\frac{12\Xi_u^3(4+\Xi_u)}{(1+\Xi_u)^4}~~;\qquad
u=i,j~~;
\end{lefteqnarray}
\begin{leftsubeqnarray}
\slabel{eq:wiHa}
&& w^{({\rm int})}(\eta)=-128y^\dagger\left\{\frac12\frac1{(y^\dagger-1)^4}
\left[-\frac{(y^\dagger-1)^2y^\dagger\eta}{(y^\dagger\eta+1)^2}+\frac
{2(y^\dagger-1)\eta}{1+\eta}\right.\right. \nonumber \\
&& \left.+\frac{(y^\dagger-1)(y^\dagger+3)y^\dagger\eta}{y^\dagger\eta+
1}+2(2y^\dagger+1)\ln\frac{\eta+1}{y^\dagger\eta+1}\right] \nonumber \\
&& \left.-\frac12\frac1{(1+\Xi_j)^2}\frac1{(y^\dagger)^2}\left[1-\frac
{2y^\dagger\eta+1}{(y^\dagger\eta+1)^2}\right]\right\}~~;\qquad y^\dagger
\ne1~~; \\
\slabel{eq:wiHb}
&& w^{({\rm int})}(\eta)=-128\left\{\frac1{12}\left[-\frac{4\eta+1}
{(\eta+1)^4}+1\right]\right. \nonumber \\
&& \left.-\frac12\frac1{(1+\Xi_j)^2}\frac{\eta^2}{(\eta+1)^2}
\right\}~~;\qquad y^\dagger=1~~;
\label{seq:wiH}
\end{leftsubeqnarray}
\begin{leftsubeqnarray}
\slabel{eq:weHa}
&& w^{({\rm ext})}(\eta)=-128\left\{-\frac12\frac1{(y^\dagger-1)^4}
\left[\frac{(y^\dagger-1)^2\eta}{(\eta+1)^2}+\frac
{2(y^\dagger)^2(y^\dagger-1)\eta}{1+y^\dagger\eta}\right.\right. \nonumber \\
&& \left.+\frac{(y^\dagger-1)(3y^\dagger+1)\eta}{\eta+
1}-2y^\dagger(y^\dagger+2)\ln\frac{y^\dagger\eta+1}{\eta+1}\right] \nonumber \\
&& \left.-\frac12\frac1{(1+\Xi_i)^2}
\frac{\eta^2}{(\eta+1)^2}\right\}~~;\qquad y^\dagger\ne1~~; \\
\slabel{eq:weHb}
&& w^{({\rm ext})}(\eta)=128\left\{\frac1{12}\left[\frac{4\eta+1}
{(\eta+1)^4}-1\right]\right. \nonumber \\
&& \left.+\frac12\frac1{(1+\Xi_i)^2}\frac{\eta^2}{(\eta+1)^2}
\right\}~~;\qquad y^\dagger=1~~;
\label{seq:weH}
\end{leftsubeqnarray}
using Eqs.\,(\ref{seq:nuuv}) and
(\ref{eq:numH})-(\ref{seq:weH}), the actual HH
(AHH) macrogas equation of state is obtained
from the particularization of Eqs.\,(\ref
{seq:Xr}) to the case of interest for
the domain, $y\ge1$.
The extension
to the domain, $0\le y\le1$, can be
done following the procedure outlined
above in dealing with  Eq.\,(\ref{eq:phic}).

In absence of truncation radius,
$\Xi\to+\infty$, $\eta\to+\infty$,
and Eqs.\,(\ref{eq:FH})-(\ref{seq:weH})
reduce to (CV08):
\begin{lefteqnarray}
\label{eq:FlH}
&& \lim_{\Xi_u\to+\infty}F_u(\xi_u)=\frac8{(1+\xi_u)^2}~~;\qquad u=i,j~~; \\
\label{eq:lumH}
&& \lim_{\Xi_u\to+\infty}(\nu_u)_{\rm mas}=12~~;\qquad u=i,j~~; \\
\label{eq:lusH}
&& \lim_{\Xi_u\to+\infty}(\nu_u)_{\rm sel}=12~~;\qquad u=i,j~~;
\end{lefteqnarray}
\begin{leftsubeqnarray}
\slabel{eq:wilHa}
&& \lim_{\eta\to+\infty}w^{({\rm int})}(\eta)=-\frac{64y^\dagger}
{(y^\dagger-1)^4}\left[-2(2y^\dagger+1)\ln y^\dagger+(y^\dagger-1)
(y^\dagger+5)\right]~~; \nonumber \\
&& y^\dagger\ne1~~; \\
\slabel{eq:wilHb}
&& \lim_{\eta\to+\infty}w^{({\rm int})}(\eta)=-\frac{32}3~~;\qquad
y^\dagger=1~~;
\label{seq:wilH}
\end{leftsubeqnarray}
\begin{leftsubeqnarray}
\slabel{eq:welHa}
&& \lim_{\eta\to+\infty}w^{({\rm ext})}(\eta)=-\frac{64}
{(y^\dagger-1)^4}\left[2y^\dagger(y^\dagger+2)\ln y^\dagger-(y^\dagger-1)
(5y^\dagger+1)\right]~~; \nonumber \\
&& y^\dagger\ne1~~; \\
\slabel{eq:welHb}
&& \lim_{\eta\to+\infty}w^{({\rm ext})}(\eta)=-\frac{32}3~~;\qquad
y^\dagger=1~~;
\label{seq:welH}
\end{leftsubeqnarray}
using Eqs.\,(\ref{seq:nuuv}) and
(\ref{eq:lumH})-(\ref{seq:welH}), the AHH
macrogas equation of state in the
special situation under discussion, is
obtained from the particularization of
Eqs.\,(\ref{seq:Xr}) to the case of interest.

The ideal situation, where the interaction
terms are omitted, is obtained using
Eqs.\,(\ref{seq:Xi}) instead of
Eqs.\,(\ref{seq:Xr}).   More specifically,
the ideal HH (IHH) macrogas equation of state 
is derived by the combination of 
Eqs.\,(\ref{eq:Xre}), (\ref{eq:Xib}),
(\ref{eq:numH}), and
(\ref{eq:nusH}).   The result is:
\begin{equation}
\label{eq:XpHi}
X_{\rm p}=\frac{4+\Xi_i}{4+\Xi_j}\frac{X_{\rm T}}{X_{\rm V}}~~;
\end{equation}
which represents a hyperbola with different
axes (unless $\Xi_i=\Xi_j$), for fixed $X_{\rm T}$.

\begin{figure*}[t]
\begin{center}
\includegraphics[scale=0.8]{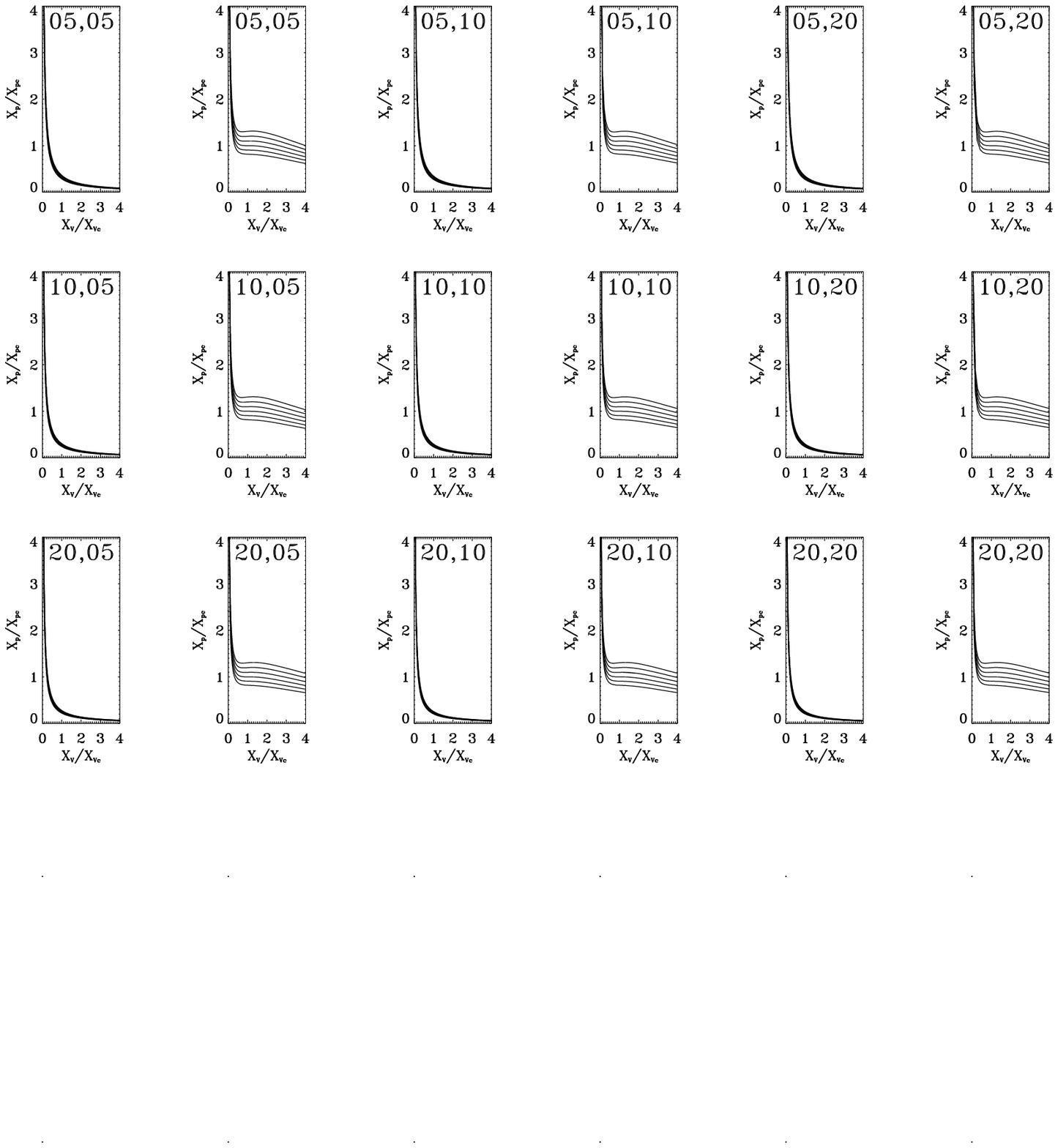}
\caption{Macroisothermal curves
($\sX_{\rm p}=X_{\rm p}/X_{p_c}$ vs. $\sX_{\rm V}=X_{\rm V}/X_{V_c}$)
related to IHH (left panels)
and AHH (right panels) macrogases, respectively, for
different values of scaled truncation radii,
$(\Xi_i,\Xi_j)$, labelled on each panel.   Macroisothermal
curves (from bottom to top) correspond to $\sX_{\rm T}=X_{\rm T}/X_{T_{\rm c}}=$
0.90, 0.95, 1.00, 1.05, 1.10, 1.15.   The limit, $(\Xi_i,\Xi_j)
\to(+\infty,+\infty)$, makes only little changes.   Owing to
Eqs.\,(\ref{eq:yma}) and (\ref{eq:XXC}), $X_{\rm V}^\dagger=X_{\rm V}
y^\dagger/y$ and $X_{\rm p}^\dagger=X_{\rm p}(m^\dagger/m)^2$,
which implies $\sX_{\rm V}^\dagger=\sX_{\rm V}$ and $\sX_{\rm p}^\dagger=
\sX_{\rm p}$.}
\label{f:hhso}
\end{center}
\end{figure*}

Macroisothermal curves ($\sX_{\rm p}=X_{\rm p}/X_{p_c}$
vs. $\sX_{\rm V}=X_{\rm V}/X_{V_c}$) related to IHH
(tidal potential energy excluded) and AHH
(tidal potential energy included)
macrogases, are plotted in
Fig.\,\ref{f:hhso}, left and right panels, for
different values of scaled truncation radii,
$(\Xi_i,\Xi_j)$, labelled on each panel, and
same values of the reduced macrotemperature,
$\sX_{\rm T}=X_{\rm T}/X_{T_{\rm c}}=$ 0.90, 0.95, 1.00,
1.05, 1.10, 1.15, from bottom to top.
The limit, $(\Xi_i,\Xi_j)\to(+\infty,+\infty)$, makes only
little changes.   The comparison with ideal and VDW gases,
plotted in Fig.\,\ref{f:viso}, shows a similar but reversed
trend.   More specifically, extremum points occur above,
instead of below, the critical macroisothermal curve.   A complete
analogy can be obtained using the transformations, $X_{\rm V}\to1/X_{\rm V}$,
$X_{\rm p}\to1/X_{\rm p}$, $X_{\rm T}\to1/X_{\rm T}$.

\begin{figure*}[t]                                                                                  
\begin{center}                                                                                      
\includegraphics[scale=0.8]{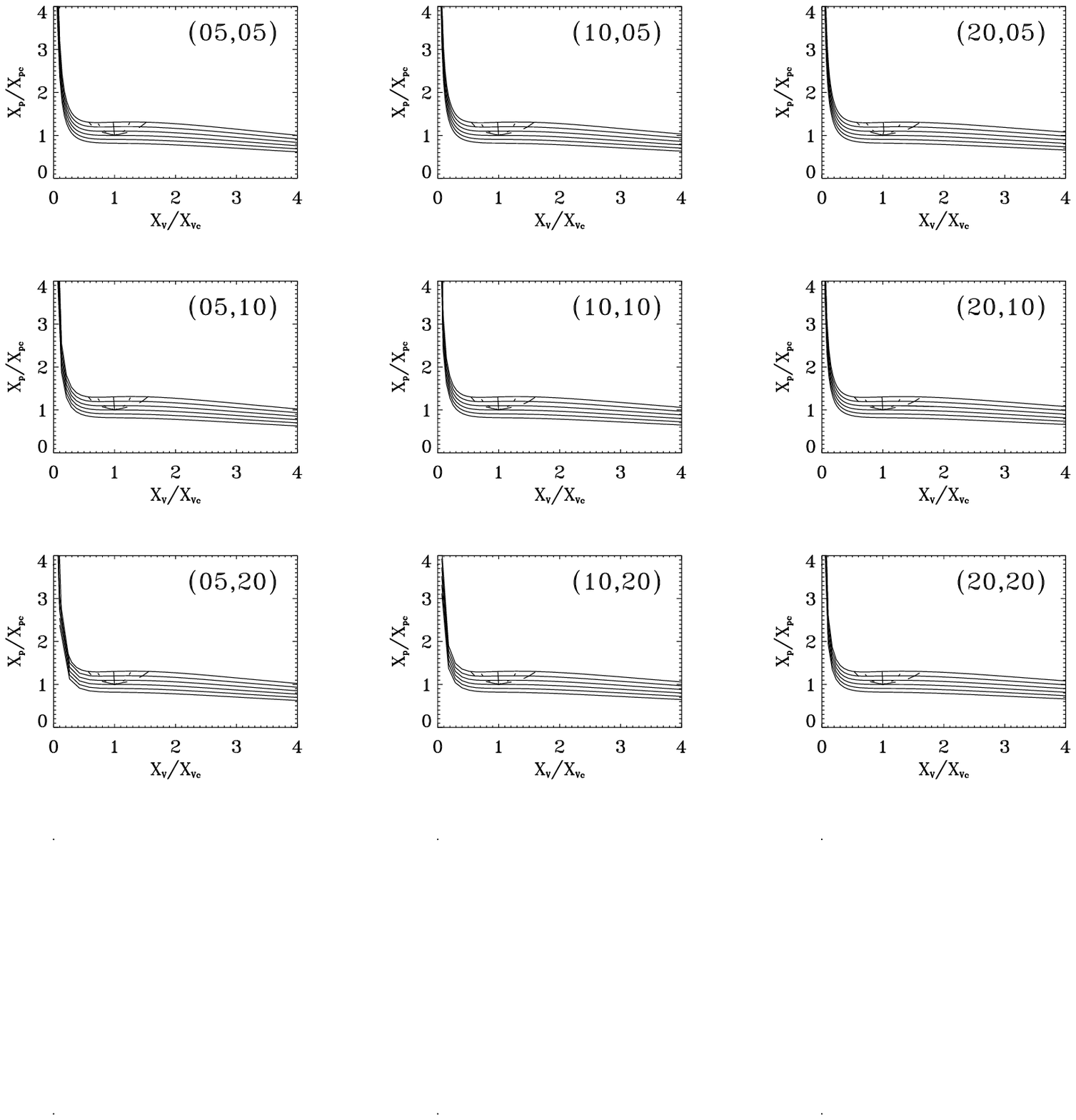}                                                               
\caption{AHH macroisothermal curves
($\sX_{\rm p}=X_{\rm p}/X_{p_c}$ vs. $\sX_{\rm V}=X_{\rm V}/X_{V_c}$)
for different choices of scaled truncation radii, $(\Xi_i,\Xi_j)$,
labelled on each panel.   Macroisothermal curves (from bottom to
top) correspond to $\sX_{\rm T}=X_{\rm T}/X_{T_{\rm c}}=$ 0.90, 0.95, 1.00,
1.05, 1.10, 1.15.   The limit, $(\Xi_i,\Xi_j)\to(+\infty,+\infty)$,
makes only little changes.   RHH macroisothermal curves, when different from
their AHH counterparts, lie within the larger reversed
bell-shaped region in each panel.   The loci of intersections
between AHH and RHH macroisothermal curves are represented as
trifid curves, where the left branch corresponds to $\sX_{V_{\rm A}}$,
the right branch to $\sX_{V_{\rm E}}$, and the middle branch to
$\sX_{V_{\rm C}}$.   The critical point is the common origin.
The loci of AHH macroisothermal curve extremum points are represented
as dotted curves starting from the critical point, where the left
branch corresponds to minimum points and the right branch to maximum
points.   Owing to
Eqs.\,(\ref{eq:yma}) and (\ref{eq:XXC}), $X_{\rm V}^\dagger=X_{\rm V}
y^\dagger/y$ and
$X_{\rm p}^\dagger=X_{\rm p}(m^\dagger/m)^2$,
which implies $\sX_{\rm V}^\dagger=\sX_{\rm V}$ and $\sX_{\rm p}^\dagger=
\sX_{\rm p}$.}
\label{f:hhis}
\end{center}
\end{figure*}

\begin{figure*}[t]                                                                                  
\begin{center}                                                                                      
\includegraphics[scale=0.8]{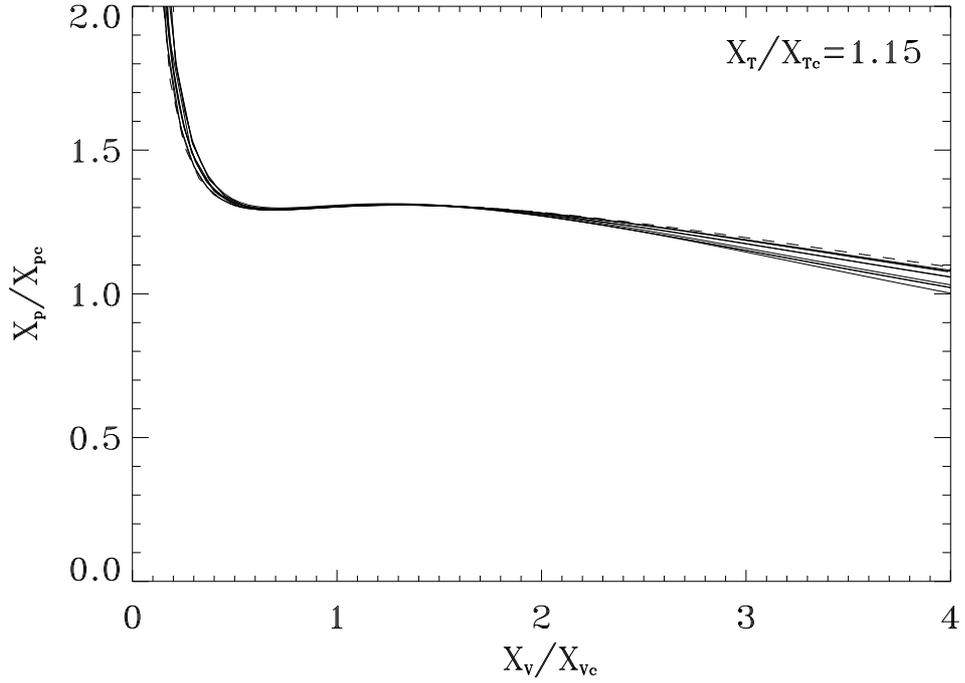}                                                               
\caption{AHH macroisothermal curves
for scaled truncation radii, $(\Xi_i,\Xi_j)$, as in
Fig.\,\ref{f:hhis}, with regard to the special case,
$\sX_{\rm T}=X_{\rm T}/X_{T_{\rm c}}=1.15$.   The limit,
$(\Xi_i,\Xi_j)\to(+\infty,+\infty)$, makes only little
changes, as shown by comparison with the dashed curve.
Owing to Eqs.\,(\ref{eq:yma}) and (\ref{eq:XXC}),
$X_{\rm V}^\dagger=X_{\rm V}y^\dagger/y$ and $X_{\rm p}^\dagger=X_{\rm p}
(m^\dagger/m)^2$, which implies $\sX_{\rm V}^\dagger=\sX_{\rm V}$ and
$\sX_{\rm p}^\dagger=\sX_{\rm p}$.}
\label{f:hhar}
\end{center}
\end{figure*}

The existence of a phase transition moving along a selected
macroisothermal curve, where the path is a horisontal line
instead of a curve including the extremum points, must necessarily
be assumed as a working hypothesis, due to the analogy between VDW
isothermal curves and AHH macroisothermal curves.
As in the case of UU macrogases, AHH macroisothermal curves must be
numerically determined, following the same procedure outlined in
Subsection \ref{UU}.
Characteristic loci of
AHH macroisothermal curves plotted in Fig.\,\ref{f:hhso}
(right panels), are represented
in Fig.\,\ref{f:hhis}.
The loci of intersections between AHH and
real HH (RHH) macroisothermal curves, are represented
in Fig.\,\ref{f:hhis} as trifid curves, where the left branch
corresponds to $\sX_{V_{\rm A}}$, the right branch to $\sX_{V_
{\rm E}}$, and the middle branch to $\sX_{V_{\rm C}}$.  The
critical point is the common starting point.   The loci of AHH
macroisothermal curve extremum points are represented in
Fig.\,\ref{f:hhis} as dotted curves starting from the critical
point, where the left branch corresponds to minimum points and
the right branch to maximum points.   The  RHH macroisothermal
curves, when different from their AHH counterparts, lie within
the larger bell-shaped regions.   The limit,
$(\Xi_i,\Xi_j)\to(+\infty,+\infty)$, makes only little changes.
Values of parameters, $\sX_{\rm T}$, $\sX_{V_{\rm A}}$, $\sX_{V_{\rm B}}$,
$\sX_{V_{\rm C}}$, $\sX_{V_{\rm D}}$, $\sX_{V_{\rm E}}$, $\sX_{p_{\rm B}}$,
$\sX_{p_{\rm C}}$, $\sX_{p_{\rm D}}$, are listed in Tab.\,\ref{t:hhspo}
within the range, $1.00<\sX_{\rm T}\le1.15$, using a step, $\Delta\sX_{\rm T}=0.01$,
in the limit, $(\Xi_i,\Xi_j)\to(+\infty,+\infty)$.   The AHH macroisothermal
curves are plotted in Fig.\,\ref{f:hhar} for scaled truncation radii,
$(\Xi_i,\Xi_j)$, as in Fig.\,\ref{f:hhis}, with regard to the special
case, $\sX_{\rm T}=1.15$.   The limit, $(\Xi_i,\Xi_j)\to(+\infty,+\infty)$,
makes only little changes, as shown by comparison with the dashed curve.
\begin{table}
\caption{Values of parameters, $\sX_{\rm T}$, $\sX_{V_{\rm A}}$, $\sX_{V_{\rm B}}$,
$\sX_{V_{\rm C}}$, $\sX_{V_{\rm D}}$, $\sX_{V_{\rm E}}$, $\sX_{p_{\rm B}}$,
$\sX_{p_{\rm C}}$, $\sX_{p_{\rm D}}$,
within the range, $1.00<\sX_{\rm T}\le1.15$, using a step,
$\Delta\sX_{\rm T}=0.01$, except near the critical point, where convergence
was not attained.   All values equal unity at the critical point.   Results
are related to infinitely extended configurations,
$(\Xi_i,\Xi_j)\to(+\infty,+\infty)$.   Index captions: A, C, E -
intersections
between AHH and RHH macroisothermal curves; B - extremum point
of minimum; D - extremum point of maximum.   Extremum points
are related to AHH macroisothermal curves, while their RHH counterparts
are flat within the range, $\sX_{V_{\rm A}}\le\sX_{\rm V}\le\sX_{V_{\rm E}}$.
For aesthetical reasons, 01 on head columns stands for unity.}
\label{t:hhspo}
\begin{center}
\begin{tabular}{|l|l|l|l|l|l|l|l|l|} \hline
$\sX_{\rm T}$ & $10\sX_{V_{\rm A}}$ & $10\sX_{V_{\rm B}}$ & $10\sX_{V_{\rm C}}$ & $01\sX_{V_{\rm D}}$
& $01\sX_{V_{\rm E}}$ & $01\sX_{p_{\rm B}}$ & $01\sX_{p_{\rm C}}$ & $01\sX_{p_{\rm D}}$ \\
\hline
 1.014 & 8.3150 & 8.9774 & 9.9858 & 1.1016 & 1.1854 & 1.0264 & 1.0266 & 1.0268 \\
 1.015 & 8.2595 & 8.9420 & 9.9849 & 1.1051 & 1.1921 & 1.0283 & 1.0285 & 1.0287 \\
 1.02  & 8.0099 & 8.7810 & 9.9803 & 1.1211 & 1.2231 & 1.0378 & 1.0381 & 1.0384 \\
 1.03  & 7.6038 & 8.5136 & 9.9711 & 1.1475 & 1.2756 & 1.0568 & 1.0575 & 1.0580 \\
 1.04  & 7.2739 & 8.2912 & 9.9620 & 1.1694 & 1.3203 & 1.0759 & 1.0770 & 1.0778 \\
 1.05  & 6.9927 & 8.0979 & 9.9533 & 1.1884 & 1.3600 & 1.0951 & 1.0966 & 1.0978 \\
 1.06  & 6.7463 & 7.9255 & 9.9446 & 1.2054 & 1.3961 & 1.1144 & 1.1165 & 1.1181 \\
 1.07  & 6.5262 & 7.7690 & 9.9359 & 1.2207 & 1.4293 & 1.1338 & 1.1365 & 1.1385 \\
 1.08  & 6.3269 & 7.6252 & 9.9277 & 1.2348 & 1.4604 & 1.1533 & 1.1566 & 1.1591 \\
 1.09  & 6.1447 & 7.4920 & 9.9192 & 1.2478 & 1.4895 & 1.1730 & 1.1770 & 1.1799 \\
 1.10  & 5.9767 & 7.3676 & 9.9112 & 1.2600 & 1.5171 & 1.1927 & 1.1975 & 1.2009 \\
 1.11  & 5.8209 & 7.2508 & 9.9031 & 1.2714 & 1.5434 & 1.2126 & 1.2182 & 1.2222 \\
 1.12  & 5.6755 & 7.1406 & 9.8950 & 1.2821 & 1.5685 & 1.2325 & 1.2391 & 1.2436 \\
 1.13  & 5.5393 & 7.0362 & 9.8872 & 1.2923 & 1.5925 & 1.2526 & 1.2601 & 1.2652 \\
 1.14  & 5.4111 & 6.9370 & 9.8796 & 1.3019 & 1.6155 & 1.2728 & 1.2813 & 1.2870 \\
 1.15  & 5.2902 & 6.8425 & 9.8720 & 1.3111 & 1.6378 & 1.2931 & 1.3027 & 1.3090 \\
\hline                                                                                              
\end{tabular}                                                                                       
\end{center}                                                                                        
\end{table}                                                                                         

Values of parameters, $m^\dagger$, $m$,
$y^\dagger$, $y$, $\phi$, related to the
critical macroisothermal curve, for
selected scaled truncation radii,
$(\Xi_i,\Xi_j)$, are listed in
Tab.\,\ref{t:hhcri}.
\begin{table}
\caption{Values of the scaling fractional
mass, $m^\dagger$, the fractional mass, $m$,
the scaling fractional radius, $y^\dagger$, 
the fractional truncation radius, $y$, and the fractional
energy, $\phi$, related to the
critical point i.e. the horizontal inflexion
point on the critical macroisothermal
curve, for selected scaled truncation radii,
$(\Xi_i,\Xi_j)$, of HH density profiles.
In absence of truncation radius, $(\Xi_i,\Xi_j)\to(+\infty,+\infty)$, the
results are independent of $y/y^\dagger=
\Xi_j/\Xi_i$.}
\label{t:hhcri}
\begin{center}
\begin{tabular}{|c|c|c|c|c|c|} \hline
($\Xi_i$,$\Xi_j$)& $m^\dagger$ & $m$ & $y^\dagger$ & $y$ &
$\varphi$ \\
\hline
(05,05)             & 07.10 & 07.10 & 2.32 & 2.32    & 06.86 \\
(05,10)             & 07.72 & 09.18 & 2.43 & 4.86    & 08.17 \\
(05,20)             & 07.90 & 10.31 & 2.46 & 9.83    & 08.62 \\
(10,05)             & 11.15 & 09.37 & 3.18 & 1.59    & 09.44 \\
(10,10)             & 11.88 & 11.88 & 3.30 & 3.30    & 11.03 \\
(10,20)             & 12.10 & 13.28 & 3.33 & 6.67    & 11.57 \\
(20,05)             & 15.06 & 11.53 & 4.00 & 1.00    & 12.09 \\
(20,10)             & 15.83 & 14.42 & 3.88 & 1.94    & 13.94 \\
(20,20)             & 16.08 & 16.08 & 3.92 & 3.92    & 14.59 \\
($\infty$,$\infty$) & 20.22 & 20.22 & 4.26 & $\dots$ & 18.15 \\
\hline
\end{tabular}                                                                                       
\end{center}                                                                                        
\end{table}                                                                                         
A better method has been used with respect
to an earlier attempt (CV08), yielding
improved results.

\subsection{HN/NH macrogases}\label{HN}

A description of H density profiles has
been provided in Subsection \ref{HH}.
The remaining N density profiles also
exhibit a central cusp and null values
at infinite distances
(Navarro et al., 1995, 1996, 1997).
Mass distributions defined by an inner
H and outer N density profile were found
to be self-consistent, in an acceptable
parameter range, with regard to the
non negativity of the distribution
function (Lowenstein and White, 1999),
using a theorem stated in an earlier
attempt (Ciotti and Pellegrini, 1992).
The particularization of the general
expressions to N density profiles and
HN macrogases, yields for the quantities
of interest (CV08):
\begin{lefteqnarray}
\label{eq:fN}
&& f_u(\xi_u)=\frac4{\xi_u(1+\xi_u)^2}~~;
\qquad0\le\xi_u\le\Xi_u~~;\qquad u=N~~; \\
\label{eq:FN}
&& F_u(\xi_u)=\frac8{1+\xi_u}-\frac8{1+\Xi_u}~~;\qquad u=N~~; \\
\label{eq:numN}
&& (\nu_u)_{\rm mas}=12\left[\ln(1+\Xi_u)-\frac{\Xi_u}{1+\Xi_u}\right]~~;
\qquad u=N~~; \\
\label{eq:nusN}
&& (\nu_u)_{\rm sel}=36\frac{\Xi_u(2+\Xi_u)-2(1+\Xi_u)\ln(1+\Xi_u)}
{(1+\Xi_u)^2}~~;
\end{lefteqnarray}
\begin{leftsubeqnarray}
\slabel{eq:wiHNa}
&& w^{({\rm int})}(\eta)=-\frac{64y^\dagger}{(y^\dagger-1)^3}\left[-\frac
{(y^\dagger-1)^2}{y^\dagger}\frac{y^\dagger\eta(y^\dagger\eta+2)}
{(y^\dagger\eta+1)^2}+\frac{2y^\dagger\eta(y^\dagger-1)}{y^\dagger\eta+1}
\right. \nonumber \\
&& \phantom{w^{({\rm int})}(\eta)=}
\left.+2\ln\frac{\eta+1}{y^\dagger\eta+1}-\frac{(y^\dagger-1)^3}{1+\Xi_j}
\frac{\eta^2}{(y^\dagger\eta+1)^2}\right]~~; \nonumber \\
&& \phantom{w^{({\rm int})}(\eta)=}y^\dagger\ne1~~;\qquad i=H~~;\qquad j=N~~; \\
\slabel{eq:wiHNb}
&& w^{({\rm int})}(\eta)=-\frac{64\eta^2}{(\eta+1)^2}\left[\frac{\eta+3}
{3(\eta+1)}-\frac1{1+\Xi_j}\right]~~;\nonumber \\
&& \phantom{w^{({\rm int})}(\eta)=}y^\dagger=1~~;\qquad i=H~~;\qquad j=N~~;
\label{seq:wiHN}
\end{leftsubeqnarray}
\begin{leftsubeqnarray}
\slabel{eq:weHNa}
&& w^{({\rm ext})}(\eta)=-\frac{64}{(y^\dagger-1)^2}\left\{-\frac\eta{\eta+1}-
\frac{y^\dagger\eta}{y^\dagger\eta+1}-\frac{y^\dagger+1}{y^\dagger-1}\ln\frac
{\eta+1}{y^\dagger\eta+1}\right. \nonumber \\
&& \phantom{w^{({\rm ext})}(\eta)=}
\left.-\frac{(y^\dagger-1)^2}{(1+\Xi_i)^2}\left[-\frac\eta{\eta+1}+\ln
(\eta+1)\right]\right\}~~;\nonumber \\
&& \phantom{w^{({\rm ext})}(\eta)=}y^\dagger\ne1~~;\qquad i=H~~;\qquad j=N~~; \\
\slabel{eq:weHNb}
&& w^{({\rm ext})}(\eta)=-64\left\{\frac16\frac{\eta^2(\eta+3)}{(\eta+1)^3}
-\frac1{(1+\Xi_i)^2}\left[-\frac\eta{\eta+1}+\ln(\eta+1)\right]\right\}~~;
\nonumber \\
&& \phantom{w^{({\rm ext})}(\eta)=}y^\dagger=1~~;\qquad i=H~~;\qquad j=N~~;
\label{seq:weHN}
\end{leftsubeqnarray}
using Eqs.\,(\ref{seq:nuuv}), (\ref{eq:numH}),
(\ref{eq:nusH}), and
(\ref{eq:numN})-(\ref{seq:weHN}), the HN
macrogas equation of state is obtained
from the particularization of Eqs.\,(\ref
{seq:Xr}) to the case of interest for
the domain, $y\ge1$.   The extension
to the domain, $0\le y\le1$, can be
done following the procedure outlined
above in dealing with Eq.\,(\ref{eq:phic}).

To this aim, the counterparts of
Eqs.\,(\ref{seq:wiHN}) and (\ref{seq:weHN}),
related to NH macrogases ($i=N$, $j=H$), are
needed.   The particularization of
Eqs.\,(\ref{seq:wie}) to the case under
discussion, yields:
\begin{leftsubeqnarray}
\slabel{eq:wiNHa}
&& w^{({\rm int})}(\eta)=-\frac{64y^\dagger}{(y^\dagger-1)^2}\left\{-\frac
\eta{1+\eta}-\frac{\Xi_i}{1+\Xi_i}-\frac
{y^\dagger+1}{y^\dagger-1}\ln\frac{1+\eta}{1+\Xi_i}\right. \nonumber \\
&& \phantom{w^{({\rm int})}(\eta)=}\left.
-\frac{(y^\dagger-1)^2}{(y^\dagger)^2}\frac1{(1+\Xi_j)^2}
\left[\ln(1+\Xi_i)-\frac{\Xi_i}{1+\Xi_i}\right]\right\}~~; \nonumber \\
&& \phantom{w^{({\rm int})}(\eta)=}y^\dagger\ne1~~;\qquad i=N~~;\qquad j=H~~; \\
\slabel{eq:wiNHb}
&& w^{({\rm int})}(\eta)=-64\left\{\frac16\frac{\eta^2(3+\eta)}{(1+\eta)^3}-
\frac1{(1+\Xi_j)^2}\left[\ln(1+\eta)-\frac\eta{1+\eta}\right]\right\}~~;
\nonumber \\
&& \phantom{w^{({\rm int})}(\eta)=}y^\dagger=1~~;\qquad i=N~~;\qquad j=H~~;
\label{seq:wiNH}
\end{leftsubeqnarray}
\begin{leftsubeqnarray}
\slabel{eq:weNHa}
&& w^{({\rm ext})}(\eta)=-\frac{64}{(y^\dagger-1)^3}\left[2(y^\dagger-1)
\frac\eta{1+\eta}+(y^\dagger-1)^2\frac{\eta(2+\eta)}{(1+\eta)^2}
\right. \nonumber \\
&& \phantom{w^{({\rm ext})}(\eta)=}
\left.+2y^\dagger\ln\frac{1+\eta}{1+\Xi_i}-\frac{(y^\dagger-1)^3}{1+\Xi_i}
\frac{\eta^2}{(1+\eta)^2}\right]~~; \nonumber \\
&& \phantom{w^{({\rm ext})}(\eta)=}y^\dagger\ne1~~;\qquad i=N~~;\qquad j=H~~; \\
\slabel{eq:weNHb}
&& w^{({\rm ext})}(\eta)=-64\left[\frac13\frac{\eta^2(3+\eta)}{(1+\eta)^3}
-\frac1{1+\Xi_i}\frac{\eta^2}{(1+\eta)^2}\right]~~; \nonumber \\
&& \phantom{w^{({\rm ext})}(\eta)=}y^\dagger=1~~;\qquad i=N~~;\qquad j=H~~;
\label{seq:weNH}
\end{leftsubeqnarray}
where $\Xi_i=\Xi_N$, $\Xi_j=\Xi_H$, while the
contrary holds with regard to Eqs.\,(\ref
{seq:wiHN}) and (\ref{seq:weHN}).
Using Eqs.\,(\ref{seq:nuuv}), (\ref{eq:numH}),
(\ref{eq:nusH}), (\ref{eq:numN}), (\ref{eq:nusN}),
(\ref{seq:wiNH}), and (\ref{seq:weNH}), the NH
macrogas equation of state is obtained
from the particularization of Eqs.\,(\ref
{seq:Xr}) to the case of interest for
the domain, $y\ge1$, which corresponds to
$0\le y\le1$ for HN macrogases and vice versa.

In absence of truncation radius, $\Xi\to+\infty$,
$\eta\to+\infty$, and Eqs.\,(\ref{eq:FN})-(\ref{seq:weHN})
reduce to:
\begin{lefteqnarray}
\label{eq:FlN}
&& \lim_{\Xi_u\to+\infty}F_u(\xi_u)=\frac8{1+\xi_u}~~;\qquad u=N~~; \\
\label{eq:lumN}
&& \lim_{\Xi_u\to+\infty}(\nu_u)_{\rm mas}=+\infty~~;\qquad u=N~~; \\
\label{eq:lusN}
&& \lim_{\Xi_u\to+\infty}(\nu_u)_{\rm sel}=36~~;\qquad u=N~~;
\end{lefteqnarray}
\begin{leftsubeqnarray}
\slabel{eq:liHNa}
&& \lim_{\eta\to+\infty}w^{({\rm int})}(\eta)=-\frac{64}
{(y^\dagger-1)^3}\left[(y^\dagger)^2-1-2y^\dagger\ln
y^\dagger\right]~~;\qquad y^\dagger\ne1~~; \\
\slabel{eq:liHNb}
&& \lim_{\eta\to+\infty}w^{({\rm int})}(\eta)=-\frac{64}3~~;\qquad
y^\dagger=1~~;
\label{seq:liHN}
\end{leftsubeqnarray}
\begin{leftsubeqnarray}
\slabel{eq:leHNa}
&& \lim_{\eta\to+\infty}w^{({\rm ext})}(\eta)=-\frac{64}
{(y^\dagger-1)^2}\left[\frac{y^\dagger+1}{y^\dagger-1}\ln y^\dagger-2
\right]~~;\qquad y^\dagger\ne1~~; \\
\slabel{eq:leHNb}
&& \lim_{\eta\to+\infty}w^{({\rm ext})}(\eta)=-\frac{32}3~~;\qquad
y^\dagger=1~~;
\label{seq:leHN}
\end{leftsubeqnarray}
where the self potential-energy profile
factor remains finite, although the mass
profile factor undergoes a
logarithmic divergence.   Similarly,
Eqs.\,(\ref{seq:wiNH}) and
(\ref{seq:weNH}) reduce to:
\begin{leftsubeqnarray}
\slabel{eq:liNHa}
&& \lim_{\eta\to+\infty}w^{({\rm int})}(\eta)=-\frac{64y^\dagger}
{(y^\dagger-1)^2}\left[\frac{y^\dagger+1}{y^\dagger-1}\ln y^\dagger-2\right]~~;
\qquad y^\dagger\ne1~~; \\
\slabel{eq:liNHb}
&& \lim_{\eta\to+\infty}w^{({\rm int})}(\eta)=-\frac{32}3~~;\qquad
y^\dagger=1~~;
\label{seq:liNH}
\end{leftsubeqnarray}
\begin{leftsubeqnarray}
\slabel{eq:leNHa}
&& \lim_{\eta\to+\infty}w^{({\rm ext})}(\eta)=-\frac{64}
{(y^\dagger-1)^3}\left[(y^\dagger)^2-1-2y^\dagger\ln y^\dagger\right]~~;
\qquad y^\dagger\ne1~~; \\
\slabel{eq:leNHb}
&& \lim_{\eta\to+\infty}w^{({\rm ext})}(\eta)=-\frac{64}3~~;\qquad
y^\dagger=1~~;
\label{seq:leNH}
\end{leftsubeqnarray}
where $\Xi_i=\Xi_N$, $\Xi_j=\Xi_H$, while the contrary
holds with regard to Eqs.\,(\ref{seq:liHN}) and
(\ref{seq:leHN}).   Using Eqs.\,(\ref{seq:nuuv}),
(\ref{eq:numH}), (\ref{eq:nusH}), (\ref{eq:lumH}),
(\ref{eq:lusH}), and (\ref{eq:numN})-(\ref{seq:leNH}),  
the actual HN (AHN) and NH (ANH) macrogas equation of
state in the special situation under discussion, is
obtained from the particularization of
Eqs.\,(\ref{seq:Xr}) to the case of interest.

The ideal situation, where the interaction terms are
omitted, is obtained using Eqs.\,(\ref{seq:Xi}) instead
of (\ref{seq:Xr}).   More specifically, the ideal HN
(IHN) and NH (INH) macrogas equation of state is
obtained from the combination of Eqs.\,(\ref{eq:Xre}),
(\ref{eq:Xib}), (\ref{eq:numH}), (\ref{eq:nusH}), (\ref{eq:numN}),
and (\ref{eq:nusN}).   The result is:
\begin{equation}
\label{eq:XHNi}
X_{\rm p}=\frac13\frac{4+\Xi_H}{\Xi_H}\frac{[(1+\Xi_N)\ln(1+\Xi_N)-\Xi_N]^2}
{\Xi_N(2+\Xi_N)-2(1+\Xi_N)\ln(1+\Xi_N)}\frac{X_{\rm T}}{X_{\rm V}}~~;
\end{equation}
which represents a hyperbola with different axes for
fixed $X_{\rm T}$.   In the limit of infinite extension,
$\Xi\to+\infty$, both the left and right-hand side
of Eq.\,(\ref{eq:XHNi}) diverge, but a different
equation of state may be derived starting from
Eqs.\,(\ref{eq:Xrd}) and (\ref{eq:Xia}), following
a similar procedure.   The result is:
\begin{equation}
\label{eq:HNic}
X_{\rm p}^\dagger=\frac13\frac{\Xi_H^3(4+\Xi_H)}{(1+\Xi_H)^4}\frac{(1+\Xi_N)^2}
{\Xi_N(2+\Xi_N)-2(1+\Xi_N)\ln(1+\Xi_N)}\frac{X_{\rm T}^\dagger}{X_{\rm V}^\dagger}~~;
\end{equation}
which also represents a hyperbola with different axes for
fixed $X_{\rm T}^\dagger$.

\begin{figure*}[t]
\begin{center}
\includegraphics[scale=0.8]{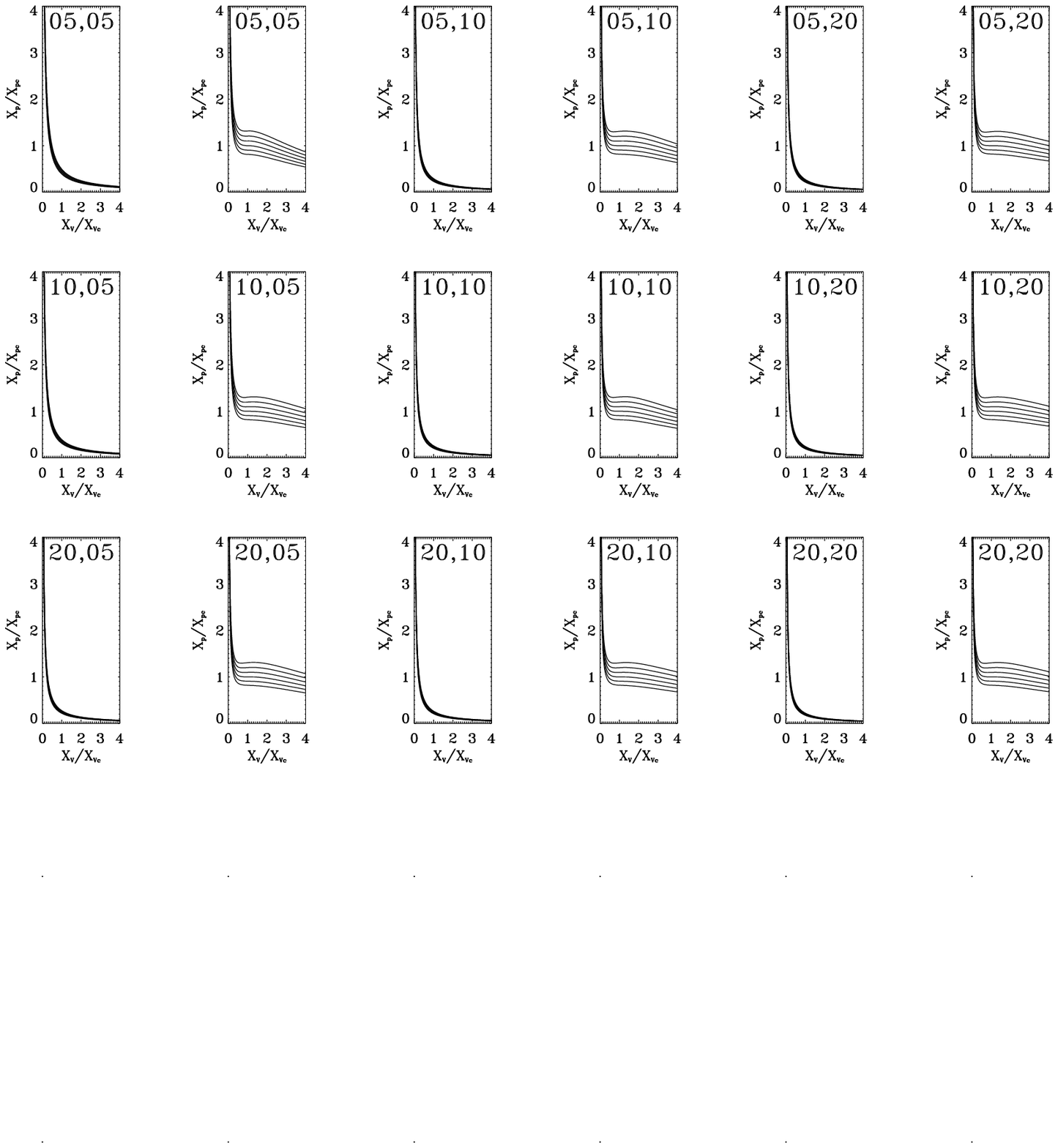}
\caption{Macroisothermal curves
($\sX_{\rm p}=X_{\rm p}/X_{p_c}$ vs. $\sX_{\rm V}=X_{\rm V}/X_{V_c}$)
related to IHN/INH (left panels)
and AHN/ANH (right panels) macrogases, respectively, for
different values of scaled truncation radii,
$(\Xi_i,\Xi_j)$, labelled on each panel.   Macroisothermal
curves (from bottom to top) correspond to $\sX_{\rm T}=X_{\rm T}/
X_{T_{\rm c}}=$
0.90, 0.95, 1.00, 1.05, 1.10, 1.15.   The limit, $(\Xi_i,\Xi_j)
\to(+\infty,+\infty)$, makes only little changes.   Owing to
Eqs.\,(\ref{eq:yma}) and (\ref{eq:XXC}), $X_{\rm V}^\dagger=X_{\rm V}
y^\dagger/y$ and $X_{\rm p}^\dagger=X_{\rm p}(m^\dagger/m)^2$,
which implies $\sX_{\rm V}^\dagger=\sX_{\rm V}$ and $\sX_{\rm p}^\dagger=
\sX_{\rm p}$.}
\label{f:hnso}
\end{center}
\end{figure*}

Macroisothermal curves ($\sX_{\rm p}=X_{\rm p}/X_{p_c}$
vs. $\sX_{\rm V}=X_{\rm V}/X_{V_c}$) related to IHN/INH
(tidal potential energy excluded) and AHN/ANH
(tidal potential energy included)
macrogases, are plotted in
Fig.\,\ref{f:hnso}, left and right panels, for
different values of scaled truncation radii,
$(\Xi_i,\Xi_j)$, labelled on each panel, and
same values of the reduced macrotemperature,
$\sX_{\rm T}=X_{\rm T}/X_{T_{\rm c}}=$ 0.90, 0.95, 1.00,
1.05, 1.10, 1.15, from bottom to top.
The limit, $(\Xi_i,\Xi_j)\to(+\infty,+\infty)$, makes only
little changes.   The comparison with ideal and VDW gases,
plotted in Fig.\,\ref{f:viso}, shows a similar but reversed
trend.   More specifically, extremum points occur above,
instead of below, the critical macroisothermal curve.   A complete
analogy can be obtained using the transformations, $X_{\rm V}\to1/X_{\rm V}$,
$X_{\rm p}\to1/X_{\rm p}$, $X_{\rm T}\to1/X_{\rm T}$.

\begin{figure*}[t]                                                                                  
\begin{center}                                                                                      
\includegraphics[scale=0.8]{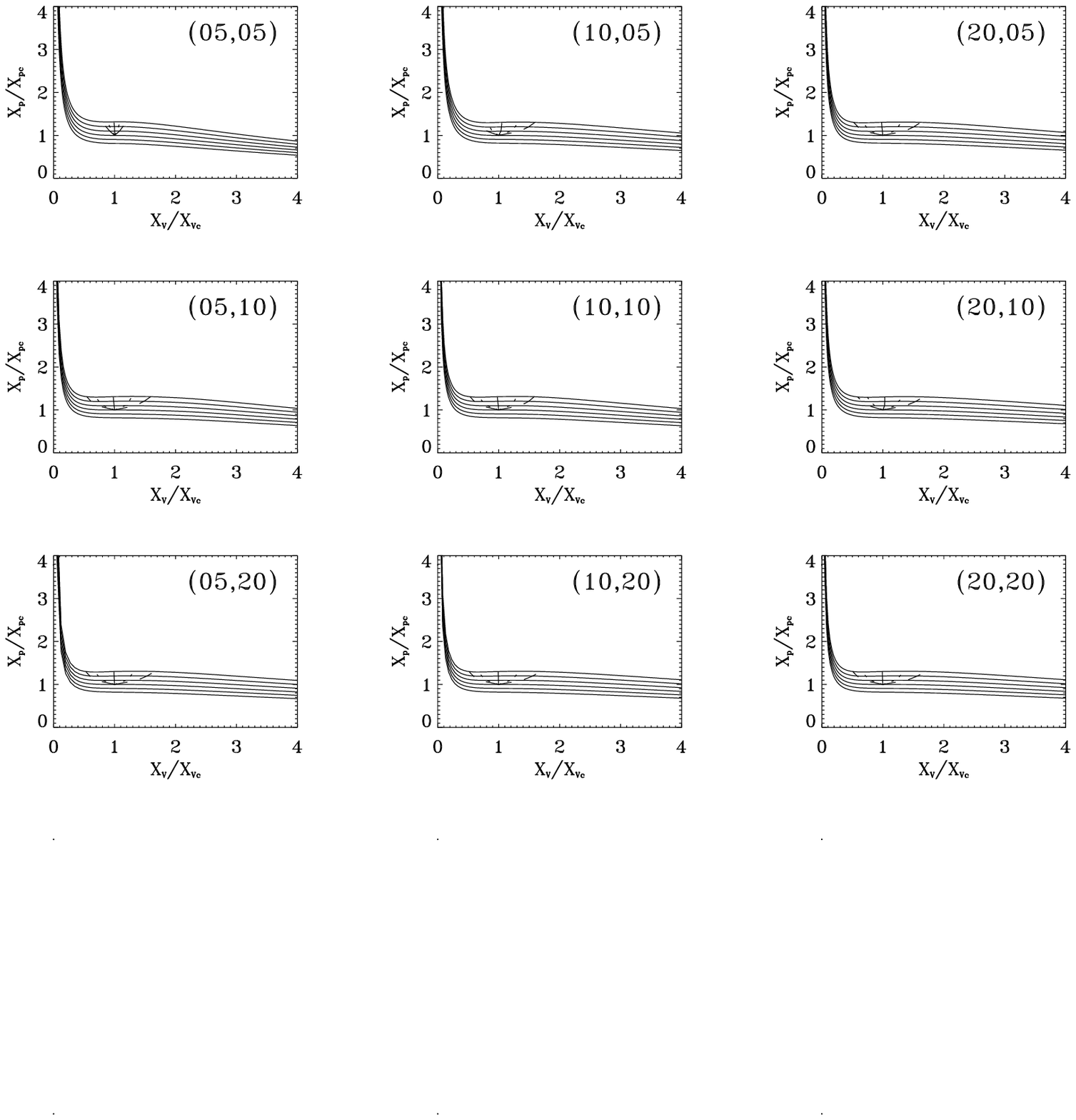}                                                               
\caption{AHN/ANH macroisothermal curves 
($\sX_{\rm p}=X_{\rm p}/X_{p_c}$ vs. $\sX_{\rm V}=X_{\rm V}/X_{V_c}$)
for different choices of scaled truncation radii, $(\Xi_i,\Xi_j)$,
labelled on each panel.   Macroisothermal curves (from bottom to
top) correspond to $\sX_{\rm T}=X_{\rm T}/X_{T_{\rm c}}=$ 0.90, 0.95, 1.00,
1.05, 1.10, 1.15.   The limit, $(\Xi_i,\Xi_j)\to(+\infty,+\infty)$,
makes only little changes.   RHN/RNH macroisothermal curves, when different
from their AHN/ANH counterparts, lie within the larger reversed
bell-shaped region in each panel.   The loci of intersections
between AHN/ANH and RHN/RNH macroisothermal curves are represented as
trifid curves, where the left branch corresponds to $\sX_{V_{\rm A}}$,
the right branch to $\sX_{V_{\rm E}}$, and the middle branch to
$\sX_{V_{\rm C}}$.   The critical point is the common origin.
The loci of AHN/ANH macroisothermal curve extremum points are represented
as dotted curves starting from the critical point, where the left
branch corresponds to minimum points and the right branch to maximum
points.   Owing to
Eqs.\,(\ref{eq:yma}) and (\ref{eq:XXC}), $X_{\rm V}^\dagger=X_{\rm V}
y^\dagger/y$ and $X_{\rm p}^\dagger=X_{\rm p}(m^\dagger/m)^2$,
which implies $\sX_{\rm V}^\dagger=\sX_{\rm V}$ and $\sX_{\rm p}^\dagger=
\sX_{\rm p}$.}
\label{f:hnis}
\end{center}
\end{figure*}

\begin{figure*}[t]                                                                                  
\begin{center}                                                                                      
\includegraphics[scale=0.8]{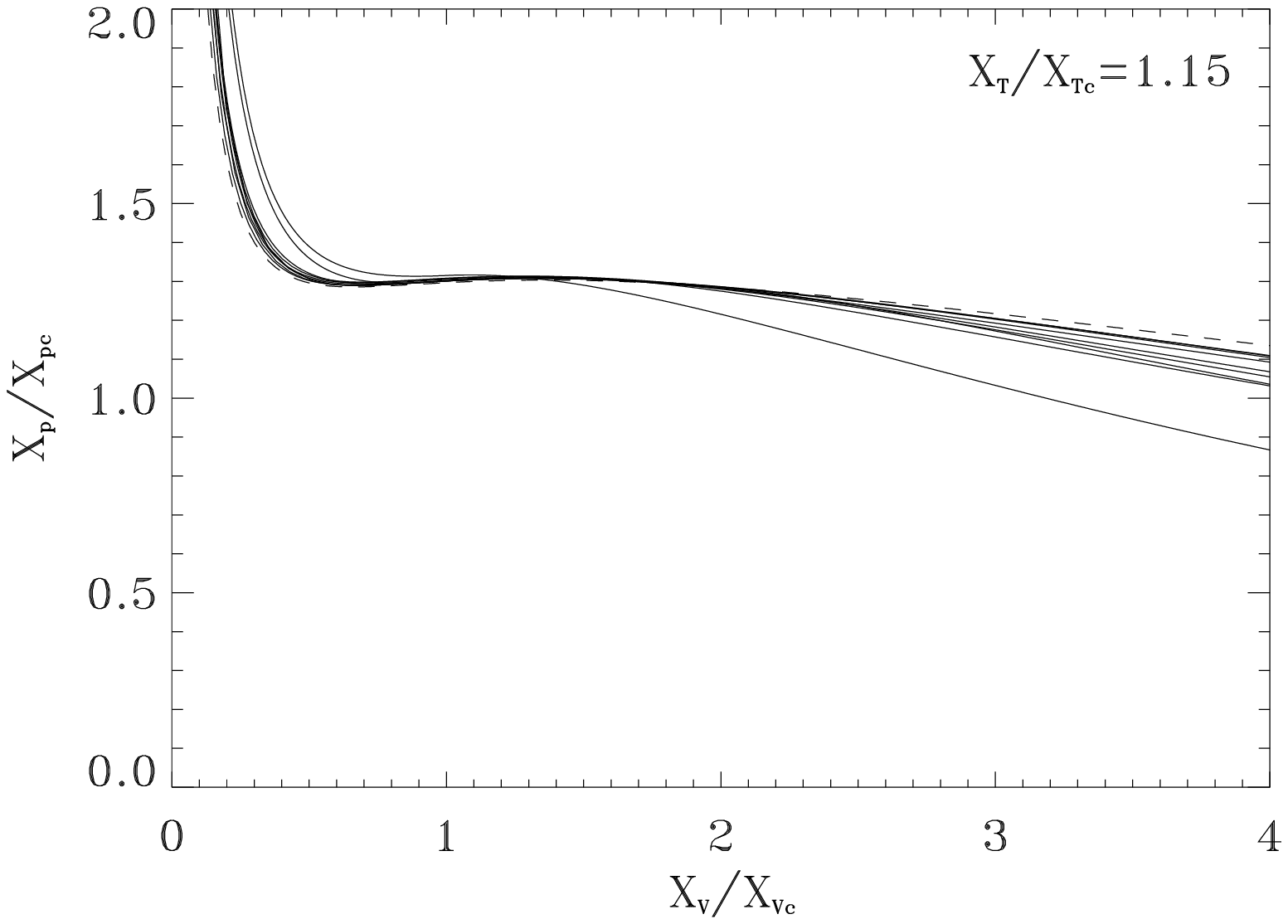}                                                               
\caption{AHN/ANH macroisothermal curves
($\sX_{\rm p}=X_{\rm p}/X_{p_c}$ vs. $\sX_{\rm V}=X_{\rm V}/X_{V_c}$)
for scaled truncation radii, $(\Xi_i,\Xi_j)$, as in
Fig.\,\ref{f:hnis}, with regard to the special case,
$\sX_{\rm T}=X_{\rm T}/X_{T_{\rm c}}=1.15$.   The limit,
$(\Xi_i,\Xi_j)\to(+\infty,+\infty)$, makes only little
changes, as shown by comparison with the dashed curve.
Owing to Eqs.\,(\ref{eq:yma}) and (\ref{eq:XXC}),
$X_{\rm V}^\dagger=X_{\rm V}y^\dagger/y$ and $X_{\rm p}^\dagger=X_{\rm p}
(m^\dagger/m)^2$, which implies $\sX_{\rm V}^\dagger=\sX_{\rm V}$ and
$\sX_{\rm p}^\dagger=\sX_{\rm p}$.}
\label{f:hnar}
\end{center}
\end{figure*}

The existence of a phase transition moving along a selected
macroisothermal curve, where the path is a horisontal line
instead of a curve including the extremum points, must necessarily
be assumed as a working hypothesis, due to analogy between VDW
isothermal curves and AHN/ANH macroisothermal curves.
As in the case of UU macrogases, AHN/ANH macroisothermal curves must be
numerically determined, following the same procedure outlined in
Subsection \ref{UU}.
The loci of
AHN/ANH macroisothermal curves plotted in
Fig.\,\ref{f:hnso}
(right panels), are represented 
in Fig.\,\ref{f:hnis}.
The loci of intersections between AHN/ANH and
 real HN/NH (RHN/RNH) macroisothermal curves, are represented
in Fig.\,\ref{f:hnis} as trifid curves, where the left branch
corresponds to $\sX_{V_{\rm A}}$, the right branch to $\sX_{V_
{\rm E}}$, and the middle branch to $\sX_{V_{\rm C}}$.  The
critical point is the common starting point.   The loci of AHN/ANH
macroisothermal curve extremum points are represented in
Fig.\,\ref{f:hnis} as dotted curves starting from the critical
point, where the left branch corresponds to minimum points and
the right branch to maximum points.   The  RHN/RNH macroisothermal
curves, when different from their AHN/ANH counterparts, lie within
the larger bell-shaped regions.   The limit,
$(\Xi_i,\Xi_j)\to(+\infty,+\infty)$, makes only little changes.
Values of parameters, $\sX_{\rm T}$, $\sX_{V_{\rm A}}$, $\sX_{V_{\rm B}}$,
$\sX_{V_{\rm C}}$, $\sX_{V_{\rm D}}$, $\sX_{V_{\rm E}}$, $\sX_{p_{\rm B}}$,
$\sX_{p_{\rm C}}$, $\sX_{p_{\rm D}}$, are listed in Tab.\,\ref{t:hnspo}
within the range, $1.00<\sX_{\rm T}\le1.15$, using a step, $\Delta\sX_{\rm T}=0.01$,
in the limit, $(\Xi_i,\Xi_j)\to(+\infty,+\infty)$.   The AHN/ANH
macroisothermal
curves are plotted in Fig.\,\ref{f:hnar} for scaled truncation radii,
$(\Xi_i,\Xi_j)$, as in Fig.\,\ref{f:hnis}, with regard to the special
case, $\sX_{\rm T}=1.15$.   The limit, $(\Xi_i,\Xi_j)\to(+\infty,+\infty)$,
makes only little changes, as shown by comparison with the dashed curve.
\begin{table}
\caption{Values of parameters, $\sX_{\rm T}$, $\sX_{V_{\rm A}}$, $\sX_{V_{\rm B}}$,
$\sX_{V_{\rm C}}$, $\sX_{V_{\rm D}}$, $\sX_{V_{\rm E}}$, $\sX_{p_{\rm B}}$,
$\sX_{p_{\rm C}}$, $\sX_{p_{\rm D}}$,
within the range, $1.01\le\sX_{\rm T}\le1.15$, using a step,
$\Delta\sX_{\rm T}=0.01$.   All values equal unity at the critical point.
Results are related to infinitely extended configurations,
$(\Xi_i,\Xi_j)\to(+\infty,+\infty)$.   Index captions: A, C, E -
intersections
between AHN/ANH and RHN/RNH macroisothermal curves; B - extremum point
of minimum; D - extremum point of maximum.   Extremum points
are related to AHN/ANH macroisothermal curves, while their RHN/RNH
counterparts
are flat within the range, $\sX_{V_{\rm A}}\le\sX_{\rm V}\le\sX_{V_{\rm E}}$.
For aesthetical reasons, 01 on head columns stands for unity.}
\label{t:hnspo}
\begin{center}
\begin{tabular}{|l|l|l|l|l|l|l|l|l|} \hline
$\sX_{\rm T}$ & $10\sX_{V_{\rm A}}$ & $10\sX_{V_{\rm B}}$ & $10\sX_{V_{\rm C}}$ & $01\sX_{V_{\rm D}}$
& $01\sX_{V_{\rm E}}$ & $01\sX_{p_{\rm B}}$ & $01\sX_{p_{\rm C}}$ & $01\sX_{p_{\rm D}}$ \\
\hline
 1.01 & 8.4036 & 9.0357 & 9.9949 & 1.0976 & 1.1768 & 1.1085 & 1.0186 & 1.0187 \\
 1.02 & 7.8103 & 8.6548 & 9.9935 & 1.1372 & 1.2535 & 1.0370 & 1.0374 & 1.0377 \\
 1.03 & 7.3771 & 8.3678 & 9.9918 & 1.1675 & 1.3139 & 1.0555 & 1.0563 & 1.0570 \\
 1.04 & 7.0273 & 8.1303 & 9.9903 & 1.1929 & 1.3656 & 1.0742 & 1.0754 & 1.0764 \\
 1.05 & 6.7309 & 7.9247 & 9.9888 & 1.2151 & 1.4118 & 1.0929 & 1.0947 & 1.0961 \\
 1.06 & 6.4724 & 7.7420 & 9.9874 & 1.2347 & 1.4540 & 1.1118 & 1.1142 & 1.1159 \\
 1.07 & 6.2427 & 7.5768 & 9.9859 & 1.2527 & 1.4930 & 1.1307 & 1.1338 & 1.1360 \\
 1.08 & 6.0356 & 7.4256 & 9.9844 & 1.2694 & 1.5295 & 1.1498 & 1.1535 & 1.1562 \\
 1.09 & 5.8488 & 7.2872 & 9.9830 & 1.2845 & 1.5637 & 1.1687 & 1.1733 & 1.1765 \\
 1.10 & 5.6738 & 7.1558 & 9.9815 & 1.2991 & 1.5967 & 1.1881 & 1.1936 & 1.1974 \\
 1.11 & 5.5137 & 7.0340 & 9.9797 & 1.3126 & 1.6279 & 1.2074 & 1.2139 & 1.2182 \\
 1.12 & 5.3647 & 6.9194 & 9.9785 & 1.3254 & 1.6577 & 1.2269 & 1.2343 & 1.2392 \\
 1.13 & 5.2256 & 6.8111 & 9.9770 & 1.3376 & 1.6864 & 1.2464 & 1.2549 & 1.2605 \\
 1.14 & 5.0952 & 6.7084 & 9.9754 & 1.3491 & 1.7139 & 1.2661 & 1.2757 & 1.2819 \\
 1.15 & 4.9726 & 6.6108 & 9.9737 & 1.3601 & 1.7406 & 1.2858 & 1.2967 & 1.3035 \\
\hline                                                                                              
\end{tabular}                                                                                       
\end{center}                                                                                        
\end{table}                                                                                         

Values of parameters, $m^\dagger$, $m$,
$y^\dagger$, $y$, $\phi$, related to the
critical macroisothermal curve, for
selected scaled truncation radii,
$(\Xi_i,\Xi_j)$, are listed in
Tab.\,\ref{t:hncri}.
\begin{table}
\caption{Values of the scaling fractional
mass, $m^\dagger$, the fractional mass, $m$,
the scaling fractional radius, $y^\dagger$, 
the fractional truncation radius, $y$, and the fractional
energy, $\phi$, related to the
critical point i.e. the horizontal inflexion
point on the critical macroisothermal
curve, for selected scaled truncation radii,
$(\Xi_i,\Xi_j)$, of HN/NH density profiles.
In absence of
truncation radius, $(\Xi_i,\Xi_j)\to(+\infty,+\infty)$, the
results are independent of $y/y^\dagger=
\Xi_j/\Xi_i$.}
\label{t:hncri}
\begin{center}
\begin{tabular}{|c|c|c|c|c|c|} \hline
($\Xi_i$,$\Xi_j$)& $m^\dagger$ & $m$ & $y^\dagger$ & $y$ &
$\varphi$ \\
\hline
(05,05)             & 02.57 & 03.54     & 1.01 & 1.01    & 03.66 \\
(05,10)             & 04.41 & 09.45     & 1.04 & 2.08    & 09.16 \\
(05,20)             & 05.14 & 15.47     & 1.14 & 4.57    & 13.42 \\
(10,05)             & 05.84 & 06.77     & 1.88 & 0.94    & 07.33 \\
(10,10)             & 06.36 & 11.45     & 1.38 & 1.38    & 11.47 \\
(10,20)             & 07.24 & 18.32     & 1.50 & 3.00    & 16.46 \\
(20,05)             & 09.40 & 09.93     & 2.13 & 0.53    & 11.30 \\
(20,10)             & 08.44 & 13.85     & 1.92 & 0.96    & 14.35 \\
(20,20)             & 09.10 & 20.99     & 1.75 & 1.75    & 19.49 \\
($\infty$,$\infty$) & 12.40 & $+\infty$ & 2.06 & $\dots$ & 35.82 \\
\hline
\end{tabular}                                                                                       
\end{center}                                                                                        
\end{table}                                                                                         
A better method has been used with respect
to an earlier attempt (CV08), yielding
improved results with the occurrence of
the critical macroisothermal curve in
all cases.   In absence of truncation
radius, $(\Xi_i,\Xi_j)\to(+\infty,+\infty)$, small differences
appear in the results, probably due to
(i) the divergence of the fractional mass,
$m\to+\infty$, as $\Xi_N\to+\infty$, and
(ii) numerical calculations have been
performed, taking $\Xi\appgeq10^{10}$.
Related results listed in Table \ref
{t:hncri} may be considered as weighted
means.

\section{Application to elliptical galaxies}
\label{aell}

The luminosity-weighted second moment of the
line-of-sight velocity distribution within
the half-light radius has recently been
determined for samples of early-type
galaxies, using integral-field spectroscopy
such as SAURON (S\,IV; S\,X).   Compared to
the central velocity dispersion, which was
sometimes used before, the above mentioned
quantity has the advantage that it is only
weakly dependent on the details of the
aperture used.   In addition, it is an
approximation to the second velocity
moment which appears in the virial equations
(e.g., Binney and Tremaine, 1987, Chap.\,4,
\S4.3).   In other words, it may be conceived
as a rms mass-weighted velocity, with a weak
dependence on the features of the orbital
distribution.  For further details refer
to the parent papers (S\,IV; S\,X).

An application of the current model to
SAURON sample objects can be performed
along the following steps: (i) select
SAURON data of interest; (ii) calculate
parameters appearing in the virial
equations; (iii) make a correspondence
between model galaxies and sample obiects;
(iv) represent model galaxies as points on
the $({\sf O}\sX_{\rm V}\sX_{\rm p})$ plane.

\subsection{Data selection}
\label{soda}

The sample used (CV08, $N=16$) is made of
elliptical galaxies, extracted from richer
samples of early-type galaxies investigated
within the SAURON project (S\,IV, $N=25$; S\,X,
$N=48$).    The selection has been restricted
to elliptical galaxies common to the above mentioned
samples for two main reasons, namely (i) the current
model best applies to elliptical galaxies and their
hosting haloes, and (ii) some physical parameters
of interest are listed in either S\,IV or S\,X.
The whole data set needed for the application
is shown in Tab.\,\ref{t:dati}.   For further
details refer to the parent papers (S\,IV, columns
2-6; S\,X, columns 7-11).

Values to be actually used in the application
\begin{table}
\caption{Data related to a sample $(N=16)$ of
elliptical galaxies, extracted from larger samples
of early-type galaxies investigated within the
SAURON project (S\,IV, $N=25$; S\,X, $N=48$), which
are used in the current paper.   Column captions:
(1) NGC number; (2) effective (half-light) radius,
$R_e$, measured in the $I$ band (S\,IV); (3) ratio between
maximum radius, $R_{\rm max}$, and effective radius,
$R_e$ (S\,IV); (4) total observed $I$ band galaxy
magnitude (S\,IV); (5) mass-luminosity ratio of the
stellar population (S\,IV); (6) galaxy distance modulus
(hats avoid confusion with the fractional mass, $m$,
and the total mass, $M$) (S\,IV); (7) luminosity-weighted
average ellipticity, $\hat{e}_\bot$, on a plane
perpendicular to the line of sight, within either
an isophote enclosing an area, $\hat{A}=\pi R_e^2$,
or the largest isophote fully contained within the
SAURON field, whichever is smaller (S\,X); (8)
luminosity-weighted squared mean velocity component,
parallel to the line of sight, within either an
ellipse of area, $\hat{A}$, ellipticity, $\hat{e}_\bot$,
and related position angle, or the largest similar
ellipse fully contained within the SAURON field,
whichever is smaller (S\,X); (9) luminosity-weighted
squared velocity dispersion, parallel to the line
of sight, within either an
ellipse of area, $\hat{A}$, ellipticity, $\hat{e}_\bot$,
and related position angle, or the largest similar
ellipse fully contained within the SAURON field,
whichever is smaller (S\,X); (10) ratio between the
square root luminosity-weighted squared mean velocity
component and dispersion velocity component,
parallel to the line of sight (S\,X); (11) kinematic
classification (F - fast rotator; S - slow rotator)
(S\,X).   For further details refer to the parent
papers (S\,IV; S\,X) and an earlier attempt (Binney,
2005).}
\label{t:dati}
\begin{center}
\begin{tabular}{|c|c|c|c|c|c|c|c|c|c|c|} \hline
NGC & $R_e$ & $\frac{R_{\rm max}}{R_e}$ & $I_T$ & $\frac{M_i}L$ & $\hat{m}-
\hat{M}$ & $<\hat{e}_\bot>$ & $<\widetilde{v}_\|^2>^{\frac12}$ & $<\sigma_\|^2>^
{\frac12}$& $\frac{<\widetilde{v}_\|^2>^{\frac12}}{<\sigma_\|^2>^{\frac12}}$ & KC
\\
 & (arcsec) & & (mag) & ($I$ band) & (mag) & & (km/s) & (km/s) & & \\
\hline
0821 & 039.0 & 0.62 & 09.47 & 2.60 & 31.85 & 0.40 & 048 & 182 & 0.26 & F \\
2974 & 024.0 & 1.04 & 09.43 & 2.34 & 31.60 & 0.37 & 127 & 180 & 0.70 & F \\
3377 & 038.0 & 0.53 & 08.98 & 1.75 & 30.19 & 0.46 & 057 & 117 & 0.49 & F \\
3379 & 042.0 & 0.67 & 08.03 & 3.08 & 30.06 & 0.08 & 028 & 198 & 0.14 & F \\
3608 & 041.0 & 0.49 & 09.40 & 2.57 & 31.74 & 0.18 & 008 & 179 & 0.05 & S \\
4278 & 032.0 & 0.82 & 08.83 & 3.05 & 30.97 & 0.12 & 044 & 228 & 0.19 & F \\
4374 & 071.0 & 0.43 & 07.69 & 3.08 & 31.26 & 0.15 & 007 & 282 & 0.03 & S \\
4458 & 027.0 & 0.74 & 10.68 & 2.27 & 31.12 & 0.12 & 010 & 084 & 0.12 & S \\
4473 & 027.0 & 0.92 & 08.94 & 2.88 & 30.92 & 0.41 & 041 & 188 & 0.22 & F \\
4486 & 105.0 & 0.29 & 07.23 & 3.33 & 30.97 & 0.04 & 007 & 306 & 0.02 & S \\
4552 & 032.0 & 0.63 & 08.54 & 3.35 & 30.87 & 0.04 & 013 & 257 & 0.05 & S \\
4621 & 046.0 & 0.56 & 08.41 & 3.12 & 31.25 & 0.34 & 052 & 207 & 0.23 & F \\
4660 & 011.0 & 1.83 & 09.96 & 2.96 & 30.48 & 0.44 & 079 & 163 & 0.49 & F \\
5813 & 052.0 & 0.53 & 09.12 & 2.97 & 32.48 & 0.15 & 032 & 223 & 0.14 & S \\
5845 & 004.6 & 4.45 & 11.10 & 2.96 & 32.01 & 0.35 & 081 & 226 & 0.36 & F \\
5846 & 081.0 & 0.29 & 08.41 & 3.33 & 31.92 & 0.07 & 007 & 240 & 0.03 & S \\
\hline
\end{tabular}
\end{center}
\end{table}
are related to: the effective (half-light)
radius, $R_e$ (S\,IV); the total observed $I$
band galaxy magnitude, $I_T$ (S\,IV); the
mass-luminosity ratio of the stellar
population, $M_i/L$ (S\,IV); the galaxy
distance modulus, $\hat{m}-\hat{M}$ (hats
avoid confusion with the fractional mass,
$m=M_j/M_i$, and the total mass, $M=M_i+M_j$)
(S\,IV); the luminosity-weighted
average ellipticity, $\hat{e}_\bot$, on a plane
perpendicular to the line of sight, within either
an isophote enclosing an area, $\hat{A}=\pi R_e^2$,
or the largest isophote fully contained within the
SAURON field, whichever is smaller (S\,X); the
luminosity-weighted squared mean velocity component,
parallel to the line of sight, within either an
ellipse of area, $\hat{A}$, ellipticity, $\hat{e}_\bot$,
and related position angle, or the largest similar
ellipse fully contained within the SAURON field,
whichever is smaller (S\,X); the luminosity-weighted
squared velocity dispersion, parallel to the line
of sight, within either an
ellipse of area, $\hat{A}$, ellipticity, $\hat{e}_\bot$,
and related position angle, or the largest similar
ellipse fully contained within the SAURON field,
whichever is smaller (S\,X).

Parameters listed to gain further insight even if
not used in the application are: the ratio between
maximum radius, $R_{\rm max}$, sampled by the
kinematical observations, and effective radius,
$R_e$ (S\,IV); the ratio between
square root luminosity-weighted squared mean velocity
component and dispersion velocity component,
parallel to the line of sight (S\,X); the kinematic
classification (F - fast rotator; S - slow rotator)
(S\,X).

For further details refer to the parent
papers (S\,IV; S\,X) and an earlier attempt (Binney,
2005).

\subsection{Determination of model parameters}
\label{mopa}

With regard to the stellar subsystem, two model
parameters can directly be inferred from the data
(CV08).   More specifically, stellar masses are
deduced from luminosities and mass-luminosity
ratios (in $I$ band), as  $M_i/{\rm M}_{10}=
(L/{\rm L}_\odot)[(M_i/L)/(10^{10}{\rm m}_\odot/
{\rm L}_\odot)]$; $L/{\rm L}_\odot=\exp_{10}\{-0.4
[I_T-(\hat{m}-\hat{M})-4.11]\}$; and scaling radii
are calculated from effective radii (in arcsec) and
distances, as 
$r_i^\dagger/{\rm kpc}=(R_e/{\rm kpc})/
1.81$; $R_e/{\rm kpc}=[(R_e/{\rm arcsec})(d/{\rm Mpc})]/
206.265$; $d/{\rm Mpc}$ $=\exp_{10}[(\hat{m}-\hat{M})
/5-5]$; where the factor, 1.81, is related to an
assumed H density profile for the inner subsystem,
and the factor, 206.265, is
related to the choice of measure units (CV08).
For further details refer to the parent paper (S\,IV).

Two additional parameters can be inferred by fitting
the data with dynamical models.   More specifically,
the inclination angle, $i$, is deduced from the best
fitting two-integral Jeans model (S\,IV), and the
anisotropy parameter, $\delta$, is determined from
the solution of the dynamical models, supposed to
be axisymmetric (S\,X).   In fact, fast rotators
show evidence of large anisotropy and axial
symmetry, while slow rotators appear to be nearly
isotropic and moderately triaxial.   For further
details refer to the parent paper (S\,X).

The intrinsic axis ratio, $\epsilon$, is deduced
from the computed inclination, under the assumption
of axisymmetric configurations (S\,X), using the
relation (Binney and Tremaine, 1987, Chap.\,4,
\S4.3):
\begin{equation}
\label{eq:ep}
1-\epsilon^2=\frac{1-\epsilon_{\rm obs}^2}{\sin^2i}~~;
\end{equation}
where $\epsilon_{\rm obs}$ is the observed axis ratio
related to an inclination angle, $i$, between
the symmetry axis and the line of sight ($i=
90^\circ$ for edge-on configurations).

The mass-weighted mean square velocity component
and dispersion velocity component, parallel to the
line of sight, for a galaxy observed (obs) at an
inclination angle, $i$, under the assumption of
axisymmetric $(a_1=a_2)$ and isotropic on the
equatorial plane $(\sigma_{11}=\sigma_{22})$
configurations (S\,X), are related to their edge-on
(edo) counterparts as (Binney and Tremaine, 1987,
Chap.\,4, \S4.3):
\begin{lefteqnarray}
\label{eq:vobs}
&& [<\widetilde{v}_\|^2>^{1/2}]_{\rm obs}=[<\widetilde{v}_\|^2>^{1/2}]_{\rm edo}
\sin i~~; \\
\label{eq:sobs}
&& [<\sigma_\|^2>^{1/2}]_{\rm obs}=[<\sigma_\|^2>^{1/2}]_{\rm edo}
(1-\delta\cos^2i)^{1/2}~~;
\end{lefteqnarray}
and the intrinsic mean rotational velocity and
velocity dispersion are expressed as:
\begin{lefteqnarray}
\label{eq:vphi}
&& <\widetilde{v_{\phi\phi}}^2>^{1/2}
=\sqrt{2}[<\widetilde{v}_\|^2>^{1/2}]_{\rm edo}=
\frac{\sqrt{2}}{\sin i}[<\widetilde{v}_\|^2>^{1/2}]_{\rm obs}~~; \\
\label{eq:sigma}
&& <\sigma^2>^{1/2}\hspace{1mm}=
(<\sigma_{11}^2>+<\sigma_{22}^2>+<\sigma_{33}^2>)^{1/2} \nonumber \\
&& \phantom{<\sigma^2>^{1/2}}=
\{2[<\sigma_\|^2>]_{\rm edo}+(1-\delta)[<\sigma_\|^2>]_{\rm edo}\}^{1/2}
\nonumber \\
&& \phantom{<\sigma^2>^{1/2}}=(3-\delta)^{1/2}[<\sigma_\|^2>^{1/2}]_{\rm edo}=
\left(\frac{3-\delta}{1-\delta\cos^2i}\right)^{1/2}[<\sigma_\|^2>^{1/2}]_
{\rm obs}~;\qquad
\end{lefteqnarray}
in terms of observed quantities, where $\delta=
1-<\sigma_{33}^2>/<\sigma_{11}^2>=1-$ $<\sigma_
{33}^2>/<\sigma_{22}^2>$ by definition (e.g.,
Binney, 2005).

\subsection{Model galaxies vs. sample objects}
\label{mgso}

The kinetic energy of the stellar subsystem is:
\begin{equation}
\label{eq:Eskin}
(E_i)_{\rm kin}=\frac12M_i\left\{<(\widetilde{v_{\phi\phi}})_i^2>
+<\sigma_i^2>\right\}~~;
\end{equation}
and the combination of Eqs.\,(\ref{seq:viruv}),
(\ref{seq:M}), (\ref{eq:Esel}), (\ref{seq:Exxx}),
(\ref{seq:nuuv}), and (\ref{eq:Eskin}) yields:
\begin{equation}
\label{eq:virs1}
M_i\left\{<(\widetilde{v_{\phi\phi}})_i^2>+<\sigma_i^2>\right\}=
\frac{(\nu_i)_{\rm sel}}{[(\nu_i)_{\rm mas}]^2}\frac{G(M_i)^2}{(a_i^\dagger)_
1}B+\frac{(\nu_{ij})_{\rm tid}}{[(\nu_i)_{\rm mas}]^2}\frac{G(M_i)^2}{(a_i^
\dagger)_1}B~~;
\end{equation}
which, after some algebra, takes the form:
\begin{equation}
\label{eq:virs2}
\frac{<\sigma_i^2>(a_i^\dagger)_1}{GM_i}\frac1B\left\{\frac{<(\widetilde
{v_{\phi\phi}})_i^2>}{<\sigma_i^2>}+1\right\}=\frac{(\nu_i)_{\rm sel}+
(\nu_{ij})_{\rm tid}}{[(\nu_i)_{\rm mas}]^2}~~;
\end{equation}
where, for an inner H density profile, the
scaling radius, $r_i^\dagger=(a_i^\dagger)_1$,
may be chosen as:
\begin{equation}
\label{eq:Heff}
r_i^\dagger=\frac{R_e}{1.81}~~;\quad M_i(r_i^\dagger)=\frac14M_i~~;\quad
M_i(R_e)=\frac12M_i~~;
\end{equation}
and the shape factor, $B$, reads (e.g.,
Chandrasekhar, 1969, Chap.\,3 \S\S17, 22; Caimmi, 2009): 
\begin{equation}
\label{eq:B}
B=2\frac{\arcsin\sqrt{1-\epsilon^2}}{\sqrt{1-\epsilon^2}}
=2\displayfrac{\arcsin\frac{\sqrt{1-(1-<\hat{e}_\bot>)^2}}{\sin i}}
{\frac{\sqrt{1-(1-<\hat{e}_\bot>)^2}}{\sin i}}~~;
\end{equation}
for axisymmetric configurations, where the last
equality is owing to Eq.\,(\ref{eq:ep}) and the
definition of ellipticity, $\hat{e}=1-\epsilon$.

The combination of Eqs.\,(\ref{eq:vphi}), (\ref{eq:sigma}),
(\ref{eq:Eskin}), (\ref{eq:virs2}), and
(\ref{eq:Heff}) yields:
\begin{leftsubeqnarray}
\slabel{eq:cija}
&& \frac{[<\sigma_\|^2>]_{\rm obs}}{2GM_i(R_e)}\frac{R_e}{1.81}\frac1B
\left\{\frac2{\sin^2i}\frac{[<\widetilde{v}_\|^2>]_{\rm obs}}{[<\sigma_\|^2>]_
{\rm obs}}+\frac{3-\delta}{1-\delta\cos^2i}\right\}=c_{ij}~~;\qquad \\
\slabel{eq:cijb}
&& c_{ij}=\frac{(\nu_i)_{\rm sel}+(\nu_{ij})_{\rm tid}}{[(\nu_i)_{\rm mas}]^2}
~~;
\label{seq:cij}
\end{leftsubeqnarray}
where the left-hand side of Eq.\,(\ref{eq:cija})
is expressed in terms of quantities which are
deduced from either observations or fitting with
dynamic models, and then may be determined for
an assigned sample object.   Conversely, the
right-hand side of Eq.\,(\ref{eq:cija}) depends
on the selected density profiles, and then may
be determined for an assigned model galaxy.
In this view, Eq.\,(\ref{eq:cija}) may be read
as a correspondence between sample objects (left)
and model galaxies (right).

The dimensionless energy, $c_{ij}$, defined by Eq.\,(\ref{eq:cijb}),
depends on four variables via Eqs.\,(\ref{seq:M})-(\ref{seq:wie}):
the scaled truncation radii, $\Xi_i$, $\Xi_j$,
the fractional mass, $m^\dagger$ (or $m$), and
the fractional radius, $y^\dagger$ (or $y$).

\subsection{Model galaxies on the $({\sf O}\sX_{\rm V}\sX_{\rm p})$ plane}
\label{corr}

For assigned density profiles and scaled
truncation radii, $\Xi_i$, $\Xi_j$, two
unknowns remain:
the fractional mass, $m^\dagger$ (or $m$), and
the fractional radius, $y^\dagger$ (or $y$).
%
%
The combination of Eqs.\,(\ref{eq:nuuva}) and
(\ref{eq:cijb}) yields:
\begin{equation}
\label{eq:dij}
w^{\rm (ext)}(\eta)=\frac89\frac1{m^\dagger}\left\{(\nu_i)_{\rm sel}-c_{ij}
[(\nu_i)_{\rm mas}]^2\right\}~~;
\end{equation}
where the function, $w^{\rm (ext)}$, depends
on $\Xi_i$ and $y^\dagger$, conformably to
Eqs.\,(\ref{eq:Mb}), (\ref{eq:nuuvc}), and
(\ref{eq:wieb}).   In the case under discussion,
Eq.\,(\ref{eq:dij}) may be conceived as a link
between the unknowns, $m^\dagger$ and $y^\dagger$.
At this stage, one additional relation is needed
(CV08).

The mere existence of a fundamental plane
(Djorgovski and Davis, 1987; Dressler et
al., 1987) indicates that structural
properties in elliptical galaxies span a narrow range,
suggesting that some self-regulating
mechanism must be at work during formation
and evolution.   In particular, projected
light profiles from elliptical galaxies exhibit large degree
of homogeneity and may well be fitted by the
$r^{1/4}$ de Vaucouleurs law.   Accordingly,
a narrow range may safely be expected
also for fractional masses of elliptical
galaxies and the
assumption, $m={\rm const}$, appears to
be a viable approximation.   This is the
reason for which
the sample used $(N=16)$ is made of
only elliptical galaxies, extracted
from larger samples $(N=25;~N=48)$ of
early-type galaxies investigated
within the SAURON project (S\,IV; S\,X).

Then the fractional radius, $y^\dagger$
(or $y$), remains as the sole unknown,
which can be determined by solving
Eq.\,(\ref{eq:dij}) with numerical
techniques.   If no solution exists,
no model galaxy corresponds to the
selected sample object or, in other
words, the related density profiles
provide no fit to the data, and some
input value has to be changed.

Values of parameters needed for representing
model galaxies on the $({\sf O}\sX_{\rm V}$ $\sX_{\rm p})$
plane, are listed in Tab.\,\ref{t:pamo}:
the galaxy stellar mass,
$M_i$; the galaxy scaling radius,
$r_i^\dagger$; the inclination angle, $i$, of the
best fitting two-integral Jeans model (S\,IV);
the anisotropy parameter, $\delta$, determined
from the solution of the dynamic models,
supposed to be axisymmetric $(a_1=a_2)$ and
isotropic on the equatorial plane $(\sigma_
{11}=\sigma_{22})$ (S\,X); the intrinsic axis
ratio, $\epsilon$, deduced from the computed
inclination, under the assumption of
axisymmetric configurations (S\,X); the
dimensionless energy, $c_{ij}$, defined by
Eqs.\,(\ref{seq:cij}).
The kinematic classification is listed again
to get more insight.
\begin{table}
\caption{Parameters calculated from the data
listed in Tab.\,\ref{t:dati} for a sample $(N=16)$ of
elliptical galaxies, extracted from larger samples
of early-type galaxies investigated within the
SAURON project (S\,IV, $N=25$; S\,X, $N=48$), which
are used in the current paper.   Column captions:
(1) NGC number; (2) galaxy stellar mass
deduced from luminosities and mass-luminosity
ratios (in $I$ band),
conformably to Eqs.\,(\ref{seq:MiRe});  (3) galaxy
scaling radius, calculated
using Eqs.\,(\ref{eq:Heff}) and (\ref{seq:Reff});
(4) inclination angle of the best fitting two-integral
Jeans model (S\,IV); (5) anisotropy parameter, 
determined from the solution of the dynamic models,
supposed to be axisymmetric (S\,X); (6) intrinsic axis
ratio, deduced from the computed inclination,
under the assumption of axisymmetric configurations (S\,X);
(7) dimensionless energy, defined by 
Eqs.\,(\ref{seq:cij}); (8) kinematic
classification (F - fast rotator; S - slow rotator)
(S\,X).   For further details refer to the parent
papers (S\,IV; S\,X) and an earlier attempt (Binney, 2005).}
\label{t:pamo}
\begin{center}
\begin{tabular}{|c|c|c|c|c|c|c|c|} \hline
NGC & $M_i$ & $r_i^\dagger$ & $i$ & $\delta_i$ & $\epsilon_i$
 & $c_{ij}$ & KC \\
 & $({\rm M}_{10})$ & (kpc) & $(^\circ)$ & & & & \\
\hline
0821 & 10.26 & 2.45 & 90 & 0.20 & 0.60 & 0.23 & F \\
2974 & 07.61 & 1.34 & 57 & 0.24 & 0.38 & 0.23 & F \\
3377 & 02.35 & 1.11 & 90 & 0.25 & 0.54 & 0.20 & F \\
3379 & 08.80 & 1.16 & 90 & 0.03 & 0.92 & 0.18 & F \\
3608 & 09.77 & 2.45 & 90 & 0.13 & 0.82 & 0.25 & S \\
4278 & 09.64 & 1.34 & 45 & 0.18 & 0.74 & 0.25 & F \\
4374 & 36.35 & 3.40 & 90 & 0.08 & 0.85 & 0.24 & S \\
4458 & 01.50 & 1.21 & 90 & 0.09 & 0.88 & 0.19 & S \\
4473 & 07.86 & 1.10 & 73 & 0.34 & 0.54 & 0.14 & F \\
4486 & 45.97 & 4.40 & 90 & 0.00 & 0.96 & 0.31 & S \\
4552 & 12.62 & 1.28 & 90 & 0.02 & 0.96 & 0.23 & S \\
4621 & 18.80 & 2.19 & 90 & 0.18 & 0.66 & 0.15 & F \\
4660 & 02.11 & 0.37 & 70 & 0.30 & 0.47 & 0.15 & F \\
5813 & 28.89 & 4.36 & 90 & 0.08 & 0.85 & 0.25 & S \\
5845 & 03.02 & 0.31 & 90 & 0.15 & 0.65 & 0.17 & F \\
5846 & 37.19 & 5.25 & 90 & 0.01 & 0.93 & 0.28 & S \\
\hline
\end{tabular}
\end{center}
\end{table}
The galaxy stellar mass within the effective radius is
calculated as:
\begin{leftsubeqnarray}
\slabel{eq:MiRea}
&& \frac{M_i}{{\rm M}_{10}}=
\frac L{{\rm L}_\odot}\frac{M_i/L}{10^{10}{\rm m}_\odot/
{\rm L}_\odot}~~; \\
\slabel{eq:MiReb}
&& \frac L{\rm L}_\odot=\exp_{10}\{-0.4
[I_T-(\hat{m}-\hat{M})-4.11]\}~~;
\label{seq:MiRe}
\end{leftsubeqnarray}
and the galaxy effective radius is
calculated as:
\begin{leftsubeqnarray}
\slabel{eq:Reffa}
&& \frac{R_e}{\rm kpc}=\frac{R_e}{\rm arcsec}\frac1{206.265}\frac d{\rm Mpc}
~~; \\
\slabel{eq:Reffb}
&& \frac d{\rm Mpc}=\exp_{10}\left[\frac{\hat{m}-\hat{M}}
5-5\right]~~;
\label{seq:Reff}
\end{leftsubeqnarray}
where the factor, 206.265, is
related to the choice of measure units.
For further details refer
to an earlier attempt (CV08).

The values of the reduced variables,
$\sX_{\rm V}$, $\sX_{\rm p}$, $\sX_{\rm T}$, are
determined via Eqs.\,(\ref{seq:phie})
and (\ref{eq:Xre}).   Both HH and HN
macrogases have been considered, for
the following values of parameters.
Scaled truncation radii (both finite
or infinite): $(\Xi_i,\Xi_j)
=(k_i,k_j)$; $k_i=5,10,20,+\infty$;
$k_j=5,10,20,+\infty$.   Fractional
masses: $m=10,20$.   Under the working
hypothesis of an analogy between VDW
gases and macrogases, the $({\sf O}
\sX_{\rm V}\sX_{\rm p})$ plane may be divided
into three parts: (i) a reversed
bell-shaped region where two phases,
gas and stars, coexist and the lower
point coincides with the critical
point (hereafter quoted as the GS
region); (ii) a region limited by the
left boundary of the reversed
bell-shaped region and the rising
side of the critical macroisothermal
curve, both branching off from the
critical point, where only stars
are present (hereafter quoted as
the S region); (iii)  a region limited
by the right boundary of the reversed
bell-shaped region and the rising
side of the critical macroisothermal
curve, both branching off from the
critical point, and the coordinate
axes, where only gas is present
(hereafter quoted as the G region).
If the density profiles are only
slightly affected in time, the
evolution of a galaxy on the $({\sf O}
\sX_{\rm V}\sX_{\rm p})$ plane is represented
by a track, starting from the G region
and ending within the GS or the S region.
The critical point is the sole which
is common to the three regions.

The position of a model galaxy
on the $({\sf O}\sX_{\rm V}\sX_{\rm p})$ plane
is affected by errors of different kind,
due to: (1) scatter around mean values
listed in Table \ref{t:dati}; (2) scatter
in fitting observed to model  density
profiles; (3) uncertainty on the
determination of the critical point.
The third contribution may safely be
neglected with respect to the other
ones.   Values related to the first
contribution are not completely found
in literature (to the knowledge of the
author).   The second contribution
could be determined using fitting
procedures, provided light distributions
are available and light traces stellar
mass in elliptical galaxies.   In summary,
error calculation on the position of
sample objects on the $({\sf O}\sX_{\rm V}\sX_{\rm p})$
plane would be cumbersome, and perhaps of
little meaning.
           
A notable simplification can be attained if
model galaxies instead of sample objects
are considered, in dealing with only
errors of the first kind mentioned above.
For fixed fractional mass, $m$, the
uncertainty on $\sX_{\rm p}=(m/m_{\rm c})^2$,
is negligible with respect to $\sX_{\rm V}=
y_{\rm c}/y$, which is determined by
solving Eq.\,(\ref{eq:cija}) for assigned
density profiles.   The combination of
Eqs.\,(\ref{eq:Heff}), (\ref{eq:B}), and
(\ref{eq:cijb}) yields an expression of
the dimensionless energy, $c_{ij}$, in terms
of observables listed in Table \ref{t:dati}, as:
\begin{leftsubeqnarray}
\slabel{eq:cijza}
&& c_{ij}=\frac1{1163.335}\frac1G\frac{\zeta_4}{\zeta_5}\exp_{10}[0.4\zeta_6-
0.2\zeta_7-6.644] \nonumber \\
&& \phantom{c_{ij}=}\cdot
\frac{[1-(1-\zeta_1)^2]^{1/2}/\sin i}{\arcsin\{[1-(1-\zeta_1)^2]^{1/2}/
\sin i\}}\left\{\frac{2\zeta_2^2}{\sin^2i}+\frac{(3-\delta)\zeta_3^2}
{1-\delta\cos^2i}\right\}~~; \\
\slabel{eq:cijzb}
&& \zeta_1=<\hat{e}_\bot>~~;\quad\zeta_2=\left[<\widetilde{v}_\|^2>^{1/2}
\right]_{\rm obs}~~;\quad\zeta_3=\left[<\sigma_\|^2>^{1/2}\right]_{\rm obs}~~;
\\
\slabel{eq:cijzc}
&& \zeta_4=\frac{R_e}{\rm arcsec}~~;\quad\zeta_5=\frac{M_i/L_i}{10^{10}{\rm m}
_\odot/{\rm L}_\odot}~~;\quad\zeta_6=I_T~~;\quad\zeta_7=\hat{m}-\hat{M}~~;
\qquad
\label{seq:cijz}
\end{leftsubeqnarray}
where some symbols have been changed to
gain simplicity.   An inspection of
Eq.\,(\ref{eq:cijza}) shows that the
dimensionless energy, $c_{ij}$, is
monotonically increasing with increasing
$\zeta_4$, $\zeta_6$, $\zeta_2$, $\zeta_3$,
and decreasing $\zeta_5$, $\zeta_7$; and
vice versa.   Establishing the trend with
the remaining variables, $\zeta_1$, $i$,
$\delta$, demands further considerations.

The function, $B(\epsilon)$, defined by
Eq.\,(\ref{eq:B}), is monotonically
decreasing in the domain, $0\le\epsilon
\le1$, where $\pi=B(0)\ge B(\epsilon)\ge
B(1)=2$.   Accordingly, $B(\epsilon)$ is
monotonically decreasing with decreasing
$\zeta_1=<\hat{e}_\bot>$ and/or decreasing
$i$, and vice versa.

With regard to the anisotropy parameter,
$\delta$, the following identity holds:
\begin{equation}
\label{eq:deli}
\frac{3-\delta}{1-\delta\cos^2i}=\left[\cos^2i+\frac{\sin^2 i-2\cos^2i}
{3-\delta}\right]^{-1}~~;
\end{equation}
and the special inclination angle, $i_0$,
which makes null the fraction within
brackets, is the solution of the equation:
\begin{equation}
\label{eq:i0eq}
\sin^2i_0-2\cos^2i_0=3\sin^2i_0-2=0~~;
\end{equation}
the result is:
\begin{equation}
\label{eq:i0}
\sin i_0=\sqrt{\frac23}~~;\qquad i_0=0.955\,316\,6=54.735\,61^\circ~~;
\end{equation}
accordingly, the fraction on the
left-hand side of Eq.\,(\ref{eq:deli})
is monotonically decreasing for increasing
$\delta$ in the range, $i_0<i\le\pi/2$,
and is monotonically increasing for increasing
$\delta$ in the range, $0\le i<i_0$,
while no dependence on $\delta$ occurs in the special
case, $i=i_0$.   Finally, Eqs.\,(\ref{eq:cija}),
(\ref{eq:deli}), and (\ref{eq:i0eq})
show that the dimensionless energy, $c_{ij}$,
is monotonically increasing or decreasing
for increasing $\delta$ according if $0\le
i<i_0$ or $i_0<i\le\pi/2$, respectively,
and vice versa, while no dependence
on $\delta$ occurs in
the special case, $i=i_0$.

The above results may be reduced to a
single relation, as:
\begin{lefteqnarray}
\label{eq:cijD}
&& c_{ij}\mp\Delta c_{ij}=\frac1{1163.335}\frac1G\frac{\zeta_4\mp\Delta
\zeta_4}{\zeta_5\pm\Delta\zeta_5} \nonumber \\
&& \phantom{c_{ij}\mp\Delta c_{ij}=}\cdot
\exp_{10}[0.4(\zeta_6\mp\Delta\zeta_6)-
0.2(\zeta_7\pm\Delta\zeta_7)-6.644] \nonumber \\
&& \phantom{c_{ij}\mp\Delta c_{ij}=}\cdot
\frac{\{1-[1-(\zeta_1\pm\Delta\zeta_1)]^2\}^{1/2}/\sin(i\pm\Delta i)}
{\arcsin\{\{1-[1-(\zeta_1\pm\Delta\zeta_1)]^2\}^{1/2}/\sin(i\pm\Delta i)\}}
\nonumber \\
&& \phantom{c_{ij}\mp\Delta c_{ij}=}\cdot
\left\{\frac{2(\zeta_2\mp\Delta\zeta_2)^2}{\sin^2(i\pm\Delta i)}+\frac
{\{3-[\delta\pm\sgn(i-i_0)\Delta\delta]\}(\zeta_3\mp\Delta\zeta_3)^2}
{1-[\delta\pm\sgn(i-i_0)\Delta\delta]\cos^2(i\pm\Delta i)}\right\};\qquad
\end{lefteqnarray}
where upper and lower signs correspond to
the lower and upper $c_{ij}$ value,
respectively, and
$\sgn$ is the sign function, defined as
$\sgn(x)=x/|x|$, $x\ne0$; $\sgn(0)=0$.   It is
worth emphasizing that Eq.\,(\ref{eq:cijD})
makes an exact formulation of the uncertainty
on the dimensionless energy, $c_{ij}$, as
defined by Eq.\,(\ref{eq:cija}).   On the
contrary, standard linear and quadratic
error propagation formulae apply to any
kind of functions allowing Taylor series
development, but are approximate instead
of being exact.

To the knowledge of the author, part of
the errors on the right-hand side of 
Eq.\,(\ref{eq:cijD}) are not available
in literature.   For the inclination
angle, $i$, and the anisotropy parameter,
$\delta$, the reason is in that they
depend on a reference dynamic model
(S\,IV; S\,X) and cannot be specified.
The following values are found:
$\Delta\zeta_4/\zeta_4=0.17$ (S\,IV),
0.20 (S\,X); $\Delta\zeta_6/\zeta_6=0.13$
(S\,IV); $\Delta\zeta_5/\zeta_5=0.10$,
strongly dependent on the assumptions
made (S\,IV); $\Delta\zeta_7=0.09$ - 0.33,
with a value listed for each sample
object (S\,IV).   In this view, it seems
better starting from assigned values
of the dimensionless energy relative
error, $\Delta c_{ij}/c_{ij}$, and
numerically evaluate the related
uncertainty on the position of a
selected model galaxy on the 
$({\sf O}\sX_{\rm V}\sX_{\rm p})$ plane,
to visualize the trend.

\subsection{Results}
\label{resu}

\begin{figure*}[t]
\begin{center}
\includegraphics[scale=0.8]{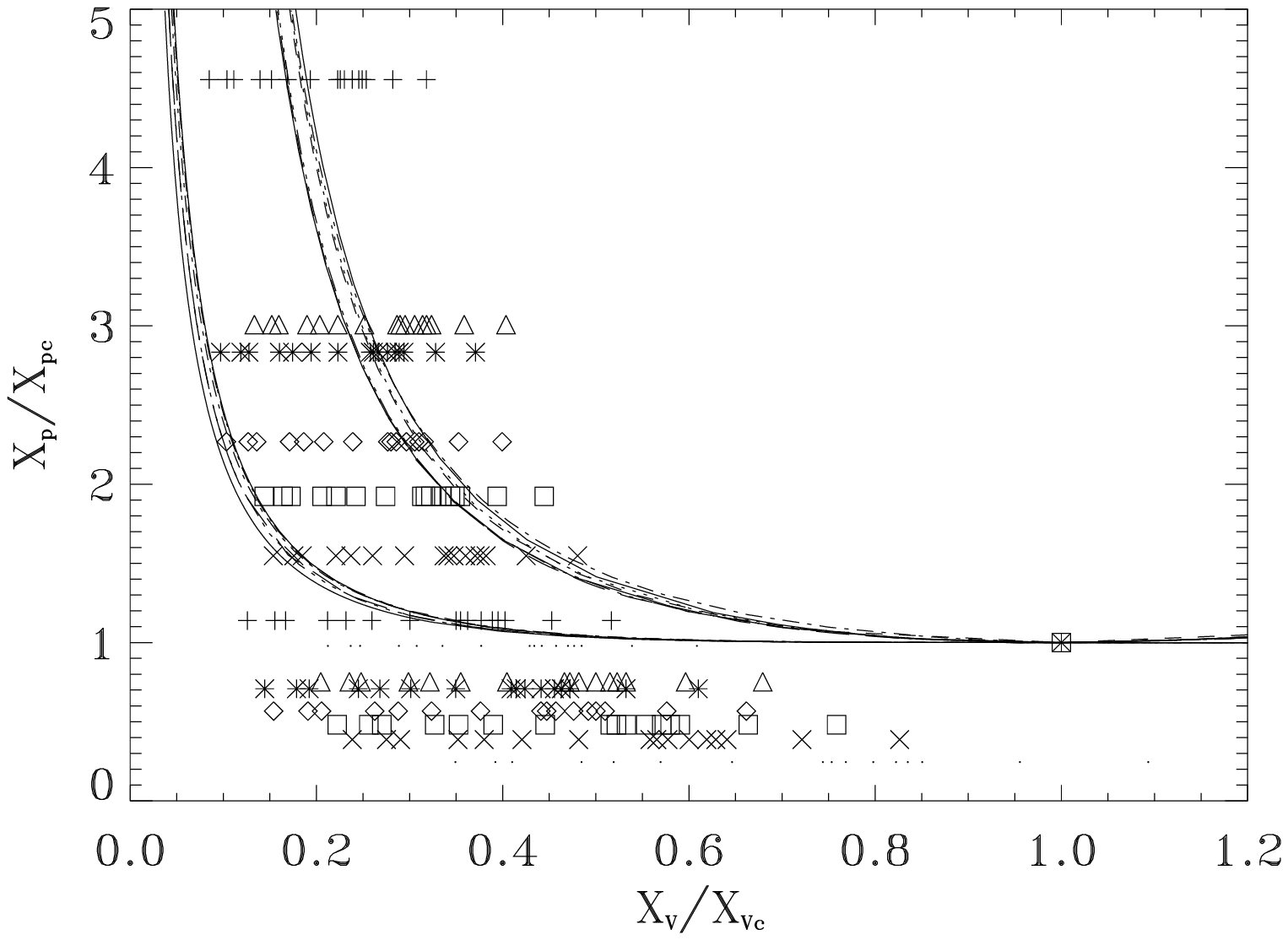}
\caption{Elliptical galaxies listed in Tables
\ref{t:dati} and \ref{t:pamo}, modelled as
HH macrogases for different choices of
scaled truncation radii, $\Xi_i$, $\Xi_j$,
and fractional mass, $m$.   The critical
macroisothermal curve (left) and the boundary
of the GS region (right) are also
plotted for each case.   Symbol caption
and line style: $(\Xi_i,\Xi_j)=(10,5)$ -
crosses, full; (10,10) - asterisks, dotted;
(10,20) - diamonds, dashed; (20,5) - triangles,
dot-dashed; (20,10) - squares, long-short-dashed;
(20,20) - St.\,Andrew's crosses, long-dashed;
$(+\infty,+\infty)$ - dots, full.   Lower and
upper symbols of the same kind are related to
$m=10,\,20$, respectively.   The composite
symbol marks the critical point.   Cases
where $\Xi_i=5$ make two galaxies unable
to be modelled, and for this reason are not
considered.}
\label{f:shhr}
\end{center}
\end{figure*}

With regard to HH macrogases, model
galaxies corresponding to sample
objects listed in Tables \ref{t:dati}
and \ref{t:pamo} are represented on
the  $({\sf O}\sX_{\rm V}\sX_{\rm p})$ plane of
Fig.\,\ref{f:shhr} for
different choices of scaled truncation
radii, $\Xi_i$, $\Xi_j$, and fractional
mass, $m$.   The critical macroisothermal
curve (left) and the boundary of the GS region
(right) are also plotted for each case.
The critical point is marked by a composite
symbol.   Two sample objects cannot be
modelled for low inner scaled truncation
radii $(\Xi_i=5)$ and, for this reason,
related cases are not considered.   Lower
and upper symbols of the same kind correspond to
$m=10,\,20$, respectively.

Under the working hypothesis of an analogy
between VDW gases and macrogases, modelled
elliptical galaxies are expected to lie in
the S region or  slightly outside the
S region within the GS region at most.   An
inspection of Fig.\,\ref{f:shhr}
shows the following: (1) model galaxies with
low fractional mass $(m=10)$ and/or no
truncation radii $(\Xi\to+\infty)$ lie
below the critical macroisothermal curve,
in the G region,
and for this reason cannot be accepted;
(2) about one half of model galaxies
with low outer scaled radii $(\Xi_j=5)$
lie well inside the GS region, and for
this reason cannot
be accepted; (3) more than one half of
model galaxies with larger scaled radii
($\Xi_i=10,\,20;$ $\Xi_j=10,\,20$) lie
within the S region,  and for this reason
are accepted.

\begin{figure*}[t]
\begin{center}
\includegraphics[scale=0.8]{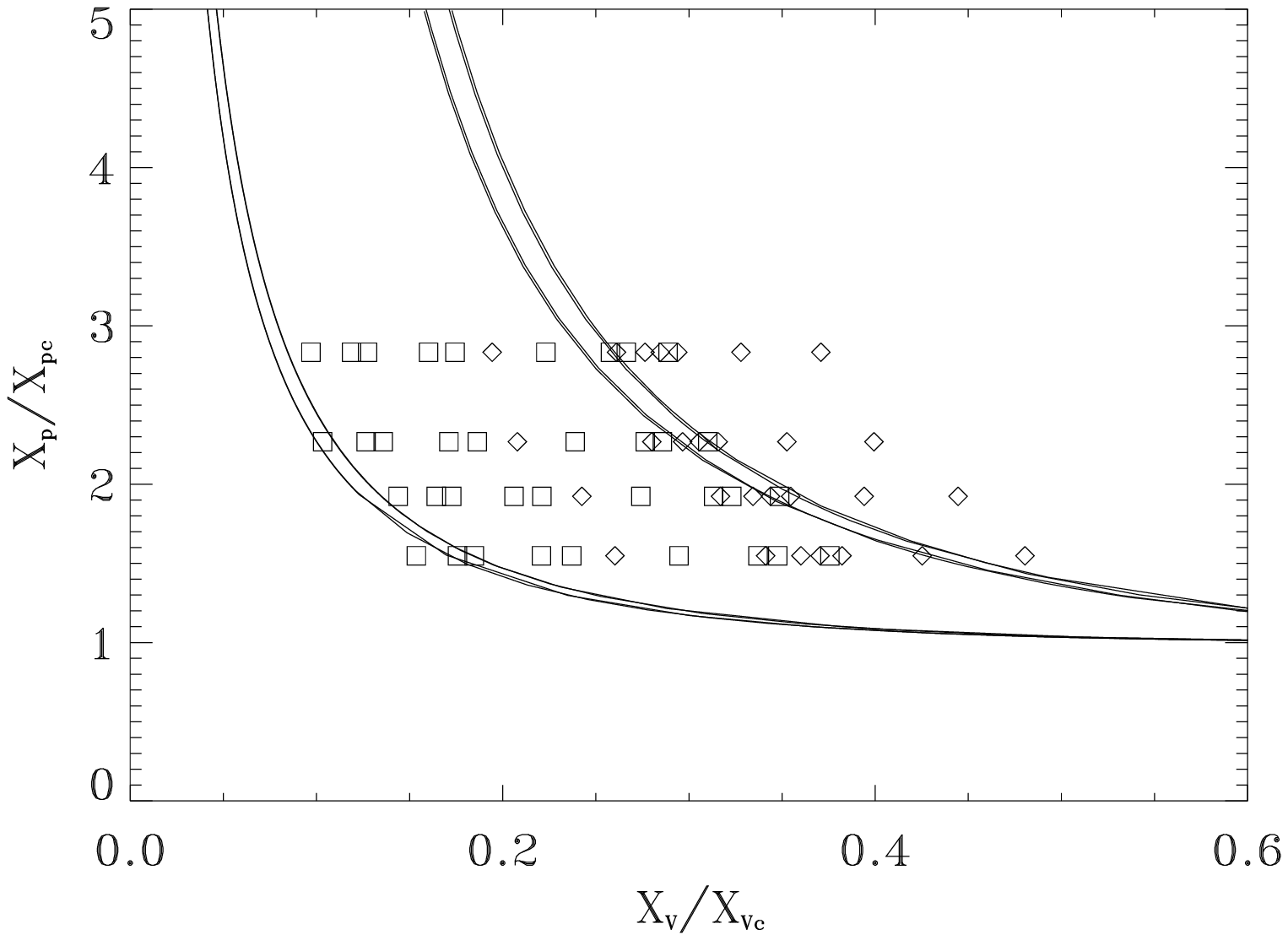}
\caption{Same as in Fig.\,\ref{f:shhr}
for scaled truncation radii, $(\Xi_i,
\Xi_j)=(10,10),$ (10, 20), (20, 10), (20, 20),
from top to bottom, and fractional mass,
$m=20$, where model galaxies are distinguished
according if their parent sample object is
classified as fast (squares) or slow (diamonds)
rotator.}
\label{f:shhb}
\end{center}
\end{figure*}

With regard to viable cases,
the plot of
Fig.\,\ref{f:shhr} is repeated in
Fig.\,\ref{f:shhb}, where model galaxies
are distinguished according if their
parent sample object is a fast (squares)
or a slow (diamonds) rotator.
The related scaled truncation radii
(from top to bottom) are $(\Xi_i,
\Xi_j)=(10, 10),$ (10, 20), (20, 10), (20, 20),
and the fractional mass is $m=20$.   The
curves are as in Fig.\,\ref{f:shhr}.

\begin{figure*}[t]
\begin{center}
\includegraphics[scale=0.8]{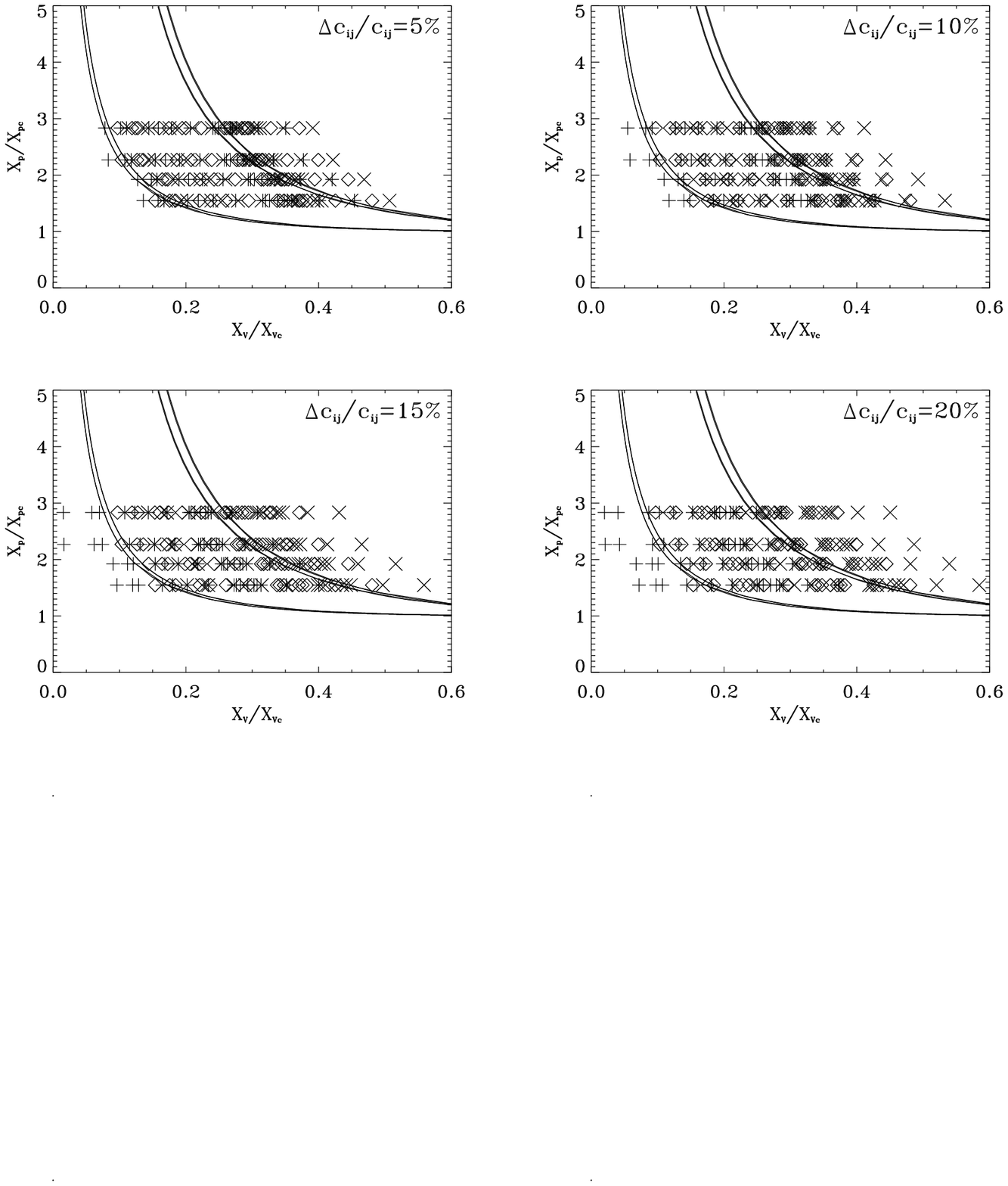}
\caption{Same as in Fig.\,\ref{f:shhr}
for different choices of
scaled truncation radii, $(\Xi_i,
\Xi_j)=(10,10)$, (10,20), (20,10),
(20,20), from top to bottom, and
fractional masses, $m=20$, where
dimensionless energies, $c_{ij}$,
defined by Eq.\,(\ref{eq:cija}),
are lowered to $c_{ij}-\Delta c_{ij}$
(Greek crosses) and increased to
$c_{ij}+\Delta c_{ij}$ (St.\,Andrew's
crosses) with respect to their original
values (diamonds), by a factor equal to
5\%, 10\%, 15\%, 20\%, respectively.
In the last case, a sample object
(NGC 4473) cannot be modelled for
lowered values, $c_{ij}-\Delta c_{ij}$,
and $(\Xi_i, \Xi_j)=(10,10)$, (10,20).}
\label{f:shhe}
\end{center}
\end{figure*}

Restricting to viable cases, the plot of
Fig.\,\ref{f:shhr} is repeated in
Fig.\,\ref{f:shhe}, where the effect of
assigned errors in dimensionless energy,
$\Delta c_{ij}/c_{ij}=5$\%, 10\%, 15\%,
20\%, labelled on each panel, on model
galaxies, is represented.
More specifically, the position of
model galaxies is marked by diamonds,
and the change due to lowered ($c_{ij}-
\Delta c_{ij}$) and increased ($c_{ij}+
\Delta c_{ij}$) dimensionless energy,
is marked by Greek and St.\,Andrew's
crosses, respectively.   Scaled
truncation radii are, from top to
bottom, $(\Xi_i, \Xi_j)=(10,10)$,
(10,20), (20,10), (20,20), and the
fractional mass is $m=20$ in all
cases.   In the special cases,  
$(\Xi_i, \Xi_j)=(10,10)$, (10,20),
and $\Delta c_{ij}/c_{ij}=20$\%, a
sample object (NGC 4473) cannot be
modelled for lowered dimensionless
energies, $c_{ij}-\Delta c_{ij}$.
In general, lowered and increased
dimensionless energies, $c_{ij}$,
make model galaxies shift on the
left and on the right, respectively,
from their position on the $({\sf O}
\sX_{\rm V}\sX_{\rm p})$ plane.   Accordingly, a
fraction of model galaxies enter the
G region and the GS region, respectively,
and the fit could be improved by
changing the input parameters,
$\Xi_i$, $\Xi_j$, and $m$.

\begin{figure*}[t]
\begin{center}
\includegraphics[scale=0.8]{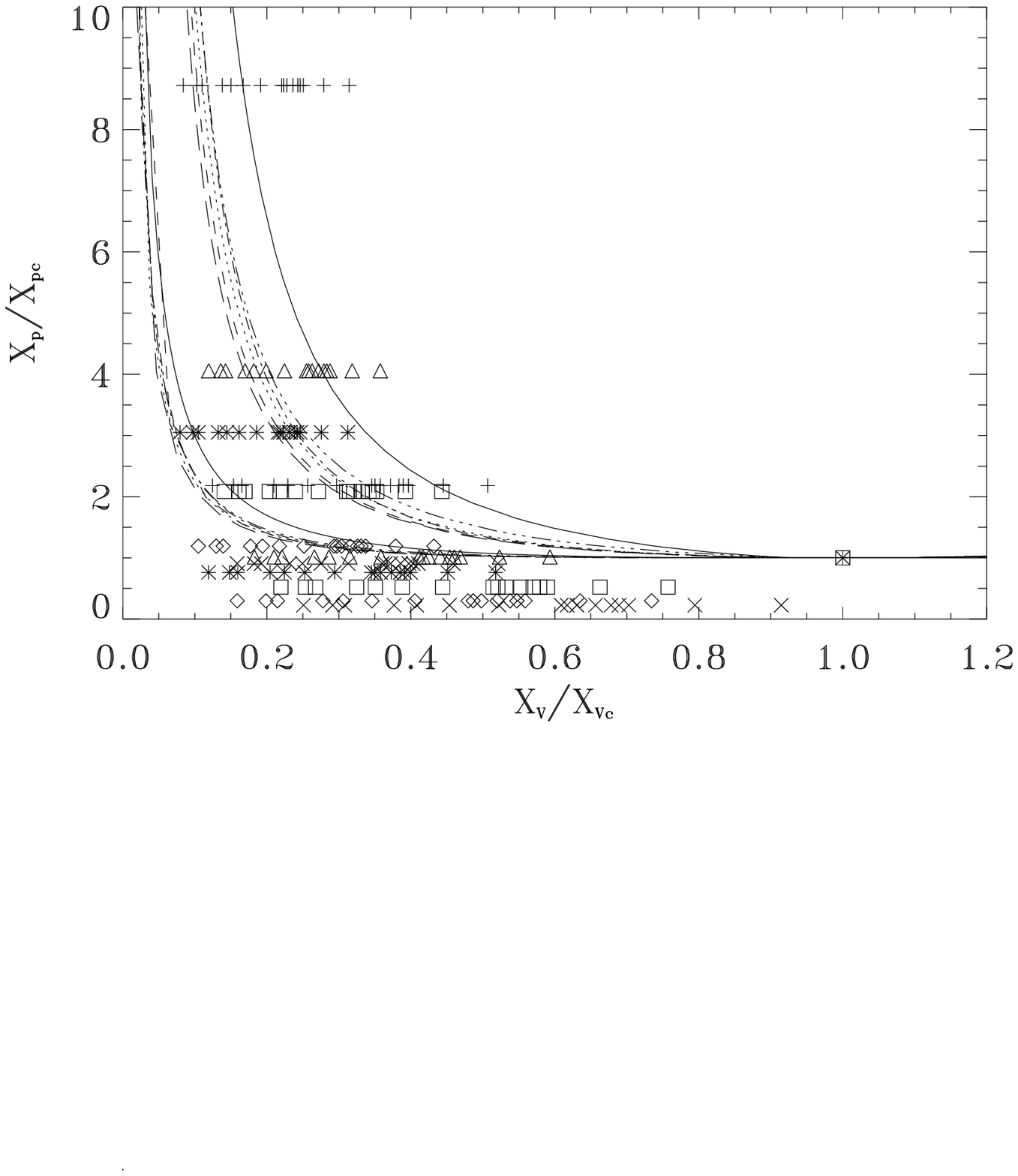}
\caption{Elliptical galaxies listed in Tables
\ref{t:dati} and \ref{t:pamo}, modelled as
HN/NH macrogases for different choices of
scaled truncation radii, $\Xi_i$, $\Xi_j$,
and fractional mass, $m$.   The critical
macroisothermal curve (left) and the boundary
of the GS region (right) are also
plotted for each case.
Symbol caption
and line style: $(\Xi_i,\Xi_j)=(10,5)$ -
crosses, full; (10,10) - asterisks, dotted;
(10,20) - diamonds, dashed; (20,5) - triangles,
dot-dashed; (20,10) - squares, long-short-dashed;
(20,20) - St.\,Andrew's crosses, long-dashed.
Lower and
upper symbols of the same kind are related to
$m=10,\,20$, respectively.   The composite
symbol marks the critical point.   Cases
where $\Xi_i=5$ make two galaxies unable
to be modelled, and for this reason are not
considered.   The same holds, to a larger
extent, for cases $(\Xi_i,\Xi_j)\to(+\infty,
+\infty)$, due to an infinite mass of the
NFW density profile.}
\label{f:shnr}
\end{center}
\end{figure*}

With regard to HN/NH macrogases, model
galaxies corresponding to sample
objects listed in Tables \ref{t:dati}
and \ref{t:pamo} are represented on
the  $({\sf O}\sX_{\rm V}\sX_{\rm p})$ plane of
Fig.\,\ref{f:shnr} for
different choices of scaled truncation
radii, $\Xi_i$, $\Xi_j$, and fractional
mass, $m$.   The critical macroisothermal
curve (left) and the boundary of the GS region
(right) are also plotted for each case.
The critical point is marked by a composite
symbol.   Two sample objects cannot be
modelled for low inner scaled truncation
radii $(\Xi_i=5)$ and, for this reason,
related cases are not considered.
The same holds, to a larger
extent, for cases $(\Xi_i,\Xi_j)\to(+\infty,
+\infty)$, due to an infinite mass of the
NFW density profile.   Lower
and upper symbols of the same kind correspond
to $m=10,\,20$, respectively.

Under the working hypothesis of an analogy
between VDW gases and macrogases, modelled
elliptical galaxies are expected to lie in
the S region or  slightly outside the
S region within the GS region at most.   An
inspection of Fig.\,\ref{f:shnr} shows
the following: (1) model galaxies
with low fractional mass $(m=10)$ and/or large
outer scaled truncation radius $(\Xi_j=20)$ lie
(at least partially)
below the critical macroisothermal curve
in the G region, and for this reason the
related cases cannot be accepted;
(2) more than one half of model galaxies
with large fractional mass $(m=20)$ and/or low
outer scaled truncation radius $(\Xi_j=5)$ lie
well inside the GS region, and for
this reason the related cases cannot
be accepted; (3) more than one half of
model galaxies with large fractional
mass $(m=20)$ and outer scaled radius
($\Xi_j=10$), or low fractional mass
$(m=10)$ and outer scaled radius ($\Xi_j=5$) lie
within the S region,  and for this reason
the related cases are accepted.

\begin{figure*}[t]
\begin{center}
\includegraphics[scale=0.8]{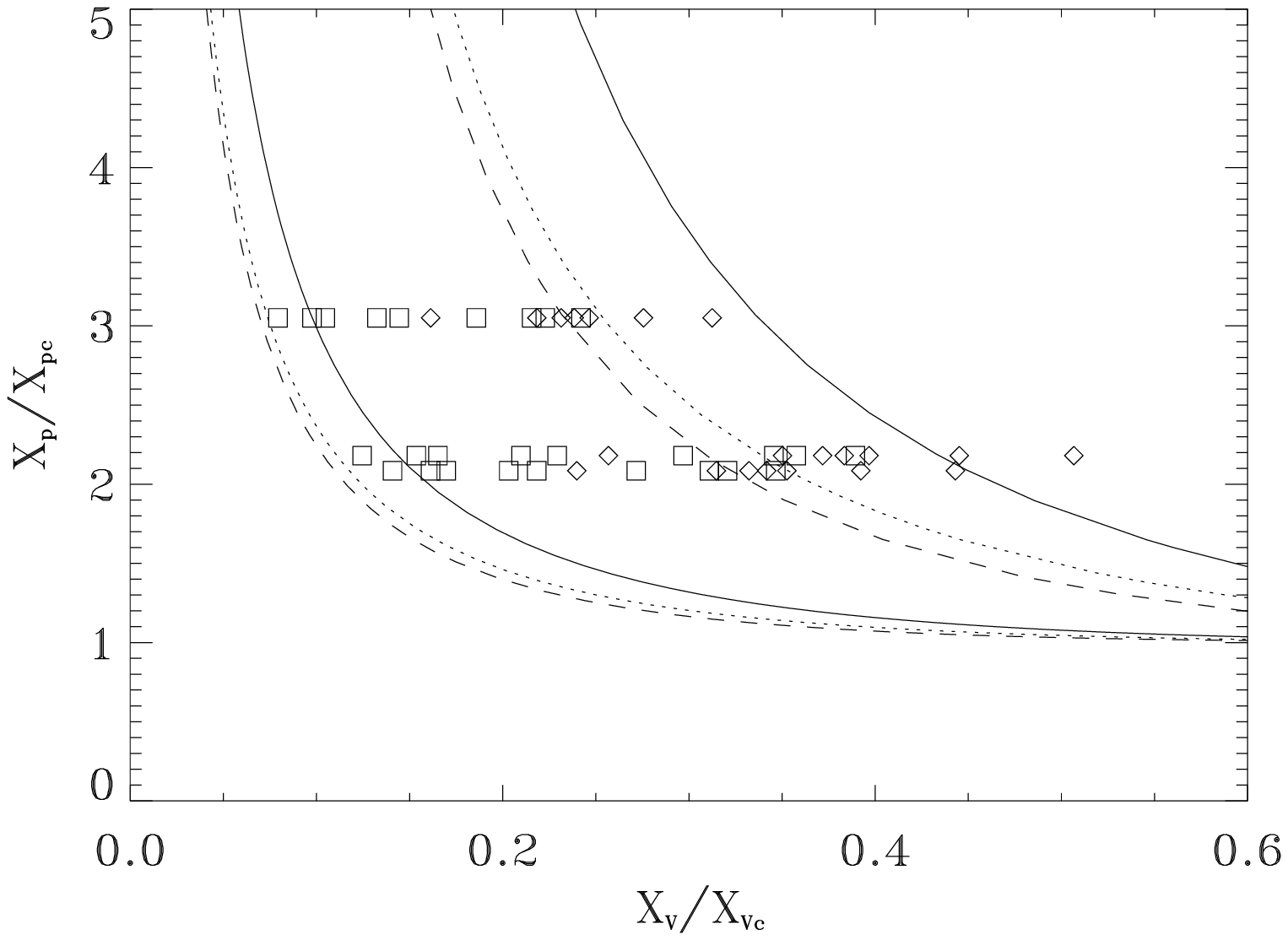}
\caption{Same as in Fig.\,\ref{f:shnr}
for scaled truncation radii, $(\Xi_i,
\Xi_j)=(10,10),$ (10, 5), (20, 10), 
from top to bottom, and fractional mass,
$m=20$, $\Xi_j=10$, and $m=10$, $\Xi_j=5$,
where model galaxies are distinguished
according if their parent sample object is
classified as fast (squares) or slow (diamonds)
rotator.}
\label{f:shnb}
\end{center}
\end{figure*}

With regard to viable cases,
the plot of
Fig.\,\ref{f:shnr} is repeated in
Fig.\,\ref{f:shnb}, where model galaxies
are distinguished according if their
parent sample object is a fast (squares)
or a slow (diamonds) rotator.
The related scaled truncation radii
(from top to bottom) are $(\Xi_i,
\Xi_j)=(10, 10),$ (10, 5), (20, 10), 
and the fractional mass is $m=20$ for
$\Xi_j=10$, and $m=10$ for $\Xi_j=5$.   The
curves are as in Fig.\,\ref{f:shnr}.

\begin{figure*}[t]
\begin{center}
\includegraphics[scale=0.8]{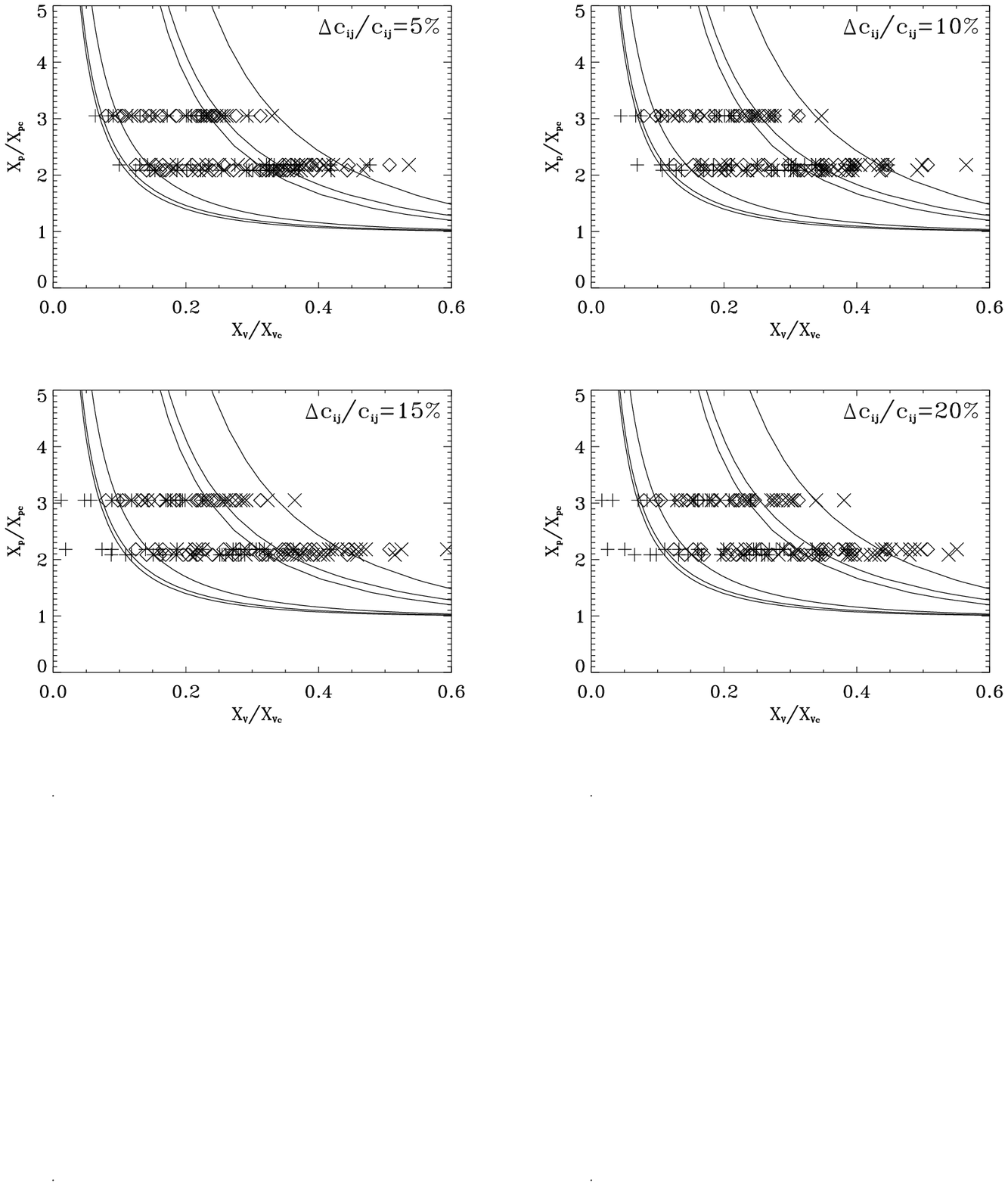}
\caption{Same as in Fig.\,\ref{f:shnr}
for different choices of
scaled truncation radii and fractional
masses, $(\Xi_i,\Xi_j,m)= (10,10,20)$,
(10,5,10), (20,10,20), from top to bottom,
where dimensionless energies, $c_{ij}$,
defined by Eq.\,(\ref{eq:cija}),
are lowered to $c_{ij}-\Delta c_{ij}$
(Greek crosses) and increased to
$c_{ij}+\Delta c_{ij}$ (St.\,Andrew's
crosses) with respect to their original
values (diamonds), by a factor equal to
5\%, 10\%, 15\%, 20\%, respectively.
In the last case, a sample object
(NGC 4473) cannot be modelled for
lowered values, $c_{ij}-\Delta c_{ij}$,
and $(\Xi_i, \Xi_j,m)=(10,5,10)$, (10,10,20),
and a sample object (NGC 4486) is out of scale
on the right for increased values, $c_{ij}+
\Delta c_{ij}$, and $(\Xi_i, \Xi_j,m)=(10,5,10)$.}
\label{f:shne}
\end{center}
\end{figure*}

Restricting to viable cases, the plot of
Fig.\,\ref{f:shnr} is repeated in
Fig.\,\ref{f:shne}, where the effect of
assigned errors in dimensionless energy,
$\Delta c_{ij}/c_{ij}=5$\%, 10\%, 15\%,
20\%, labelled on each panel, on model
galaxies, is represented.
More specifically, the position of
model galaxies is marked by diamonds,
and the change due to lowered ($c_{ij}-
\Delta c_{ij}$) and increased ($c_{ij}+
\Delta c_{ij}$) dimensionless energy,
is marked by Greek and St.\,Andrew's
crosses, respectively.   Scaled
truncation radii and fractional masses
are $(\Xi_i, \Xi_j,m)=(10,10,20)$,
(10,5,10), (20,10,20), from top to bottom.

In the special case,
$(\Xi_i, \Xi_j,m)=(10,5,10)$,
and $\Delta c_{ij}/c_{ij}=20$\%, a
sample object (NGC 4486) is out of
scale on the right for increased
dimensionless
energies, $c_{ij}+\Delta c_{ij}$.
In the special cases,
$(\Xi_i, \Xi_j,m)=(10,5,10)$, (10,10,20),
and $\Delta c_{ij}/c_{ij}=20$\%, a
sample object (NGC 4473) cannot be
modelled for lowered dimensionless
energies, $c_{ij}-\Delta c_{ij}$.
In general, lowered and increased
dimensionless energies, $c_{ij}$,
make model galaxies shift on the
left and on the right, respectively,
from their position on the $({\sf O}
\sX_{\rm V}\sX_{\rm p})$ plane.   Accordingly, a
fraction of model galaxies enter the
G region and the GS region, respectively,
and the fit could be improved by
changing the input parameters,
$\Xi_i$, $\Xi_j$, and $m$.

\subsection{Discussion}
\label{disc}

Current cosmological models imply
large-scale celestial bodies, such
as galaxies and clusters of galaxies,
are embedded within nonbaryonic dark
haloes, where the two subsystems interact
only via gravitation.   Accordingly,
large-scale celestial bodies can be
modelled as macrogases where macrovolume,
macropressure, and macrotemperature can
be defined, and a counterpart of the
VDW theory for ordinary gases can be developed.
Sufficiently steep density profiles, as
in HH and HN/NH macrogases, show a similar
trend with respect to VDW gases: the
macroisothermal curves are nonmonotonic
with two extremum points (one maximum
and one minimum) above a critical
macrotemperature, and are monotonic below
a critical macrotemperature.   The critical
macroisothermal curve is characterized by a
single extremum point, where the maximum and
the minimum coincide yielding a horisontal
inflexion point.   On the other hand,
sufficiently mild density profiles, such
as in UU and PP (CV08) macrogases, show
only nonmonotonic macroisothermal curves with
two extremum points (one maximum and one
minimum) where no critical macroisothermal
curve exists.

A generic macrogas equation of state is
formulated in terms of dimensionless variables
normalized to critical values (or conveniently
chosen in absence of the critical point).
Similar to what has been done in dealing with
the reduced VDW equation for ordinary
gases (e.g., LL67,
Chap.\,VIII, \S85), the states of two
large-scale celestial bodies with equal
$\sX_{\rm V}$, $\sX_{\rm p}$, $\sX_{\rm T}$,
or $y$, $m$, $\phi$, for assigned scaled
truncation radii, $\Xi_i$, $\Xi_j$, and
belonging to the same family of macrogases, may
be defined as corresponding states.   The
mere existence of a macrogas equation of
state yields the following result.
\begin{trivlist}
\item[\hspace\labelsep{\bf Law of corresponding
states.}] \sl
Given two large-scale celestial bodies
belonging to the same family of macrogases
with assigned scaled truncation radii, $\Xi_i$,
$\Xi_j$, the equality between two among
three reduced variables,
$\sX_{\rm V}$, $\sX_{\rm p}$, $\sX_{\rm T}$,
or $y$, $m$, $\phi$,  implies the equality between the
remaining related reduced variables i.e. the
two macrogases are in corresponding states.
\end{trivlist}
The law of corresponding states cannot be
extended to macrogases with different scaled
truncation radii, as shown in Figs.\,\ref{f:hhar}
and \ref{f:hnar}.   A possible explanation may
be the following.   Contrary to ordinary gases,
bounded by rigid walls which have no influence
on the equation of state, macrogases are
confined by ``gravitational'' walls 
appearing in the equation of state via the
potential energy terms which, in turn,
depend on the scaled truncation radii.

Ordinary gases exhibit monotonic isothermal
curves where the central part of the related
VDW isothermal curve, including the extremum
points, is replaced by a horisontal line and
a phase transition occurs therein.   With
regard to macrogases, the existence of a phase
transition and monotonic macroisothermal
curves of the kind considered, must necessarily
be assumed as a working hypothesis by analogy
with VDW gases.   The phase transition must be
conceived between gas and stars, and the
$({\sf O}\sX_{\rm V}\sX_{\rm p})$ plane may be
divided into three
parts, namely (i) the G region, where only gas
exists; (ii) the S region, where only stars
exist; (iii) the GS region, where both gas and
stars exist.   In this view, model elliptical
galaxies are expected to lie within the S region
or slightly outside the boundary between the
S and the GS region at most.

It is the case for different models related to
both HH and HN/NH macrogases, where acceptable
values of scaled truncation radii and fractional
masses are used, as shown in Figs.\,\ref{f:shhr}
and \ref{f:shnr}.   The assumption of universal
fractional mass for sample objects is not a
limit of the model, which equally holds assigning
different fractional masses to different sample
objects.

It can be seen from Figs.\,\ref{f:shhb}
and \ref{f:shnb} that fast rotators
lie within the S region, while slow
rotators are close (from both sides)
to the boundary between the S and the GS region.
This dichotomy could be interpreted
by the different nature of the two
classes of sample objects.   More
specifically, fast rotators seem
consistent with elliptical galaxies
with disky isophotes, which experienced
minor mergers and accreted a significant
amount of gas (S\,X), suddenly turned into
stars.   On the other hand, systematically
more massive (with the exception of NGC
4458 and, marginally, NGC 3608, see Table
\ref{t:pamo}) slow rotators may be related
to elliptical galaxies with boxy isophotes,
which experienced major gas-rich mergers,
or sequences of mergers, and regulation by
the feedback of a powerful central active
galactic nucleus (S\,X), in some cases
allowing gas survival as e.g., diffuse,
still undetected interstellar medium.

Accordingly, in the $({\sf O}\sX_{\rm V}\sX_{\rm p})$
plane fast rotators and a fraction of
slow rotators are expected to lie within
the S region, and the remaining part of
slow rotators to be placed in the GS
region, as shown in Figs.\,\ref{f:shhb}
and \ref{f:shnb}.
The following trend is also exhibited:
fast rotators are systematically on the
left with respect to slow rotators, with
the exception of NGC 4458, which is the
sole sample slow rotator with low mass,
possibly due to having experienced minor
instead of major mergers.   In this view,
NGC 4458 should be considered as a peculiar
fast rotator, where low rotation could be
due to special configurations e.g., still
undetected counter-rotating components,
as observed in the disk of NGC 4450 (e.g.,
S\,X).

The reduced variable, $\sX_{\rm V}=X_{\rm V}/X_
{{\rm V}_{\rm c}}$,
via Eqs.\,(\ref{eq:ymb}) and (\ref{eq:Xre}) is
proportional to the fractional
truncation radius, $1/y=R_i/R_j$.
The above mentioned dichotomy, exhibited
by fast and slow rotators in the 
$({\sf O}\sX_{\rm V}\sX_{\rm p})$ plane, implies a
larger fractional truncation radius, $y=R_j/R_i$,
for fast rotators with respect to slow
rotators.   This result could be interpreted
as due to different formation mechanisms:
minor mergers would produce larger contraction
of the baryonic matter, while major mergers
would make (possibly via active galactic nuclei)
smaller contraction of the baryonic matter,
yielding a larger or smaller fractional radius,
respectively.

\section{Conclusion}\label{conc}

In the current attempt, two-component
large-scale celestial bodies where the
subsystems interact only via
gravitation, are conceived as
macrogases bounded by ``gravitational''
walls.   The macrogas equation of state
is formulated in terms of macrovolume,
macropressure, and macrotemperature,
which are dimensionless variables.
For sufficiently steep density profiles,
which fit to observed elliptical galaxies
(or more generally, spheroid components)
and to simulated nonbaryonic dark matter
haloes, macroisothermal curves on the
$({\sf O}\sX_{\rm V}\sX_{\rm p})$ plane
show an analogy with VDW isothermal
curves exhibited by VDW gases.   More
specifically, a critical macroisothermal
curve exists, below and above which the
macroisothermal curves are monotonic and
nonmonotonic (with two extremum points,
one maximum and one minimum), respectively.
The critical macroisothermal curve is
characterized by a single extremum point
(a horisontal inflexion point), which is
the critical point.

Contrary to ordinary gases, macrogases
cannot be tested in laboratory, and for
this reason a working hypothesis is
inescapable.   By analogy with ordinary
gases, real macroisothermal curves are
supposed to occur instead of their
theoretical counterparts (deduced from
the macrogas equation of state), where
the central part containing the extremum
points is replaced by a horisontal line
along which a phase transition takes
place.   The intersection between a
selected theoretical macroisothermal
curve and its real counterpart, yields
two regions of equal area.   The phase
transition is assumed to be gas-stars
instead of vapour-liquid as in ordinary
gases.   Accordingly, the first quadrant
of the $({\sf O}\sX_{\rm V}\sX_{\rm p})$
plane is divided into three parts, namely
(i) the G region, where only gas exists;
(ii) the S region, where only stars exist;
(iii) the GS region, where both gas and
stars exist.

For selected density profiles and
scaled truncation radii, $\Xi_i$,
$\Xi_j$, the macrogas equation of
state depends on three parameters,
$\sX_{\rm V}$, $\sX_{\rm p}$,
$\sX_{\rm T}$, or the fractional
truncation radius, $y$, the fractional
mass, $m$, and the fractional energy,
$\phi$.   If elliptical galaxies and
their hosting nonbaryonic dark haloes
are conceived as macrogases, a selected
model is accepted only if a whole set
of sample objects (from which radii,
masses, and rsm velocities can be
determined) lies within the S region
or slightly outside the boundary between the
S and the GS region at most.   The sample
used (CV08, $N=16$) is extracted from
larger samples of early-type galaxies
investigated within the SAURON project
(S\,IV, $N=25$; S\,X, $N=48$).   The
position of model galaxies on the
$({\sf O}\sX_{\rm V}\sX_{\rm p})$
plane is determined through the
following steps: (i) select SAURON
data of interest; (ii) calculate the
parameters appearing in the virial
equations; (iii) make a correspondance
between model galaxies and sample
objects; (iv) represent model galaxies
on the $({\sf O}\sX_{\rm V}\sX_{\rm p})$
plane.

The main results found in the present
investigation may be summarized as
follows.
\begin{description}
\item[(1)]
A new numerical algorithm has been
used for determining the critical
point of selected HH and HN/NH
macrogases, improving earlier
results (CV08).   In particular,
the critical point exists for all
the cases considered.
\item[(2)]
A principle of corresponding states
rigorously holds for selected density
profiles and scaled truncation radii,
and to a first extent only for selected
density profiles.
\item[(3)]
The following models (on a total of 20)
can be accepted in the above mentioned
sense: $(\Xi_i,\Xi_j,m)=(10, 10, 20)$,
$(10, 20, 20)$, $(20, 10, 20)$,
$(20, 20, 20)$, with regard to HH
macrogases, and $(\Xi_i,\Xi_j,m)=
(10, 5, 10)$, $(10, 10, 20)$,
$(20, 10, 20)$, for HN/NH macrogases.
The values of model parameters may be
changed by the occurrence of systematic
errors.
\item[(4)]
Fast rotators exhibit larger fractional
truncation radii with respect to slow
rotators, which makes the former lie
within the S region and the latter
close (from both sides) to the boundary
between the S and the GS region, with
regard to acceptable models.   This
dichotomy could be interpreted in
terms of a different evolution
related to fast and slow rotators,
where gas is currently absent in
the former and can be present
(even if still undetected) in the
latter.
\end{description}

\section{Acknowledgements}
The author is indebted to an anonymous referee for helpful
comments which  improved an
earlier version of the manuscript.
Thanks are due to T. Valentinuzzi for fruitful discussions.

\appendix
\section*{Appendix}

\section{Tidal potential energy profile factors}
\label{a:prof}

The tidal potential energy for homeoidally striated
ellipsoids related to similar and similarly
placed boundaries, depends on the reference fractional mass,
$m^\dagger$, the fractional scaling radius, $y^\dagger$,
and the functions, $w^{({\rm int})}(\eta)$,
$w^{({\rm ext})}(\eta)$, $\eta=\Xi_i/y^\dagger=\Xi_j/y$,
expressed by Eqs.\,(\ref{seq:wie}).   Let the two
subsystems be denoted as A and B regardless of what
is the outer and what is the inner.   Conformingly,
Eqs.\,(\ref{seq:ym}) read:
\begin{leftsubeqnarray}
\slabel{eq:abbaa}
&& \frac{\Xi_B}{\Xi_A}=\frac{y_{BA}}{y_{BA}^\dagger}~~;\quad
\eta_{BA}=\frac{\Xi_A}{y_{BA}^\dagger}=\frac{\Xi_B}{y_{BA}}~~;\quad
y_{BA}^\dagger=\frac{r_B^\dagger}{r_A^\dagger}~~;\quad
y_{BA}=\frac{R_B}{R_A}\ge1~~;\quad \nonumber \\
&& \xi_A=y_{BA}^\dagger\xi_B~~;\quad m_{BA}^\dagger=\frac{M_B^\dagger}
{M_A^\dagger}~~;\quad m_{BA}=\frac{M_B}{M_A}~~; \\
\slabel{eq:abbab}
&& \frac{\Xi_A}{\Xi_B}=\frac{y_{AB}}{y_{AB}^\dagger}~~;\quad
\eta_{AB}=\frac{\Xi_B}{y_{AB}^\dagger}=\frac{\Xi_A}{y_{AB}}~~;\quad
y_{AB}^\dagger=\frac{r_A^\dagger}{r_B^\dagger}~~;\quad
y_{AB}=\frac{R_A}{R_B}\ge1~~;\quad \nonumber \\
&& \xi_B=y_{AB}^\dagger\xi_A~~;\quad m_{AB}^\dagger=\frac{M_A^\dagger}
{M_B^\dagger}~~;\quad m_{AB}=\frac{M_A}{M_B}~~;
\label{seq:abba}
\end{leftsubeqnarray}
according if $R_B\ge R_A$ or $R_A\ge R_B$, respectively.

In dealing with sequences of configurations where the
scaled truncation radii, $\Xi_A$ and $\Xi_B$, are kept
unchanged, the combination of Eqs.\,(\ref{eq:abbaa})
and (\ref{eq:abbab}) yields:
\begin{equation}
\label{eq:yabba}
y_{BA}^\dagger y_{AB}^\dagger=y_{BA}y_{AB}~~;
\end{equation}
for any pair of configurations belonging to opposite
sides of the sequence, with respect to $y_{BA}=y_{AB}
=1$.

In the special case of equal scaled density profiles,
$F_A=F_B=F$, and equal scaled truncation radii,
$\Xi_A=\Xi_B=\Xi$, Eqs.\,(\ref{seq:wie}) reduce to:
\begin{leftsubeqnarray}
\slabel{eq:wibbaa}
&& w^{({\rm int})}(\Xi,y_{BA}^\dagger)=\int_0^{\Xi/y_{BA}^\dagger}F(\xi)
\frac{\diff F(y_{BA}^\dagger\xi)}{\diff\xi}\xi\diff\xi~~;
\quad y_{BA}^\dagger=y_{BA}\ge1~~; \\
\slabel{eq:wibbab}
&& w^{({\rm ext})}(\Xi,y_{BA}^\dagger)=\int_0^{\Xi/y_{BA}^\dagger}
F(y_{BA}^\dagger\xi)\frac{\diff F(\xi)}{\diff\xi}\xi\diff\xi~~;
\quad y_{BA}^\dagger=y_{BA}\ge1~~; \\
\slabel{eq:wibbac}
&& w^{({\rm int})}(\Xi,y_{AB}^\dagger)=\int_0^{\Xi/y_{AB}^\dagger}F(\xi)
\frac{\diff F(y_{AB}^\dagger\xi)}{\diff\xi}\xi\diff\xi~~;
\quad y_{AB}^\dagger=y_{AB}\ge1~~; \\
\slabel{eq:wibbad}
&& w^{({\rm ext})}(\Xi,y_{AB}^\dagger)=\int_0^{\Xi/y_{AB}^\dagger}
F(y_{AB}^\dagger\xi)\frac{\diff F(\xi)}{\diff\xi}\xi\diff\xi~~;
\quad y_{AB}^\dagger=y_{AB}\ge1~~;
\label{seq:wibba}
\end{leftsubeqnarray}
in the special case where the fractional scaling
radius coincides for both configurations,
$y_{BA}^\dagger=y_{AB}^\dagger$, the combination
of Eqs.\,(\ref{eq:wibbaa}) and (\ref{eq:wibbac});
(\ref{eq:wibbab}) and (\ref{eq:wibbad}); yields:
\begin{leftsubeqnarray}
\slabel{eq:abblaa}
&& w^{({\rm int})}(\Xi,y_{BA}^\dagger)=w^{({\rm int})}(\Xi,y_{AB}^\dagger)~~;
\qquad y_{BA}^\dagger=y_{AB}^\dagger~~; \\
\slabel{eq:abblab}
&& w^{({\rm ext})}(\Xi,y_{BA}^\dagger)=w^{({\rm ext})}(\Xi,y_{AB}^\dagger)~~;
\qquad y_{BA}^\dagger=y_{AB}^\dagger~~;
\label{seq:abbla}
\end{leftsubeqnarray}
regardless of the scaled truncation radius, $\Xi$.

If, on the other hand, $\Xi_A\ne\Xi_B$, the upper
integration limits are $\Xi_A/y_{BA}^\dagger$ and
$\Xi_B/y_{AB}^\dagger$ for Eqs.\,(\ref{eq:wibbaa})
and (\ref{eq:wibbab}); (\ref{eq:wibbac}) and
(\ref{eq:wibbad}); respectively.   Then $y_{BA}^
\dagger=y_{AB}^\dagger$ implies equal integrands
but different upper integration limits, while
$\Xi_A/y_{BA}^\dagger=\Xi_B/y_{AB}^\dagger$ implies
equal upper integration limits but different integrands,
with regard to Eqs.\,(\ref{eq:wibbaa}) and
(\ref{eq:wibbac}); (\ref{eq:wibbab}) and
(\ref{eq:wibbad}); respectively.   Accordingly,
Eqs.\,(\ref{eq:abblaa}) and (\ref{eq:abblab}) no
longer hold in the case under discussion, unless
$\Xi_A\to+\infty$, $\Xi_B\to+\infty$, which erases
the dependence on $\Xi_A$ or $\Xi_B$, regardless
of the value of $\lim_{(\Xi_A,\Xi_B)\to+\infty}
(\Xi_B/\Xi_A)$.

More specifically, Eqs.\,(\ref{seq:wibba}) where
$\Xi=\Xi_A,\Xi_B$; $y^\dagger=y_{BA}^\dagger,
y_{AB}^\dagger$; reduce to:
\begin{leftsubeqnarray}
\slabel{eq:wiibaa}
&& w^{({\rm int})}_\infty(y^\dagger)=\lim_{\Xi\to+\infty}w^{({\rm int})}
(\Xi,y^\dagger)=\int_0^{+\infty}F(\xi)
\frac{\diff F(y^\dagger\xi)}{\diff\xi}\xi\diff\xi~~; \nonumber \\
&& \phantom{w^{({\rm int})}_\infty(y^\dagger)=\lim_{\Xi\to+\infty}
w^{({\rm int})}(\Xi_A,y^\dagger)=}
y^\dagger=y\ge1~~;\qquad \\
\slabel{eq:wiibab}
&& w^{({\rm ext})}_\infty(y^\dagger)=\lim_{\Xi\to+\infty}w^{({\rm ext})}
(\Xi,y^\dagger)=\int_0^{+\infty}F(y^\dagger\xi)\frac{\diff F(\xi)}
{\diff\xi}\xi\diff\xi~~; \nonumber \\
&& \phantom{w^{({\rm ext})}_\infty(y^\dagger)=\lim_{\Xi_A\to+\infty}
w^{({\rm int})}(\Xi_A,y^\dagger)=}
y^\dagger=y\ge1~~;\qquad
\label{seq:wiiba}
\end{leftsubeqnarray}
replacing $y^\dagger$ with $1/y^\dagger$ yields:
\begin{leftsubeqnarray}
\slabel{eq:widbaa}
&& w^{({\rm int})}_\infty\left(\frac1{y^\dagger}\right)=\int_0^{+\infty}F(\xi)
\frac{\diff F(\xi/y^\dagger)}{\diff\xi}\xi\diff\xi~~;\qquad
y^\dagger=y\ge1~~; \\
\slabel{eq:widbab}
&& w^{({\rm ext})}_\infty\left(\frac1{y^\dagger}\right)
=\int_0^{+\infty}F\left(\frac\xi{y^\dagger}\right)\frac{\diff F(\xi)}
{\diff\xi}\xi\diff\xi~~;\qquad y^\dagger=y\ge1~~;\qquad
\label{seq:widba}
\end{leftsubeqnarray}
which, choosing $\xi/y^\dagger$ as integration variable, and
keeping in mind that integrals are independent of integration
variables, is equivalent to:
\begin{leftsubeqnarray}
\slabel{eq:wiebaa}
&& w^{({\rm int})}_\infty\left(\frac1{y^\dagger}\right)=y^\dagger\int_0^
{+\infty}F(y^\dagger\xi)
\frac{\diff F(\xi)}{\diff\xi}\xi\diff\xi~~;~\qquad
y^\dagger=y\ge1~~; \\
\slabel{eq:wiebab}
&& w^{({\rm ext})}_\infty\left(\frac1{y^\dagger}\right)
=y^\dagger\int_0^{+\infty}F\left(\xi\right)\frac{\diff F(y^\dagger\xi)}
{\diff\xi}\xi\diff\xi~~;\qquad y^\dagger=y\ge1~~;\qquad
\label{seq:wieba}
\end{leftsubeqnarray}
and the combination of Eqs.\,(\ref{eq:wiibaa}) and
(\ref{eq:wiebab}); (\ref{eq:wiibab}) and (\ref{eq:wiebaa});
produces:
\begin{leftsubeqnarray}
\slabel{eq:wifbaa}
&& w^{({\rm int})}_\infty\left(\frac1{y^\dagger}\right)=y^\dagger
w^{({\rm ext})}_\infty(y^\dagger)~~; \\
\slabel{eq:wifbab}
&& w^{({\rm ext})}_\infty\left(\frac1{y^\dagger}\right)=y^\dagger
w^{({\rm int})}_\infty(y^\dagger)~~;
\label{seq:wifba}
\end{leftsubeqnarray}
where, on the other hand, $1/y^\dagger=1/y\le1$ is
outside the domain, and the role of the two
components should be interchanged therein,
according to Eqs.\,(\ref{eq:wibbac}) and
(\ref{eq:wibbad}), which makes the above
result only mathematically relevant.

Turning to the special case, $F_A=F_B=F$,
$\Xi_A=\Xi_B=\Xi$, the last implying
$(\nu_A)_{\rm sel}=(\nu_B)_{\rm sel}=
\nu_{\rm sel}$, and taking, in addition,
$y_{BA}^\dagger=y_{AB}^\dagger=1$,
Eqs.\,(\ref{seq:wibba}) reduce to:
\begin{equation}
\label{eq:wigba}
w^{({\rm int})}(\Xi,1)=w^{({\rm ext})}(\Xi,1)=w(\Xi,1)~~;
\end{equation}
and the fractional virial potential energy,
expressed by Eq.\,(\ref{eq:phiea}), reduces to:
\begin{leftsubeqnarray}
\slabel{eq:phi1a}
&& \phi=m^\dagger\frac{m^\dagger\nu_{\rm sel}-(9/8)w(\Xi,1)}
{\nu_{\rm sel}-(9/8)m^\dagger w(\Xi,1)}~~;\qquad y^\dagger=1~~; \\
\slabel{eq:phi1b}
&& \nu_{\rm sel}=\frac9{16}\int_0^\Xi F^2(\xi)\diff\xi~~; \\
\slabel{eq:phi1c}
&& w(\Xi,1)=\int_0^\Xi F(\xi)\frac{\diff F}{\diff\xi}\xi\diff\xi~~; \\
\slabel{eq:phi1d}
&& F(\Xi)=0~~;
\label{seq:phi1}
\end{leftsubeqnarray}
where Eqs.\,(\ref{eq:phi1b}) and (\ref{eq:phi1d})
follow from the definition of $\nu_{\rm sel}$ and
$F(\xi)$, respectively.   For further details
refer to earlier attempts (e.g., Roberts, 1962;
Caimmi and Secco, 1992; CV08).

Integrating by parts Eq.\,(\ref{eq:phi1c}) and
combining with (\ref{eq:phi1b}) and (\ref{eq:phi1d}),
yields:
\begin{equation}
\label{eq:wnu}
w(\Xi,1)=-\frac89\nu_{\rm sel}~~;
\end{equation}
finally, substituting Eq.\,(\ref{eq:wnu}) into
(\ref{eq:phi1a}) produces:
\begin{equation}
\label{eq:phim}
\phi=m^\dagger~~;\qquad y^\dagger=1~~;
\end{equation}
which is a general result for subsystems
with equal scaled density profiles, $F_A
(\xi)=F_B(\xi)=F(\xi)$, $0\le\xi\le\Xi$.

\section{Dimensional macrogas equation of state}
\label{a:dime}

Let macrogases be defined as large-scale
collisionless fluids with the following properties:
(i) particles are identical mass points; (ii)
particle number is extremely large; (iii)
particle motions obey Newton laws of mechanics;
(iv) particle collisions are absent; (v) particle
interactions obey Newton law of gravitation.

Under the assumption of homeoidally striated
density profiles, by use of Eq.\,(\ref{eq:Esel}),
the virial theorem reads:
\begin{equation}
\label{eq:vis1}
2E_{\rm kin}-\nu_{\rm sel}\frac{G(M^\dagger)^2}{a_1^\dagger}B=0~~;
\end{equation}
let the macrovolume, $V_{\rm M}$, the macropressure,
$p_{\rm M}$, the macrotemperature, $T_{\rm M}$, and the
mass weighted rms velocity, $\sigma_{\rm M}$, be
defined as:
\begin{lefteqnarray}
\label{eq:VM}
&& V_{\rm M}=\frac{4\pi}3a_1a_2a_3~~; \\
\label{eq:pM}
&& p_{\rm M}=\frac{GM^2}{a_1^2a_2a_3}~~; \\
\label{eq:TM}
&& kT_{\rm M}=\frac23\frac1NE_{\rm kin}~~; \\
\label{eq:sM}
&& \sigma_{\rm M}=\left(\frac{2E_{\rm kin}}M\right)^{1/2}~~;
\end{lefteqnarray}
where $N$ is the total number of particles,
$k$ the Boltzmann's constant, and the index,
M, means macrogas.   In particular, Eq.\,(\ref
{eq:TM}) discloses that the macrotemperature,
$T_{\rm M}$, coincides with the temperature of an
ideal gas with particle number, $N$, and
translational kinetic energy, $E_{\rm kin}$.

Typical values for galaxies are $M=3\cdot10^{11}
{\rm m}_\odot$, $N=3\cdot10^{11}$, $\sigma_{\rm M}=
100\sqrt{3}\,{\rm km\,s}^{-1}$, which yields via
Eqs.\,(\ref{eq:TM}) and (\ref{eq:sM}):
$kT_{\rm M}=M\sigma_{\rm M}^2/(3N)=10^{14}\cdot1.99
\cdot10^{33}{\rm erg}=1.99\cdot10^{47}{\rm erg}$, and
$T_{\rm M}=1.99\cdot10^{47}/(1.38\cdot10^{-16}){\rm K}
=1.44\cdot10^{63}{\rm K}=1440\,{\rm K}_{\rm M}$, where
where ${\rm K}_{\rm M}=10^{60}{\rm K}$ is the macrodegree,
assumed as macrotemperature unit.

The combination of Eqs.\,(\ref{seq:rho})-(\ref{eq:Esel})
and (\ref{eq:vis1})-(\ref{eq:sM}) yields:
\begin{equation}
\label{eq:cof}
\frac{p_{\rm M}V_{\rm M}}{NkT_{\rm M}}=4\pi\frac{GM}{a_1\sigma_{\rm M}^2}=
4\pi\frac{(\nu_{\rm mas})^2}{\Xi\nu_{\rm sel}B}~~;
\end{equation}
which may be conceived as a compressibility factor.

In presence of a similar, similarly placed, homeoidally
striated density profile, by use of
Eqs.\,(\ref{eq:Esel})-(\ref{seq:Exxx}), the virial
theorem for the subsystem under consideration reads:
\begin{equation}
\label{eq:vir2T}
2(E_u)_{\rm kin}-\frac{G(M_u^\dagger)^2}{(a_u^\dagger)_1}(\nu_u)_{\rm sel}
\left[1+\frac{(\nu_{uv})_{\rm tid}}{(\nu_u)_{\rm sel}}\right]B~~;
\end{equation}
and the macrogas equation of state is:
\begin{leftsubeqnarray}
\slabel{eq:eqs2a}
&& (p_u)_{\rm M}(V_u)_{\rm M}=\frac{4\pi[(\nu_u)_{\rm mas}]^2}{\Xi_u
(\nu_u)_{\rm sel}B}\left[1+\frac{(\nu_{uv})_{\rm tid}}{(\nu_u)_{\rm sel}}
\right]^{-1}N_uk(T_u)_{\rm M}~~;  \\
\slabel{eq:eqs2b}
&& u=i,j~~;\qquad v=j,i~~;
\label{seq:eqs2}
\end{leftsubeqnarray}
where $i$ and $j$ denote the inner and the outer
subsystem, respectively.   The compressibility
factor is:
\begin{leftsubeqnarray}
\slabel{eq:cof2a}
&& \frac{(p_u)_{\rm M}(V_u)_{\rm M}}{N_uk(T_u)_{\rm M}}=4\pi\frac{GM_u}
{(a_u)_1[(\sigma_u)_{\rm M}]^2}=4\pi\frac{[(\nu_u)_{\rm mas}]^2}
{\Xi_u(\nu_u)_{\rm sel}B}\left[1+\frac{(\nu_{uv})_{\rm tid}}{(\nu_u)_
{\rm sel}}\right]^{-1}~~;\qquad  \\
\slabel{eq:cof2b}
&& u=i,j~~;\qquad v=j,i~~;
\label{seq:cof2}
\end{leftsubeqnarray}
in all cases, the effect of the tidal potential
is expressed by the sum within square brakets.
If the two subsystems were infinitely distant
one from the other, $(\nu_{uv})_{\rm tid}=0$,
and the above results reduce to their one-component
counterparts.

In this view, one-component macrogases should be
conceived as ``ideal'' and two-component
macrogases as ``VDW'', the related equations
of state resembling ideal and VDW gas equation of
state, respectively.   In fact, the mass ratio,
$m=M_j/M_i$, and the axis ratio, $y=(a_j)_1/(a_i)_1
=(a_j)_2/(a_i)_2=(a_j)_3/(a_i)_3$, appear in the
explicit expression of $(\nu_{uv})_{\rm tid}$.
Owing to Eqs.\,(\ref{eq:VM}) and (\ref{eq:pM}),
the following relations hold:
\begin{lefteqnarray}
\label{eq:rV}
&& \frac{(V_j)_{\rm M}}{(V_i)_{\rm M}}=\frac{(a_j)_1(a_j)_2(a_j)_3}
{(a_i)_1(a_i)_2(a_i)_3}=y^3~~; \\
\label{eq:rp}
&& \frac{(p_j)_{\rm M}}{(p_i)_{\rm M}}=\frac{(M_j)^2}{(M_i)^2}\frac
{(a_i)_1^2(a_i)_2(a_i)_3}{(a_j)_1^2(a_j)_2(a_j)_3}=\frac{m^2}{y^4}~~; \\
\label{eq:vphi1}
&& \frac{N_j(T_j)_{\rm M}}{N_i(T_i)_{\rm M}}=\frac{(E_j)_{\rm kin}}
{(E_i)_{\rm kin}}=\phi~~;
\label{eq:rT}
\end{lefteqnarray}
which show the profile factor, $(\nu_{uv})_{\rm tid}$,
depends on the fractional macrovolume and the fractional
macropressure, via $m$ and $y$.

The combination of Eqs.\,(\ref{eq:TM}), (\ref{seq:eqs2}),
(\ref{eq:rV}), and (\ref{eq:rp}) yields:
\begin{equation}
\label{eq:vphi2}
\frac{m^2}y=\frac{\Xi_i}{\Xi_j}\left[\frac{(\nu_j)_{\rm mas}}{(\nu_i)_
{\rm mas}}\right]^2\frac{(\nu_i)_{\rm sel}+(\nu_{ij})_{\rm tid}}
{(\nu_j)_{\rm sel}+(\nu_{ji})_{\rm tid}}\phi~~;
\end{equation}
which, using Eqs.\,(\ref{seq:wie}), may be cast under the equivalent form:
\begin{leftsubeqnarray}
\slabel{eq:rega}
&& X_{\rm p}X_{\rm V}\displayfrac{1-\frac98\frac{\Xi_i}{\Xi_j}\frac{(\nu_j)_
{\rm mas}}{(\nu_i)_{\rm mas}}\frac1{(\nu_j)_{\rm sel}}\frac{w^{({\rm int})}
(X_{\rm V},\Xi_j)}{X_{\rm p}^{1/2}X_{\rm V}^{1/3}}}{1-\frac98\frac{(\nu_i)_
{\rm mas}}{(\nu_j)_{\rm mas}}\frac1{(\nu_i)_{\rm sel}}X_{\rm p}^{1/2}X_
{\rm V}^{2/3}w^{({\rm ext})}(X_{\rm V},\Xi_j)}=K(\Xi_i,\Xi_j)X_{\rm T}~~;
\qquad \\
\slabel{eq:regb}
&& K(\Xi_i,\Xi_j)=\frac{\Xi_i}{\Xi_j}\left[\frac{(\nu_j)_{\rm mas}}
{(\nu_i)_{\rm mas}}\right]^2\frac{(\nu_i)_{\rm sel}}{(\nu_j)_{\rm sel}}~~; \\
\slabel{eq:regc}
&& X_{\rm p}=\frac{(p_j)_{\rm M}}{(p_i)_{\rm M}}=\frac{m^2}{y^4}~~;\quad
X_{\rm V}=\frac{(V_j)_{\rm M}}{(V_i)_{\rm M}}=y^3~~;\quad
X_{\rm T}=\frac{N_j(T_j)_{\rm M}}{N_i(T_i)_{\rm M}}=\phi~~;\qquad
\label{seq:reg}
\end{leftsubeqnarray}
that is the macrogas fractional equation of state.
A simpler and more intuitive choice, adopted
in the text, is $X_{\rm p}=m^2$, $X_{\rm V}=1/y$, but the
connection with the fractional macropressure
and fractional macrovolume is lost in this case.

\end{document}